\documentclass[twocolumn,superscriptaddress,showpacs,nofootinbib]{revtex4}

\usepackage{bm,color}

\usepackage{graphicx}
\usepackage{amsmath}
\usepackage{amssymb}
\usepackage{amsfonts}
\usepackage{bm}
\usepackage{color}

\def\be{\begin{equation}}
\def\ee{\end{equation}}
\def\bea{\begin{eqnarray}}
\def\eea{\end{eqnarray}}

\begin{document}

\title{Models of dark matter halos based on statistical mechanics: \\
I. The classical King model}

\author{Pierre-Henri Chavanis}
\affiliation{Laboratoire de Physique Th\'eorique, Universit\'e Paul
Sabatier, 118 route de Narbonne 31062 Toulouse, France}
\author{Mohammed Lemou}
\affiliation{CNRS and IRMAR, Universit\'e de Rennes 1 and INRIA-Rennes Bretagne Atlantique, France}
\author{Florian M\'ehats}
\affiliation{CNRS and IRMAR, Universit\'e de Rennes 1 and INRIA-Rennes Bretagne Atlantique, France}

\begin{abstract}
We consider the possibility that dark matter halos are described by the
Fermi-Dirac distribution at finite temperature. This is the case if dark matter
is a self-gravitating quantum gas made of massive neutrinos at statistical
equilibrium. This is also the case if dark matter can be treated  as a
self-gravitating  collisionless gas experiencing Lynden-Bell's type of violent
relaxation. In order to avoid the infinite mass problem and carry out a rigorous
stability analysis, we consider the fermionic King model. In this paper, we
study the non-degenerate limit leading to the classical King model. This model
was initially introduced to describe globular clusters. We propose to apply it
also to large dark matter halos where quantum effects are negligible. We
determine the caloric curve and study the thermodynamical stability of the
different configurations.  Equilibrium states exist only above a critical energy
$E_c$ in the microcanonical ensemble and only above a critical temperature $T_c$
in the canonical ensemble. For $E<E_c$, the system undergoes a gravothermal
catastrophe and, for $T<T_c$, it undergoes an isothermal collapse. We compute
the profiles of density, circular velocity, and velocity dispersion. We compare
the prediction of the classical King model to the observations of large dark
matter halos. Because of collisions and evaporation, the central density
increases while the slope of the halo density profile decreases until an
instability
takes place. We show that large dark matter halos are relatively well-described
by the
King model at, or close to, the point of marginal microcanonical stability. At
that point, the
King model generates a density profile that  can be approximated by the modified
Hubble profile. This profile has a flat core and decreases as $r^{-3}$ at large
distances, like the observational Burkert profile. Less 
steep halos are unstable. For large halos, the flat
core is due to finite temperature effects, not to quantum mechanics.  We argue 
that statistical mechanics may provide a good description of dark matter halos. 
We interpret the  discrepancies as a result of incomplete relaxation like in the
case of stellar systems.

\end{abstract}



\maketitle


\section{\label{intro}Introduction}

According to contemporary cosmology,  the universe is made of about $70\%$ dark energy, $25\%$ dark matter, and $5\%$ baryonic (visible) matter \cite{bt}. Thus, the overwhelming preponderance of matter and energy in the universe is believed to be dark, i.e. unobservable by telescopes.  The dark energy is responsible for the accelerated expansion of the universe. Its origin is mysterious and presumably related to the cosmological constant or to some form of exotic fluid with negative pressure such as the Chaplygin gas \cite{copeland}. On the other hand, dark matter is necessary to account for the observed flat rotation curves of galaxies \cite{persic}. Its nature is one of the most important puzzles in particle physics and cosmology. Many candidates for dark matter have been proposed, the most popular ones being the axions and the weakly interacting massive particles (WIMPs) \cite{overduin}.

Dark matter is usually modeled as a cold classical collisionless gas with vanishing pressure. In the cold dark matter (CDM) model, primordial density fluctuations are generated during the inflation and become the seeds of the bottom-up structure formation model. The CDM model with a cosmological constant ($\Lambda$CDM) successfully describes the
accelerated expansion of the universe, the temperature fluctuations of the cosmic microwave background (CMB), and the large-scale structures of the universe \cite{ratra}. However, it seems to encounter many problems at the scale of galactic or sub-galactic structures. Indeed, CDM simulations \cite{nfw} lead to $r^{-1}$ cuspy density profiles at galactic centers (in the scales of the order of $1$ kpc and smaller) while most rotation curves indicate a smooth core density  \cite{observations}. On the other hand, the predicted number of satellite galaxies around each galactic halo is far beyond what we see around the Milky Way  \cite{satellites}.

These problems might be solved, without altering the virtues of the CDM model, if the dark matter is composed of quantum particles such as fermions (e.g. massive neutrinos) or bosons (e.g. axions). The wave properties of the dark matter may stabilize the system against gravitational collapse providing halo cores instead of cuspy profiles. In these models, the formation of dark matter structures at small scales is suppressed by quantum mechanics.  Therefore, quantum mechanics could be a way to solve the problems of the CDM model such as the cusp problem and the missing satellite problem.

Some authors have proposed that dark matter is a gas of bosons at $T=0$ forming Bose-Einstein condensates (BECs).  In this scenario, dark matter halos may be understood as the ground state of some gigantic bosonic atom where the boson particles are condensed in a single macroscopic quantum state $\psi({\bf r})$.   At the scale of galaxies, gravity can be treated with the Newtonian framework so the evolution of the wave function $\psi({\bf r},t)$ is governed by the Gross-Pitaevskii-Poisson (GPP) system. Using the Madelung  \cite{madelung} transformation,
 the Gross-Pitaevskii  (GP) equation  \cite{gross,pitaevskii} turns out to be equivalent to hydrodynamic (Euler) equations involving an isotropic pressure due to short-range
interactions (scattering) and an anisotropic quantum pressure arising
from the Heisenberg uncertainty principle. At large scales, quantum
effects are negligible and one recovers the classical hydrodynamic
equations of the CDM model which are remarkably
successful in explaining the large-scale structures of the universe.
At small scales, gravitational collapse is prevented by the repulsive
scattering of the bosons or by the uncertainty principle. This model could solve the cusp problem and the missing satellites problem.

The possibility that dark matter could be in the form of BECs has a long history
\cite{prd1,prd2,bookspringer}. In some works
\cite{baldeschi,membrado,sin,schunckpreprint,matosguzman,guzmanmatos,hu,mu,
arbey1,silverman1,matosall,silverman,bmn,sikivie,mvm,lee09,ch1,lee,mhh,ch2,ch3},
it is assumed that the bosons have no self-interaction. In that case,
gravitational collapse is prevented by the Heisenberg uncertainty principle
which is equivalent to a quantum pressure. This leads to dark matter halos with
a mass-radius relation $MR=9.95\hbar^2/Gm^2$ \cite{rb,membrado,prd2}. In order
to account for the mass and size of dark matter halos (typically $M=3\, 10^{11}
M_{\odot}$ and $R=10\, {\rm kpc}$), the mass of the bosons must be extremely
small, of the order of $m\sim 10^{-24}\, {\rm eV}/c^2$ \cite{baldeschi}.
Ultralight scalar fields
like axions may have such small masses (multidimensional string theories predict
the existence of bosonic particles down to masses of the order of $m\sim
10^{-33}\, {\rm eV}/c^2$). This corresponds to ``fuzzy cold dark matter''
\cite{hu}. In other works
\cite{leekoh,peebles,goodman,arbey,lesgourgues,bohmer,briscese,harko,abril,
pires,rmbec,rindler,lora,lensing,glgr1},  it is assumed that the bosons have a
repulsive self-interaction measured by a scattering length $a>0$. In that case,
gravitational collapse is prevented by the pressure arising from the scattering.
In the Thomas-Fermi (TF) approximation, which amounts to neglecting the quantum
pressure, the resulting structure is equivalent to a polytrope of index $n=1$.
The radius of the halo is given by $R=\pi(a\hbar^2/Gm^3)^{1/2}$, independent on
its mass $M$ \cite{goodman,arbey,bohmer,prd1}. For $a\sim 10^6\, {\rm fm}$,
corresponding to the values of the scattering length observed in terrestrial BEC
experiments \cite{revuebec}, the mass and size of dark matter halos are
reproduced if the bosons have a mass $m\sim 1\, {\rm eV/c^2}$ \cite{bohmer}.
This mass is much
larger than the mass $m\sim 10^{-24}\, {\rm eV}/c^2$ required in the absence of
self-interaction.  This may be more realistic from a particle physics point of
view. The general mass-radius relation of self-gravitating BECs at $T=0$ with an
arbitrary scattering length $a$, connecting the non-interacting limit ($a=0$) to
the TF limit ($G M^2 m a/\hbar^2\gg 1$), has been determined analytically and
numerically in \cite{prd1,prd2}. These papers also provide the general density
profile of dark matter halos interpreted as self-gravitating BECs at $T=0$.

However, the BEC scenario encounters serious problems. In the
non-interacting case, the mass of the bosons must be extremely small, of the
order of $m\sim 10^{-24}\, {\rm eV}/c^2$, in order to reproduce the properties
of dark matter halos. The existence of particles with such small masses remains
dubious (although not impossible a priori). Furthermore, the mass of the halo
decreases with the radius which is in contradiction with the observations that
reveal that the mass increases with the radius. On the other hand, for
self-interacting BECs in the TF approximation, the radius of the halos turns out
to be independent on their mass, and fixed by the properties of the bosons
(their mass and scattering length). This is a major drawback of the BEC model
because it implies that all the halos should have the same radius (unless the
characteristics of the bosons change from halo to halo), which is clearly not
the case. It is possible that the BEC model at $T=0$ describes only dwarf dark
matter halos. In order to describe large halos, finite temperature effects
should be taken into account (see also footnote 4). Finite temperature effects
in the self-gravitating Bose gas have been studied in
\cite{ir,ingrosso,nikolic,msepl,madarassy,slepian,harko3,rm} using different
approaches. When temperature effects are included in the model, the system takes
a core-halo structure with a small condensed core (equivalent to a BEC at $T=0$)
surrounded by an extended isothermal atmosphere of non-condensed bosons.
These structures may be more realistic to describe dark matter halos.

Another possible scenario is that dark matter is made of fermions (such as
massive neutrinos) instead of bosons.  This model also solves the cusp problem
and the missing satellite problem. In that case, gravitational collapse is
prevented by the Pauli exclusion principle. The distribution function satisfies
$f\le \eta_0^{Pauli}\equiv g m^4/h^3$ where $g=2s+1$ is the spin multiplicity of
the quantum states (in the numerical applications, we shall take $s=1/2$ and
$g=2$). The fact that the distribution function is bounded implies that the
density cannot diverge. At $T=0$, the halos are completely degenerate and,
except for a matter of scales, they are similar to classical white dwarf stars
where gravitational collapse is prevented by the quantum pressure of the
electrons \cite{chandra,st}. Their mass-radius relation is $MR^3=1.49\,
10^{-3}\, h^6/(G^3 m^8)$ \cite{chandra}. This model could describe dwarf dark
matter halos. However, in order to describe large halos, like in the case of the
bosonic scenario, it may be necessary to consider the Fermi gas at finite
temperature. Indeed, the mass of a self-gravitating Fermi gas at $T=0$ decreases
with its size which is not consistent with observations. A detailed study of
phase transitions in the self-gravitating Fermi gas  at finite temperature has
been performed by Chavanis \cite{pt,dark,rieutord,ptd,ijmpb}. This study shows
how a degenerate compact object forms as the energy and the temperature are
reduced. Originally, the self-gravitating Fermi gas at finite temperature with 
neutrino masses in the $\sim {\rm eV}/c^2$  range  was proposed as a model for
dark matter halos (e.g. $M=10^{12} M_{\odot}$ and $R= 100\, {\rm kpc}$) and
clusters of galaxies \cite{baldeschi,gr,stella,mr,gao,mr2,merafina}. Then, it
was suggested that degenerate superstars composed of weakly interacting fermions
in the $\sim 10\, {\rm keV}/c^2$ range  could be an alternative to the
supermassive black holes  that are reported to exist at the center of galaxies
(e.g. $M= 2.6 \, 10^{6} M_{\odot}$ and $R= 18 \,{\rm mpc}$ in our Galaxy)
\cite{bilic1,tv,bilic2,bv,bilic3}. Finally, 
it was argued that a weakly interacting fermionic gas at finite temperature
could  provide a self-consistent model of dark matter that describes both the
center and the halo of the galaxies \cite{viollier,bilic4}. In that model, the
system has a core-halo structure with a small condensed core (equivalent to a
fermion ball at $T=0$) surrounded by an extended isothermal atmosphere. Since
the density of a self-gravitating isothermal gas decreases as $r^{-2}$ at large
distances \cite{chandra}, this model is consistent with the flat rotation curves
of the galaxies. On the other hand, since the core is degenerate in the sense of
quantum mechanics (Pauli exclusion principle), it leads to flat density profiles
at the center and  avoids the cusp problem of CDM models. In addition, the
gravitational collapse of fermionic matter leads to a compact object (fermion
ball) at the center of galaxies  that could be an alternative to a central black
hole \cite{viollier}.\footnote{These results can be transposed to bosonic dark
matter where the BEC nucleus (soliton) is the counterpart of the fermion ball
\cite{torres2000,guzmanbh}.}

One  difficulty with the finite temperature self-gravitating Bose and  Fermi
gases is to explain how the particles have thermalized and how they have reached
a statistical equilibrium state. Indeed, the collisional relaxation time of a
self-gravitating halo is usually very large
and exceeds the age of the universe by many orders of magnitude \cite{bt}. To
solve this timescale
problem,\footnote{The relaxation time can be shorter if the system is coupled to
a thermal bath instead of being isolated. However, as discussed in Appendix B of
Paper II, it is unlikely that dark matter halos are coupled to a thermostat.} we
propose that dark matter halos can be treated  as a  collisionless gas having 
experienced a form of violent relaxation.\footnote{A spatially homogeneous
collisionless self-gravitating system described by the Vlasov-Poisson system
undergoes gravitational collapse (Jeans instability) and forms regions of over
density. When the density has sufficiently grown, these regions collapse under
their own gravity at first in free fall. Then, as nonlinear gravitational
effects become important at higher densities, these configurations undergo
damped oscillations and phase mixing. They heat up and finally
settle into a quasi stationary state (QSS) with a core-halo structure on a
coarse-grained scale.} This process, introduced by Lynden-Bell \cite{lb} in
stellar dynamics, and worked out in \cite{phd,csr,csmnras,dubrovnik}, leads to 
a distribution function similar to the Fermi-Dirac distribution function. The
coarse-grained distribution function satisfies $\overline{f}\le \eta_0^{LB}$
where $\eta_0^{LB}$ is the initial value of the distribution function before
mixing. In that case, the origin of ``degeneracy'' is due to dynamical
constraints (Liouville's theorem) instead of quantum mechanics (Pauli's
principle). This theory was initially developed to  describe collisionless
stellar systems such as elliptical galaxies for which the non-degenerate limit
may be the most relevant \cite{lb}. However, this approach with dynamical
degeneracy retained could also apply to dark matter halos \cite{kull,csmnras}.
In that  case, gravitational collapse is prevented by Lynden-Bell's type of
exclusion principle. Furthermore, this scenario  provides a much more  efficient
relaxation mechanism than the fermionic scenario. Indeed, the violent relaxation
of collisionless systems (leading to the Lynden-Bell statistics) takes place on
a few dynamical times while the collisional relaxation of fermions (leading to
the Fermi-Dirac statistics) is very long and possibly exceeds the age of the
universe by many orders of magnitude. Therefore, it is not clear how the
fermions have thermalized in the whole cluster. In addition, in the fermionic
scenario, the thermodynamical temperature $T$ is expected to be very low  so
that the halos would be completely degenerate and would appear very different
from what is observed (except in the case of dwarf halos).  By contrast, in
Lynden-Bell's theory, the temperature is an effective out-of-equilibrium
temperature $T_{eff}$ that can be much larger than the thermodynamical
temperature. This could account for the value of the temperature inferred from
the rotation curves of the galaxies by using the virial theorem. Therefore, the
Lynden-Bell theory predicts a large effective temperature (even if $T=0$
initially), a density profile decreasing as $r^{-2}$ at large distances 
consistent with the flat rotation curves of galaxies, and an effective exclusion
principle at short distances that could avoid the cusp problem and lead to
fermion balls mimicking black holes, just like in the fermionic
scenario \cite{csmnras,dubrovnik}. As a
result, the Lynden-Bell theory has the same
properties as the fermionic theory while  solving the timescale problem and the
temperature problem \cite{csmnras,dubrovnik}. This makes this scenario very
attractive.\footnote{There exist a process similar to violent relaxation
in
dark matter made of condensed bosons called  gravitational cooling
\cite{seidel}. A system of bosons at $T=0$ described by
the
Schr\"odinger-Poisson equation undergoes gravitational collapse (Jeans
instability), oscillates, and settles into a compact bosonic object through the
radiation of a complex scalar field. As a result, the system reaches a QSS made
of a solitonic core surrounded by a halo made of scalar radiation. The halo is
similar to a thermal halo so this process may explain how self-gravitating
bosons can ``thermalize'' and acquire a large effective temperature $T_{eff}$
even if $T=0$ formally. The presence of the radiative halo may also explain why
the mass of the halos increase with their radius. In the analogy between bosons
and fermions, the soliton
corresponds to the fermion ball and the halo  made of scalar radiation
corresponds to the isothermal halo predicted by Lynden-Bell's theory.}

The dark matter halos formed by Jeans instability and violent relaxation can
merge and create bigger structures. This is called hierarchical
clustering.\footnote{This process shares some analogies with the process of
two-dimensional decaying turbulence in hydrodynamics \cite{houches}.} This is
also a process of violent relaxation. If dark
matter is collisionless, a large halo should not evolve anymore after having
reached  a virialized state. As a result, it cannot have very high densities.
In order to be more general, and because very little is known
concerning the nature of dark matter, we consider the possibility that the
core of dark matter halos can be  collisional
\cite{spergel}. This seems to be necessary to explain the presence of black
holes\footnote{We shall argue later that black holes at the center of galaxies
are favored over fermion balls \cite{nature,reid}.} at the center of large
dark matter halos as proposed by Balberg {\it et al.} \cite{balberg}. When
collisional effects are taken into account, dark matter halos behave similarly
to globular clusters. However, the collisions between particles do not
correspond to two-body encounters as in globular clusters but rather to
collisions similar to those in a gas.\footnote{The relaxation time due to strong
short-range collisions is large (of the order of the Hubble time) but, still,
much smaller than the relaxation time due to weak long-range encounters.}  On
the other hand, in fermionic dark matter halos,  the Pauli exclusion principle
must be taken into account. As a result, collisions tend to establish a 
Fermi-Dirac distribution at finite temperature. This distribution is not very
different from Lynden-Bell's distribution but collisions allow the central
concentration of the system to evolve in time towards large values.

In a recent series of papers, de Vega and Sanchez
\cite{vega,vega2,vega3,vega4,vega5} compared the predictions of the finite
temperature self-gravitating Fermi gas  with observations of dark matter halos.
They argued that small halos are degenerate quantum objects while large halos
are non degenerate classical objects. Assuming that the smallest known halos are
completely degenerate, they found that the mass of the fermions must be of the
order of $2\, {\rm keV}/c^2$ corresponding possibly to sterile neutrinos.
Concerning the rotation curves, they obtained encouraging results showing that
the description of dark matter halos in terms of the Fermi-Dirac distribution
may be a good starting point.\footnote{There remains, however, quantitative
discrepancies with observations (Burkert profile) indicating that more elaborate
models are required. In particular, the density profile of isothermal systems
decreases as $r^{-2}$ \cite{chandra} at large distances while the density
profile of dark matter halos decreases as $r^{-3}$ \cite{nfw,observations}.} de
Vega and Sanchez  justify the Fermi-Dirac distribution by quantum mechanics (for
a system of fermions at statistical equilibrium) although, as explained above,
it may be due to Lynden-Bell's form of relaxation. We shall consider the two
possibilities since they lead to similar distribution functions \cite{ijmpb}.
Actually, quantum degeneracy and Lynden-Bell's type of  degeneracy compete with 
each other \cite{kull,csmnras}.\footnote{It is likely that the mixing process
giving rise to a Fermi-Dirac distribution at finite temperature is due to
violent collisionless relaxation (Lynden-Bell) while the maximum accessible
distribution function $\eta_0$ is fundamentally fixed by quantum mechanics
(Pauli). Indeed, the maximum distribution function of the spatially homogeneous
collisionless gas prior to violent relaxation is $f_0=
(1/2)\eta_0^{Pauli}=(g/2)m^4/h^3$ since the gas has a  relativistic Fermi
distribution $f=\eta_0^{Pauli}/(1+e^{p c/k_B T})$ \cite{tg}. Therefore, the
Lynden-Bell bound is equal to half the Pauli bound: $\eta_0^{LB}=f_0=
\eta_0^{Pauli}/2=(g/2)m^4/h^3$.} In their study, de Vega and Sanchez  use the
usual Fermi-Dirac distribution. However, when coupled to gravity, this
distribution has infinite mass  so that it cannot constitute a physical model.
Furthermore, this infinite mass problem precludes the possibility of studying
the stability of the cluster (except if we enclose the cluster within an
artificial  ``box''). For these reasons, we propose, as a next step, to describe
dark matter halos by the fermionic King model which is a truncated Fermi-Dirac
distribution. This model was introduced independently by Ruffini and Stella
\cite{stella} and Chavanis \cite{mnras}. It can be viewed as a generalization of
the classical King model to the case of fermions. This model has a finite mass
so it is more realistic than the Fermi-Dirac distribution.  The fermionic King
model can be derived  \cite{mnras}  from a kinetic equation (the fermionic
Landau equation) assuming that the particles leave the system when their energy
overcomes  a critical escape energy $\epsilon_m$. This derivation is valid both
for quantum particles (fermions) and for  collisionless self-gravitating systems
undergoing Lynden-Bell's  form of violent relaxation. In the non-degenerate
limit, the fermionic King model reduces to the classical King model.

The classical King model  \cite{king} was introduced in the context of stellar systems in
order to describe globular clusters made of classical point mass stars. On the
basis of thermodynamics, we would expect that a system of classical point mass
stars in gravitational interaction reaches a statistical equilibrium state
described by the Boltzmann distribution. However, it is well-known that no
statistical equilibrium state exist for self-gravitating systems because the
Boltzmann entropy has no maximum in an unbounded domain (the isothermal sphere,
corresponding to the Boltzmann distribution coupled to the Poisson equation, has
infinite mass) \cite{bt}. Therefore, the statistical mechanics of
self-gravitating systems is essentially an out-of-equilibrium problem
\cite{aakin}. The absence of statistical equilibrium state is related to the
fact that self-gravitating systems such as globular clusters have the tendency
to evaporate. However, evaporation is a slow process and a globular cluster can
be found, for intermediate times, in a quasi stationary state close to the
Michie-King distribution \cite{michie,king} which is a truncated Boltzmann
distribution\footnote{Since the isothermal sphere has an
infinite mass, the basic idea of Michie \cite{michie} and King \cite{king} is to introduce a bound on
the energy of the stars so that, if a star has a too large energy, it escapes
the system. The energy bound introduces automatically a bound on the radius of
the system that is interpreted as a tidal radius beyond which the stars are lost
by the cluster. The Michie-King distribution can be derived from the classical Landau equation. The King distribution \cite{king} simply amounts
to subtracting a constant from the Boltzmann factor so that the distribution
function vanishes when the energy reaches its maximum value. The Michie \cite{michie} distribution includes, in addition, effects of anisotropy.} with parameters
slowly changing with time.

The caloric curve of the King model was determined
by Katz \cite{katzking}. Like in the case of box-confined isothermal spheres
\cite{antonov,lbw}, equilibrium states exist only above a critical energy $E_c$.
These configurations are metastable but their lifetime is considerable since it
scales as $e^N$ (except close to the critical point) \cite{metastable}.  For
globular clusters, for which $N\sim 10^{6}$, this lifetime is so large that
metastable states can be considered as stable states.  In this sense, we can say
that self-gravitating systems with $E>E_c$ do reach a
statistical equilibrium state described by a truncated Boltzmann distribution (even if there is
no statistical equilibrium state in a strict sense).
However, because of evaporation, the energy of a self-gravitating system slowly
decreases.  For $E<E_c$, there is no equilibrium state anymore. Therefore, when
the energy passes below this threshold, the system undergoes a gravothermal
catastrophe \cite{lbw} and experiences core collapse \cite{cohn,lbe,heggie}.
This corresponds to a saddle-node bifurcation. For classical self-gravitating
systems, such as globular clusters, core collapse leads to the formation of a
binary star surrounded by a hot halo (at the collapse time, the singular density profile has infinite
central density but zero central mass) \cite{henonbinaries,henonbinaries2}. The
binary can release sufficient energy to stop the collapse and even drive a
re-expansion of the cluster in a post-collapse regime \cite{inagaki}. This is
followed by a series of gravothermal oscillations \cite{oscillations,hr}. It is
estimated that about $80\%$ of globular clusters are described by the King model
while $20\%$ have undergone core collapse \cite{bt}. For self-gravitating systems made of
fermions (white dwarfs, neutron stars, dark matter halos), the collapse stops
when the core of the system becomes degenerate in virtue of the Pauli exclusion
principle. In that case, we have to take quantum mechanics into account. To
treat dark matter, we propose to use the fermionic King model \cite{stella,mnras}.

The fermionic King model was studied by Ruffini and Stella \cite{stella} who
determined the density profiles of dark matter halos for various values of the
central potential. Our series of papers is intended to complete their study in
the following directions. We determine the caloric curves of the fermionic
King model for arbitrary values of the degeneracy parameter and study in detail
the phase transitions that may occur between a gaseous phase unaffected by
quantum mechanics and a condensed phase dominated by quantum mechanics. In this
way, we generalize the study of phase transitions in the self-gravitating
Fermi gas at finite temperature confined within a box performed by Chavanis
\cite{pt,dark,rieutord,ptd,ijmpb}.  This generalization is important since the
fermionic King model provides a more relevant description of dark matter halos
than box models. Therefore, we obtain realistic caloric
curves of dark matter halos. We also determine the density profile, the velocity
dispersion profile, and the circular velocity profile of the different
configurations in order to compare the predictions of the fermionic King  model
to the observations of dark matter halos.  In the present paper (Paper I), we
consider the non-degenerate limit corresponding to the classical King model. The
non-degenerate limit is expected to be valid for large dark matter halos so that
it is a good starting point. In our companion paper \cite{cml} (paper II), we
consider the fermionic King model for arbitrary values of the degeneracy
parameter. Degeneracy effects are expected to be important for dwarf and
intermediate-size dark matter halos. A short account of our results is given in
\cite{henonbook}.

The paper is organized as follows. In Sec. \ref{sec_sm}, we discuss models of
dark matter halos based on statistical mechanics and we introduce the classical
and fermionic King models. In Sec.  \ref{sec_gen}, we  formulate the
general problem of determining the structure of a spherically symmetric
self-gravitating system described by a distribution function of the form
$f=f(\epsilon)$ with $f'(\epsilon)<0$, where $\epsilon=v^2/2+\Phi({\bf r})$ is
the individual energy of the particles. We introduce the ``generalized entropy''
associated to this distribution and discuss its physical interpretation. In
Secs. \ref{sec_kingclass}-\ref{sec_improper}, we apply this general formalism to
the classical King model and compute several quantities of interest. We show
that the King model leads to configurations with an isothermal core, an isothermal halo, and a polytropic
envelope of index $n=5/2$. In Sec. \ref{sec_comp}, we compare the predictions of
the classical King model to the observations of dark matter halos.
Because of collisions and evaporation, the central density increases while the slope of the halo profile decreases until an instability takes place.
We show that dark matter halos are relatively well-described by
a  King model at, or close to, the point of marginal microcanonical stability.
At that point, the King model generates a density profile that can be
approximated by the modified Hubble profile \cite{bt}. This profile has a flat core and
decreases as $r^{-3}$ at large distances, like the observational Burkert \cite{observations}
profile. Less steep halos are unstable. The flat core is due to
finite temperature effects, not to quantum
mechanics. On the other hand, the large distance behavior of the density profile
is due to the polytropic nature of the King
distribution at high energies that departs from the  isothermal Boltzmann
distribution. We conclude that statistical mechanics provides a good description of dark
matter halos when evaporation is taken  into account. The agreement is very good in the core of the system that is well-relaxed. The discrepancies that remain in the halo may be interpreted as a
result of an incomplete relaxation, like in the case of stellar systems.

Although our results are exposed in the context of dark matter halos, our study of the classical King model presented in this paper also applies to globular clusters.

\section{Models of dark matter halos based on statistical mechanics}
\label{sec_sm}

We  consider the possibility that dark matter halos can be described by the Fermi-Dirac distribution
\begin{equation}
f=\frac{\eta_0}{1+e^{\beta\epsilon+\alpha}},
\label{sm1}
\end{equation}
where $f({\bf r},{\bf v})$ gives the mass density of particles with position ${\bf r}$ and velocity ${\bf v}$,  $\rho({\bf r})=\int f({\bf r},{\bf v})\, d{\bf v}$ gives the mass density of particles with position ${\bf r}$, $\Phi({\bf r})$ is the gravitational potential determined by the Poisson equation $\Delta\Phi=4\pi G\rho$, $\eta_0$ is the maximum accessible value of the distribution function, $\epsilon=v^2/2+\Phi({\bf r})$ is the individual energy of the particles by unit of mass,  $\beta$ is the inverse temperature, and $\epsilon_F=-\alpha/\beta$ is the chemical potential (Fermi energy). In the non-degenerate limit $\alpha\rightarrow +\infty$, we can make the approximation $e^{\beta\epsilon+\alpha}\gg 1$ implying $f\ll\eta_0$ and the Fermi-Dirac distribution reduces to the Boltzmann distribution
\begin{equation}
f=\eta_0 e^{-(\beta\epsilon+\alpha)}.
\label{sm2}
\end{equation}

As recalled in the Introduction, the Fermi-Dirac distribution may have two origins: (i) It may describe a gas of fermions at statistical equilibrium in which case $\eta_0=g m^4/h^3$ is the maximum accessible value of the distribution function fixed by the Pauli exclusion principle; (ii) It may result from the violent relaxation of a collisionless system of particles (classical or quantum) as described by Lynden-Bell \cite{lb} and worked out in \cite{phd,csr,csmnras,dubrovnik}. In that case, Eqs. (\ref{sm1}) and (\ref{sm2}) are valid for the coarse-grained distribution function (usually denoted $\overline{f}$) and $\eta_0$ is the maximum value of the fine-grained distribution function.\footnote{In the general case, the theory of violent relaxation leads to a coarse-grained distribution function $\overline{f}$ that is a superposition of ``Fermi-Dirac'' distributions \cite{lb,csr}. The single ``Fermi-Dirac'' distribution function (\ref{sm1}) is obtained when the fine-grained distribution function takes only two values $f=\eta_0$ and $f=0$. It may also provide an approximation of more general cases where $\eta_0$ represents the maximum value of the fine-grained distribution function. The fine-grained distribution function coincides with the initial distribution function before the system has mixed.}  We shall consider the two possibilities since the distribution functions are formally the same. In the quantum interpretation, $\beta={m}/{k_B T}$ where $T$ is the thermodynamical temperature. In Lynden-Bell's interpretation, $\beta={\eta_0}/{T_{eff}}$ where $T_{eff}$ is a generalized (out-of-equilibrium) ``temperature''. In order to unify the notations, we write $\beta=1/T$ where $T$ has the dimension of an energy by unit of mass.

When coupled to the gravity through the Poisson equation, the Fermi-Dirac distribution (\ref{sm1}) has an infinite mass since it reduces to the Boltzmann distribution (\ref{sm2}) at large distances where the system is diluted (non-degenerate). As a result, the density decreases as $\rho\sim r^{-2}$ for $r\rightarrow +\infty$  \cite{chandra} which is not normalizable. In order to avoid the infinite mass problem, we shall use a truncated  Fermi-Dirac distribution. Specifically, we use the fermionic King model that we write as (see Paper II):
\begin{equation}
f=A\frac{e^{-\beta(\epsilon-\epsilon_m)}-1}{1+\frac{A}{\eta_0}e^{-\beta(\epsilon-\epsilon_m)}} \quad {\rm if} \quad \epsilon\le \epsilon_m,
\label{sm3}
\end{equation}
\begin{equation}
f=0 \quad {\rm if} \quad \epsilon\ge\epsilon_m,
\label{sm4}
\end{equation}
where $\epsilon_m$ is the escape energy above which the particles are lost by
the system and $\mu\equiv \eta_0/A$ is a dimensionless parameter  that measures
the importance of degeneracy. The chemical potential (Fermi energy) is related
to the escape energy by
$\epsilon_F\equiv -\alpha/\beta=\epsilon_m-(1/\beta)\ln(\eta_0/A)$ (see Paper
II). For $\epsilon\ll \epsilon_m$, we can make 
the approximation $e^{-\beta(\epsilon-\epsilon_m)}\gg 1$ and we recover the
Fermi-Dirac distribution (\ref{sm1}). The fermionic King model was introduced
heuristically by Ruffini and Stella \cite{stella} as a natural extension of the
classical King model to fermions in order to describe dark matter halos made of
massive neutrinos. This distribution function was independently introduced by
Chavanis \cite{mnras} where it was derived from a kinetic equation (the
fermionic Landau equation) assuming that the particles leave the system when
they reach a maximum energy $\epsilon_m$. The kinetic derivation given in
\cite{mnras} is valid either for quantum particles (fermions) or for
collisionless self-gravitating systems (classical or quantum) experiencing
Lynden-Bell's type of relaxation. This derivation can also be extended to 
non-condensed bosons by simply replacing $1-f/\eta_0$ by $1+f/\eta_0$ in the
kinetic equation. This leads to the bosonic King model
\begin{equation}
f=A\frac{e^{-\beta(\epsilon-\epsilon_m)}-1}{1-\frac{A}{\eta_0}e^{-\beta(\epsilon-\epsilon_m)}} \quad {\rm if} \quad \epsilon\le \epsilon_m,
\label{sm5}
\end{equation}
\begin{equation}
f=0 \quad {\rm if} \quad \epsilon\ge\epsilon_m.
\label{sm6}
\end{equation}

In the non degenerate limit $\mu=\eta_0/A\rightarrow +\infty$, we can make the approximation $(A/\eta_0)e^{-\beta(\epsilon-\epsilon_m)}\ll 1$ and we recover the classical King model
\begin{equation}
f=A \left\lbrack {e^{-\beta(\epsilon-\epsilon_m)}-1}\right\rbrack\quad {\rm if} \quad \epsilon\le \epsilon_m,
\label{sm7}
\end{equation}
\begin{equation}
f=0 \quad {\rm if} \quad \epsilon\ge\epsilon_m.
\label{sm8}
\end{equation}
For $\epsilon\ll \epsilon_m$, we can make the additional approximation $e^{-\beta(\epsilon-\epsilon_m)}\gg 1$ and we recover the Boltzmann distribution (\ref{sm2}). The classical King model describes globular clusters and, possibly, large dark matter halos for which degeneracy effects (due to the Pauli exclusion principle for fermions or due to the Liouville theorem for collisionless systems undergoing violent relaxation) are negligible.

\section{The general formulation of the problem}
\label{sec_gen}

Before studying specifically the classical King model in Sec. \ref{sec_kingclass}, we formulate here the problem for a general distribution function of the form $f=f(\epsilon)$ with $f'(\epsilon)<0$ describing spherical clusters.  This will allow us to extend our study to various situations in future works without having to recall the general formalism at each time. We emphasize that the scalings derived below (for the energy, the temperature, the tidal radius...) are ``universal'', i.e. they do not depend on the precise form of the considered distribution function.

\subsection{Variational principles}
\label{sec_vp}

For any functional of the form
\begin{equation}
S=-\int C(f)\, d{\bf r}d{\bf v},
\label{vp1}
\end{equation}
where $C(f)$ is a convex function (i.e. $C''>0$), we consider the following maximization problems
\begin{equation}
S(E,M)=\max_f \lbrace S[f]\, |\, E[f]=E,\quad M[f]=M\rbrace
\label{vp2}
\end{equation}
and
\begin{equation}
J(\beta,M)=\max_f \lbrace J[f]=S[f]-\beta E[f]\, |\,  M[f]=M\rbrace,
\label{vp3}
\end{equation}
where
\begin{equation}
E=\frac{1}{2}\int f v^2\, d{\bf r}d{\bf v}+\int \rho\Phi\, d{\bf r}=K+W
\label{vp4}
\end{equation}
is the energy ($K$ is the kinetic energy and $W$ is the potential energy) and
\begin{equation}
M=\int \rho\, d{\bf r}
\label{vp5}
\end{equation}
is the mass.

The critical points of the maximization problem (\ref{vp2}) are determined by the variational principle
\begin{equation}
\delta S-\beta\delta E-\alpha\delta M=0,
\label{vp6}
\end{equation}
where $\beta$ and $\alpha$ are Lagrange multipliers associated with the constraints $E$ and $M$. The critical points of the maximization problem (\ref{vp3}) are determined by the variational principle
\begin{equation}
\delta J-\alpha\delta M=0,
\label{vp7}
\end{equation}
where $\alpha$ is a Lagrange multiplier associated with the constraint $M$. Obviously, the maximization problems (\ref{vp2}) and (\ref{vp3}) have the same critical points (canceling the first order variations). They are given by the equation
\begin{equation}
C'(f)=-\beta\epsilon-\alpha,
\label{vp8}
\end{equation}
where $\epsilon=v^2/2+\Phi({\bf r})$ is the individual energy of the particles by unit of mass. Since $C$ is convex, this equation can be reversed to give $f=F(\beta\epsilon+\alpha)$ where $F(x)=(C')^{-1}(-x)$. We note that $f'(\epsilon)=-\beta/C''(f)$, so that $f'(\epsilon)$ keeps the same sign everywhere. Since $f(\epsilon)$  is positive and vanishes at the escape energy $\epsilon_m$, we must have  $f'(\epsilon)<0$ close to the escape energy. Therefore,  $f'(\epsilon)<0$ everywhere and, consequently, $\beta>0$. In conclusion, the temperature is positive and the distribution function decreases monotonically with $\epsilon$ until it vanishes at $\epsilon_m$. For future convenience, we write $F(x)=A\, {\cal F}(x)$, where $A$ is a constant with the dimension of a distribution function and ${\cal F}$ is a dimensionless function (it can still depend on $A$ and on other ``external'' parameters). This amounts to writing $C(f)=A\, {\cal C}(f/A)$ and ${\cal F}(x)=({\cal C}')^{-1}(-x)$. Then, the critical points of the maximization problems (\ref{vp2}) and (\ref{vp3}) are given by
\begin{equation}
f=A \, {\cal F}(\beta\epsilon+\alpha).
\label{vp9}
\end{equation}
We assume that both $A$ and ${\cal F}$ (or equivalently $A$ and ${\cal C}$) are given. In the maximization problem (\ref{vp2}),  $\beta$ and $\alpha$ must be related to $E$ and $M$. In the maximization problem (\ref{vp3}),  $\beta$ is prescribed  and $\alpha$ must be related to $M$.

A distribution function of the form of Eq. (\ref{vp9}) is a (local) maximum of $S$ at fixed $E$ and $M$ if, and only if,
\begin{eqnarray}
\delta^2 G\equiv -\int C''(f)\frac{(\delta f)^2}{2}\, d{\bf r}d{\bf v}-\frac{1}{2}\beta\int\delta\rho\delta\Phi\, d{\bf r}<0
\label{vp10}
\end{eqnarray}
for all perturbations $\delta f$ that conserve mass and energy at first order,
i.e. $\delta E=\delta M=0$. A distribution of the form of Eq. (\ref{vp9}) is a
(local) maximum of $J$ at fixed $M$ if, and only if, the inequality of Eq.
(\ref{vp10}) is satisfied for all perturbations $\delta f$ that conserve mass,
i.e. $\delta M=0$. The derivation of these results can be found in \cite{cc}.

To study the maximization problems (\ref{vp2}) and (\ref{vp3}), we shall use a thermodynamical analogy. We call $S$ the entropy, $J$ the free energy\footnote{The free energy is usually defined by $F=E-TS$ so that $J=-\beta F$. The function $J$ is sometimes called the Massieu function. To simplify the terminology we will call it here the free energy.}, $\beta=1/T$ the inverse temperature, and $-\alpha/\beta$ the chemical potential. The maximization problem  (\ref{vp2}) in which the energy and the mass are fixed is associated to the microcanonical ensemble (MCE) and the maximization problem  (\ref{vp3}) in which the temperature and the mass are fixed is associated to the canonical ensemble (CE). We shall be interested by local and global maxima of entropy at fixed mass and energy in MCE, and by local and global maxima of free energy at fixed mass in CE. Different interpretations of the variational problems (\ref{vp2}) and (\ref{vp3}) are discussed in Appendix A of Paper II.

\subsection{The fundamental differential equation}
\label{sec_fde}

The maximization problems (\ref{vp2}) and (\ref{vp3}) determine distribution functions of the form $f=f(\epsilon)$ with $f'(\epsilon)<0$. Such distribution functions, that depend only on the individual energy $\epsilon$ of the particles, describe spherically symmetric self-gravitating systems \cite{bt}. Inversely, any distribution function of the form $f=f(\epsilon)$ with $f'(\epsilon)<0$ is a critical point of the maximization problems (\ref{vp2}) and (\ref{vp3}) for a specific entropy of the form of Eq. (\ref{vp1}). In practice, it is convenient to prescribe a form of distribution function $f$, determine the corresponding entropy $S$, and consider the variational problems (\ref{vp2}) and (\ref{vp3}). This is how we shall proceed in Sec.  \ref{sec_kingclass} and in Paper II. However, for the moment, we remain very general.

As we have seen, a distribution function $f=f(\epsilon)$ with $f'(\epsilon)<0$ can always be written in the form of Eq. (\ref{vp9}). We assume furthermore that $f(\epsilon)$ vanishes at some escape energy $\epsilon_m$ and that $f=0$ for $\epsilon\ge \epsilon_m$. Therefore ${\cal F}(\beta\epsilon_m+\alpha)=0$. If $x_0$ denotes the zero of  ${\cal F}(x)$, we have $\beta\epsilon_m+\alpha=x_0$. This relation shows that $\epsilon_m$ is not a new parameter but that it is equivalent to the Lagrange multiplier $\alpha$ (for a given value of $\beta$). In the following, we shall work in terms of $\epsilon_m$ and $\beta$ instead of $\alpha$ and $\beta$. Introducing the shifted function ${\cal F}_s(x)={\cal F}(x+x_0)$, satisfying
${\cal F}_s(0)=0$, we can write $f(\epsilon)$ in the form
\begin{equation}
f=A \, {\cal F}_s\lbrack \beta(\epsilon-\epsilon_m)\rbrack \quad {\rm if} \quad \epsilon\le \epsilon_m,
\label{fde1}
\end{equation}
\begin{equation}
f=0 \quad {\rm if} \quad \epsilon\ge\epsilon_m.
\label{fde2}
\end{equation}

The local density is defined by
\begin{equation}
\rho=\int f\, d{\bf v}.
\label{fde3}
\end{equation}
Substituting Eqs. (\ref{fde1}) and (\ref{fde2}) in Eq. (\ref{fde3}), we get
\begin{equation}
\rho=A\int_0^{v_m(r)}{\cal F}_s\left\lbrack \beta\left (\frac{v^2}{2}+\Phi(r)-\epsilon_m\right )\right \rbrack 4\pi v^2\, dv,
\label{fde4}
\end{equation}
where  $v_m(r)=\sqrt{2(\epsilon_m-\Phi(r))}$ is the local escape velocity. These expressions are valid only for $r\le R$, where $R$ is the radius of the cluster such that $v_m(R)=0$, i.e. $\Phi(R)=\epsilon_m$. This is the distance at which the density vanishes: $\rho(R)=0$. For $r>R$, we have $\rho=0$. In the King model, $R$ represents the tidal radius. Making the change of variables $w=(\beta/2)^{1/2}v$, we obtain
\begin{eqnarray}
\rho&=& 4\pi \left (\frac{2}{\beta}\right )^{3/2} A\nonumber\\
&\times&\int_0^{\sqrt{\beta(\epsilon_m-\Phi(r))}}{\cal F}_s(w^2+\beta\Phi(r)-\beta\epsilon_m) w^2\, dw.\qquad
\label{fde5}
\end{eqnarray}
Defining $\chi(r)=\beta(\epsilon_m-\Phi(r))$ and $k=\beta(\epsilon_m-\Phi_0)$, where the index $0$ refers to the center of the cluster, the foregoing equation can be rewritten as
\begin{equation}
\rho=4\pi \left (\frac{2}{\beta}\right )^{3/2} A\int_0^{\sqrt{\chi(r)}}{\cal F}_s(w^2-\chi(r)) w^2\, dw.
\label{fde6}
\end{equation}
At that point, it is convenient to introduce the family of functions
\begin{equation}
I_n(z)=4\pi \int_0^{\sqrt{z}}{\cal F}_s(w^2-z) w^{2n}\, dw \quad (z\ge 0).
\label{fde7}
\end{equation}
For future reference, we note the identity
\begin{equation}
I_n'(z)=\frac{1}{2}(2n-1)I_{n-1}(z)
\label{fde}
\end{equation}
that can be established by a simple integration by parts. In terms of these functions, the density profile can be written as
\begin{equation}
\rho=A\left (\frac{2}{\beta}\right )^{3/2}  I_1(\chi).
\label{fde8}
\end{equation}
The central density is
\begin{equation}
\rho_0=A\left (\frac{2}{\beta}\right )^{3/2}  I_1(k).
\label{fde9}
\end{equation}
Therefore, we obtain
\begin{equation}
\rho=\rho_0\frac{I_1(\chi)}{I_1(k)}.
\label{fde10}
\end{equation}
Substituting these results in the Poisson equation
\begin{equation}
\Delta\Phi=4\pi G\rho
\label{fde10b}
\end{equation}
and introducing the rescaled distance
\begin{equation}
\zeta=r/r_0,
\label{fde12}
\end{equation}
where
\begin{equation}
r_0=\frac{1}{(4\pi G\beta\rho_0)^{1/2}}
\label{fde12b}
\end{equation}
is the core radius, we obtain the fundamental ordinary differential equation
\begin{equation}
\frac{1}{\zeta^2}\frac{d}{d\zeta}\left (\zeta^2\frac{d\chi}{d\zeta}\right )=-\frac{I_1(\chi)}{I_1(k)}
\label{fde13}
\end{equation}
with the boundary conditions
\begin{equation}
\chi(0)=k,\qquad \chi'(0)=0.
\label{fde14}
\end{equation}
This differential equation is defined for $\zeta\le \zeta_1$ where
\begin{equation}
\zeta_1=R/r_0=(4\pi G\beta\rho_0)^{1/2}R
\label{fde15}
\end{equation}
is the dimensionless radius of the cluster determined by the condition
$\chi(\zeta_1)=0$. The function $\chi(\zeta)$ decreases monotonically with $\zeta$. The differential equation (\ref{fde13}) defines a
one-parameter family of
density profiles with parameter $k$ (the normalized central potential) going
from $0$ to $+\infty$. The
dimensionless radius $\zeta_1$ is a function of $k$. We also note that $f=A{\cal F}_s\left\lbrack w^2-\chi(\zeta)\right\rbrack$.

\subsection{The equation of state}
\label{sec_eosb}

For a spherically symmetric distribution function $f(\epsilon)$, the local pressure is defined by
\begin{equation}
p=\frac{1}{3}\int f v^2\, d{\bf v}.
\label{ene2}
\end{equation}
Substituting Eqs. (\ref{fde1}) and (\ref{fde2}) in Eq. (\ref{ene2}), and introducing the variables defined in Sec. \ref{sec_fde}, we obtain
\begin{equation}
p=\frac{1}{3} A\left (\frac{2}{\beta}\right )^{5/2}  I_2(\chi).
\label{ene3}
\end{equation}
We note that the density $\rho({\bf r})$ and the pressure  $p({\bf r})$ are functions  of $\chi({\bf r})$ and $T$ (for a given $A$):  $\rho=\rho\lbrack \chi({\bf r}),T\rbrack$ and $p=p\lbrack \chi({\bf r}),T\rbrack$. Eliminating $\chi({\bf r})$ between Eqs. (\ref{fde8}) and (\ref{ene3}) we find that the cluster is described by a barotropic equation of state $p=p_T(\rho)$ parameterized by the temperature $T$ (for a given $A$). We note the universal scaling $p=\beta^{-5/2}\phi(\beta^{3/2}\rho)$. Furthermore, one can easily check that the condition of hydrostatic equilibrium $\nabla p+\rho\nabla\Phi={\bf 0}$ is automatically satisfied for a system described by a distribution function of the form $f=f(\epsilon)$ (see Appendix \ref{sec_hydro}). Therefore, the differential equation (\ref{fde13}) may be derived equivalently from the fundamental equation of hydrostatic equilibrium \cite{chandra} with the equation of state specified above (see Appendix  \ref{sec_hydro}).

\subsection{The normalized temperature}
\label{sec_temp}

Using Eqs. (\ref{fde10}) and (\ref{fde12}), the mass profile $M(r)=\int_0^r \rho(r') 4\pi {r'}^2\, dr'$ is given by
\begin{equation}
M(r)=4\pi\rho_0r_0^3\int_0^{\zeta}\frac{I_1(\chi)}{I_1(k)}\, \zeta^2 d\zeta.
\label{temp1}
\end{equation}
Combining this equation with the differential equation (\ref{fde13}), we get
\begin{equation}
M(r)=-4\pi\rho_0 r_0^3 \zeta^2 \chi'(\zeta).
\label{temp2}
\end{equation}
Applying this equation at $r=R$, and using Eqs. (\ref{fde12b}) and (\ref{fde15}), we obtain
\begin{equation}
\eta\equiv \frac{\beta GM}{R}=-\zeta_1(k)\chi'[\zeta_1(k)].
\label{temp3}
\end{equation}
This relation can also be derived from the Gauss theorem $d\Phi/dr=GM(r)/r^2$ applied at $r=R$. The parameter $\eta$ is the dimensionless inverse temperature normalized by the size $R$ of the system. This is the correct dimensionless parameter when we work in a box of fixed radius $R$ \cite{antonov,lbw}. However, in the present problem, the size of the configuration $R$ is not a fixed parameter. The fixed parameter is $A$, not $R$. We need therefore to normalize the inverse temperature by $A$. Combining Eqs. (\ref{fde9}) and (\ref{fde15}), we find that
\begin{equation}
R^2=\frac{\zeta_1^2\beta^{1/2}}{8\pi\sqrt{2}GI_1(k)A}.
\label{temp4}
\end{equation}
Substituting this relation in Eq. (\ref{temp3}), we obtain
\begin{equation}
\tilde\beta\equiv \beta G^2 M^{4/3}(8\pi A\sqrt{2})^{2/3}=\frac{\lbrack -\zeta_1^2\chi'(\zeta_1)\rbrack^{4/3}}{I_1(k)^{2/3}}.
\label{temp5}
\end{equation}
This equation relates the normalized inverse temperature $\tilde\beta$ to the parameter $k$.

\subsection{The normalized energy}
\label{sec_ene}

The total energy $E=K+W$ can be computed as follows. Using the virial theorem $2K+W=0$ we have
\begin{equation}
E=-K,
\label{ene1}
\end{equation}
so  we just need to compute the kinetic energy. The kinetic energy can be written in terms of the pressure defined by Eq. (\ref{ene2}) as
\begin{equation}
K=\frac{3}{2}\int p\, d{\bf r}.
\label{ene4}
\end{equation}
Substituting Eq. (\ref{ene3}) in Eq. (\ref{ene4}), and using Eqs. (\ref{fde12}) and (\ref{fde15}), we obtain
\begin{equation}
K=\frac{1}{2}A \left (\frac{2}{\beta}\right )^{5/2}\frac{R^3}{\zeta_1^3}\int_0^{\zeta_1} I_2[\chi(\zeta)] 4\pi\zeta^2\, d\zeta.
\label{ene5}
\end{equation}
According to Eqs. (\ref{temp3}), (\ref{ene1}), and (\ref{ene5}), the total energy normalized by $R$ is
\begin{equation}
\epsilon\equiv \frac{ER}{GM^2}=-\frac{1}{\zeta_1(k) \eta^2(k) I_1(k)}\int_0^{\zeta_1(k)} I_2[\chi(\zeta)]\zeta^2\, d\zeta.
\label{ene6}
\end{equation}
This is the proper normalization of the energy when we work in a box of fixed radius $R$ \cite{antonov,lbw}. However, in the present problem, as explained previously, we must normalize the energy by $A$, not by $R$. Using Eqs. (\ref{temp3})-(\ref{temp5}), we obtain
\begin{eqnarray}
\tilde E&\equiv& \frac{E}{G^2M^{7/3}(8\sqrt{2}\pi A)^{2/3}}\nonumber\\
&=&-\frac{1}{\tilde\beta^{7/4}(k) I_1(k)^{3/2}}\int_0^{\zeta_1(k)} I_2[\chi(\zeta)]\zeta^2\, d\zeta.
\label{ene7}
\end{eqnarray}
This equation relates the normalized energy $\tilde E$ to the parameter $k$.

\subsection{The normalized radius}
\label{sec_rad}

According to Eqs. (\ref{temp4}) and (\ref{temp5}), the radius of the cluster normalized by $A$ is given by
\begin{equation}
\tilde R\equiv R G M^{1/3}(8\pi A\sqrt{2})^{2/3}=\frac{\zeta_1(k)\tilde\beta^{1/4}(k)}{I_1(k)^{1/2}}.
\label{rad1}
\end{equation}
This equation relates the normalized radius $\tilde R$ to the parameter $k$. The normalized distance is then given by
\begin{equation}
G M^{1/3}(8\pi A\sqrt{2})^{2/3}r=\frac{{\tilde R}(k)}{\zeta_1(k)}\zeta.
\label{rad1b}
\end{equation}

\subsection{The chemical potential}
\label{sec_chem}

The escape energy is related to the radius of the cluster and to its mass by
\begin{equation}
\epsilon_m=\Phi(R)=-\frac{GM}{R}.
\label{chem1}
\end{equation}
Using Eq. (\ref{temp3}), we find that $-\beta\epsilon_m=\eta(k)$. Therefore, the
chemical potential (times $-\beta$)  is equal to
$-\beta\epsilon_F=\alpha=x_0+\eta(k)$. On the other hand, using Eq.
(\ref{temp5}), the normalized escape energy varies along the series of
equilibria according to
\begin{equation}
\tilde\epsilon_m\equiv \frac{\epsilon_m}{G^2 M^{4/3}(8\pi A\sqrt{2})^{2/3}}=-\frac{\eta(k)}{\tilde\beta(k)}.
\label{chem2}
\end{equation}

\subsection{The normalized density}
\label{sec_centraldensity}

According to Eqs. (\ref{fde8}) and (\ref{temp5}), the normalized density profile is given by
\begin{equation}
\frac{\rho(r)}{32\pi A^2 G^3 M^2}=\frac{I_1[\chi(\zeta)]}{\tilde\beta^{3/2}(k)}.
\label{centraldensity0}
\end{equation}
The normalized central density is related to $k$ by
\begin{equation}
\tilde\rho_0\equiv \frac{\rho_0}{32\pi A^2 G^3 M^2}=\frac{I_1(k)}{\tilde\beta^{3/2}(k)}.
\label{centraldensity1}
\end{equation}
In general, the central density is a monotonically increasing function of $k$.
Therefore, the parameter $k$ can be interpreted as a measure of the central
density. We shall call it the concentration parameter. It can also be
interpreted as the normalized central potential (with the opposite sign).

\subsection{The normalized circular velocity}
\label{sec_circb}

The circular velocity is defined by \cite{bt}:
\begin{equation}
v_c^2(r)=\frac{GM(r)}{r}.
\label{circb3}
\end{equation}
Using Eq. (\ref{temp2}), we obtain
\begin{equation}
v_c^2(r)=-4\pi G\rho_0 r_0^2 \zeta\chi'(\zeta).
\label{circb4}
\end{equation}
According to Eqs. (\ref{fde12b}) and (\ref{temp5}), the normalized circular velocity profile  is given by
\begin{equation}
\frac{v_c^2(r)}{G^2 M^{4/3}(8\pi A\sqrt{2})^{2/3}}=-\frac{\zeta\chi'(\zeta)}{\tilde\beta(k)}.\label{circb5}
\end{equation}
We also note that $v_c^2(R)=GM/R=-\epsilon_m$.

\subsection{The normalized velocity dispersion}
\label{sec_disb}

The local velocity dispersion (in one direction) of a spherically symmetric distribution function $f(\epsilon)$ is defined by
\begin{equation}
\sigma^2(r)=\frac{p(r)}{\rho(r)}=\frac{1}{3\rho}\int f v^2\, d{\bf v}.
\label{dis1}
\end{equation}
Using Eqs. (\ref{fde10}) and (\ref{ene3}) we obtain
\begin{equation}
\sigma^2(r)=\frac{2}{3\beta}\frac{I_2[\chi(\zeta)]}{I_1[\chi(\zeta)]}.
\label{dis2}
\end{equation}
The central velocity dispersion is therefore
\begin{equation}
\sigma_0^2=\frac{2}{3\beta}\frac{I_2(k)}{I_1(k)}.
\label{dis3}
\end{equation}
According to Eq. (\ref{temp5}), the normalized velocity dispersion profile  is given by
\begin{equation}
\frac{\sigma^2(r)}{G^2 M^{4/3}(8\pi A\sqrt{2})^{2/3}}=\frac{2}{3\tilde\beta(k)}\frac{I_2[\chi(\zeta)]}{I_1[\chi(\zeta)]}.\label{dis4}
\end{equation}
The normalized central velocity dispersion is related to $k$ by
\begin{equation}
{\tilde\sigma}_0^2\equiv \frac{\sigma^2_0}{G^2 M^{4/3}(8\pi A\sqrt{2})^{2/3}}=\frac{2}{3\tilde\beta(k)}\frac{I_2(k)}{I_1(k)}.\label{dis5}
\end{equation}

\subsection{The parameter ${\cal K}$}
\label{sec_k}

Instead of working with $k$ it is sometimes convenient to work in terms of the parameter
\begin{equation}
{\cal K}=-\frac{\Phi_0}{\sigma_0^2}
\label{k1}
\end{equation}
that is more directly accessible to observations and numerical simulations. For example, this parameter was used by Katz \cite{katzking} and Cohn \cite{cohn} in their studies of globular clusters. Since $k=\beta(\epsilon_m-\Phi_0)$ with $\epsilon_m=\Phi(R)=-GM/R$, we get $\Phi_0=-(k+\eta)/\beta$ where $\eta(k)$ is given by Eq. (\ref{temp3}). Combining this relation with Eq. (\ref{dis3}), we obtain
\begin{equation}
{\cal K}=\frac{3}{2}\lbrack k+\eta(k)\rbrack \frac{I_1(k)}{I_2(k)}.
\label{k2}
\end{equation}
For the classical King model, ${\cal K}$ is a monotonically increasing function of $k$ so it can be used equivalently to parameterize the series of equilibria. For an extended classical King cluster (large $k$), we have $\beta\sim 1/\sigma_0^2$ and $\Phi_0\gg\Phi(R)=\epsilon_m$ so that $k\sim -\Phi_0/\sigma_0^2\sim {\cal K}$.

\subsection{Kinetic and thermodynamic specific heats}
\label{sec_heat}

If we define the kinetic temperature $T_{kin}$ through the relation $K=(3/2)Nk_B T_{kin}$ (where $K$ denotes the kinetic energy), we find from the virial theorem (\ref{ene1}) that the kinetic caloric curve is simply given by $E=-(3/2)Nk_BT_{kin}$. Therefore, the kinetic specific heat is
\begin{equation}
C_{kin}=\frac{dE}{dT_{kin}}=-\frac{3}{2}Nk_B<0.
\label{heat1}
\end{equation}
It has a constant negative value.  However, $T_{kin}$ is not the thermodynamic temperature in the present case. The thermodynamic temperature is $T=1/\beta$ and the thermodynamic specific heat is
\begin{equation}
C=\frac{dE}{dT}.
\label{heat1bis}
\end{equation}
When the distribution function is non-Boltzmannian, the kinetic and thermodynamic caloric curves $T_{kin}(E)$ and $T(E)$ can be very different.\footnote{In particular, the thermodynamic specific heat is necessarily positive in CE while the kinetic specific heat may be positive or negative in CE \cite{prl}.} As we shall see in Sec.  \ref{sec_kingclass} for the  King model, the thermodynamic specific heat is not constant and differs from Eq. (\ref{heat1}).

\subsection{Ensemble inequivalence and Poincar\'e theory on the linear series of equilibria}
\label{sec_poincare}

The maximization problems   (\ref{vp2}) and (\ref{vp3}) have the same critical points. They correspond to the distribution function (\ref{vp9}). However, these maximization problems may not be equivalent. The stability of the distribution function (\ref{vp9}) may differ in MCE and CE. As a result, the set of solutions of (\ref{vp2}) may not coincide with the set of solutions of (\ref{vp3}). It can be shown that the solution of a maximization problem is always the solution of a more constrained dual maximization problem \cite{ellis}. Therefore, a solution of (\ref{vp3}) with given $\beta$ is always a solution of (\ref{vp2}) with the corresponding $E$. In the thermodynamical analogy, this means that ``canonical stability implies microcanonical stability'': $(\ref{vp3}) \Rightarrow (\ref{vp2})$.\footnote{This can be checked at the level of the second order variations. Indeed, if inequality (\ref{vp10}) is satisfied for all perturbations $\delta f$ that conserve mass (canonical stability criterion), it is {\it a fortiori} satisfied for all perturbations that conserve mass {\it and} energy at  first order (microcanonical stability criterion).} However, the converse in wrong: a solution of (\ref{vp2}) is not necessarily a solution of (\ref{vp3}). When this happens, we speak of ensemble inequivalence. Ensemble inequivalence is generic for systems with long-range interactions but it is not compulsory.

In order to determine the stability of a distribution function according to the maximization problems   (\ref{vp2}) and (\ref{vp3}) we can use the theory of Poincar\'e on the linear series of equilibria \cite{poincare}. This is a powerful graphical method that just requires to determine the critical points of  (\ref{vp2}) and (\ref{vp3}) and plot the series of equilibria $\beta(E)$. This theory uses the fact that $\beta=\partial S/\partial E$  in MCE (the inverse temperature is the conjugate of the energy with respect to the entropy) and $E=-\partial J/\partial\beta$ in CE (minus the energy is the conjugate of the inverse temperature with respect to the free energy). It can be shown that a change of stability can occur only at a turning point or at a bifurcation point of the series of equilibria. In this paper and in Paper II, we shall only encounter the case of turning points. If we plot $\beta$ as a function of $-E$, we have the following results. In MCE, a change of stability can only occur at a turning point of energy where $d\beta/dE=\infty$. A mode of stability is lost if the curve rotates clockwise and gained if it rotates anti-clockwise. In CE, a change of stability can only occur at a turning point of temperature where $d\beta/dE=0$. A mode of stability is lost if the curve rotates clockwise and gained if it rotates anti-clockwise. We refer to Katz \cite{katzpoincare} and Chavanis \cite{ijmpb} for an application of the Poincar\'e theory to the case of self-gravitating systems.

\section{The classical King model}
\label{sec_kingclass}

In this section, we apply the general formalism developed previously to the case of the classical King model.

\subsection{The distribution function}
\label{sec_df}

\begin{figure}[!h]
\begin{center}
\includegraphics[clip,scale=0.3]{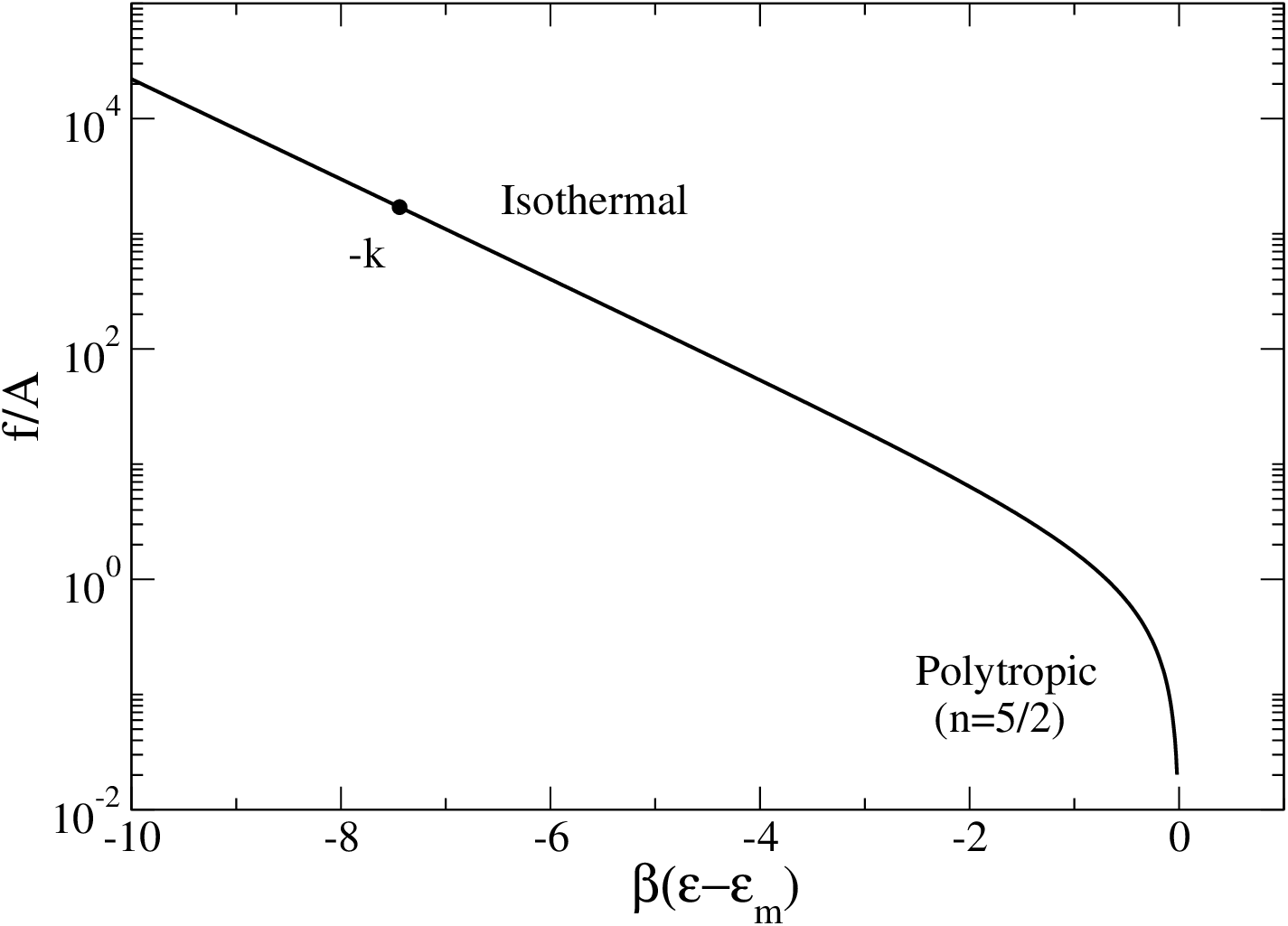}
\caption{The distribution function $f(\epsilon)$ in scaled variables showing the
isothermal core, the isothermal halo, and the polytropic envelope.}
\label{feps}
\end{center}
\end{figure}

The classical King model is defined by
\begin{equation}
f=A \left \lbrack e^{-\beta(\epsilon-\epsilon_m)}-1\right \rbrack \quad {\rm if} \quad \epsilon\le \epsilon_m,
\label{df1}
\end{equation}
\begin{equation}
f=0 \quad {\rm if} \quad \epsilon\ge\epsilon_m,
\label{df2}
\end{equation}
where $\epsilon_m$ is the escape energy at which the particles leave the system
($f=0$). For $\epsilon\rightarrow -\infty$, the King distribution reduces to the
Boltzmann distribution $f\sim A e^{-\beta(\epsilon-\epsilon_m)}$ and, for
$\epsilon\rightarrow \epsilon_m^{-}$, it reduces to  $f\sim
A\beta(\epsilon_m-\epsilon)$ corresponding to a  polytropic distribution of
index $n=5/2$ \cite{bt}. Therefore, the King model generically describes a
cluster with an
isothermal core, an isothermal halo, and a   polytropic envelope of index
$n=5/2$. The proportion of these different regions depends on the concentration
parameter $k$ as shown in the sequel. The distribution
function $f(\epsilon)$ is represented in Fig. \ref{feps}.

The King distribution is of the form of Eqs. (\ref{fde1}) and (\ref{fde2}) with
\begin{equation}
{\cal F}_s(x)=e^{-x}-1.
\label{df3}
\end{equation}
The corresponding entropy is given by Eq. (\ref{vp1}) with (see Paper II):
\begin{equation}
C(f)=A \left \lbrack \left (1+\frac{f}{A}\right ) \ln \left (1+\frac{f}{A}\right )-\frac{f}{A}\right\rbrack-\ln\left (\frac{\eta_0}{A}\right )f.
\label{df4}
\end{equation}

\begin{figure}[!h]
\begin{center}
\includegraphics[clip,scale=0.3]{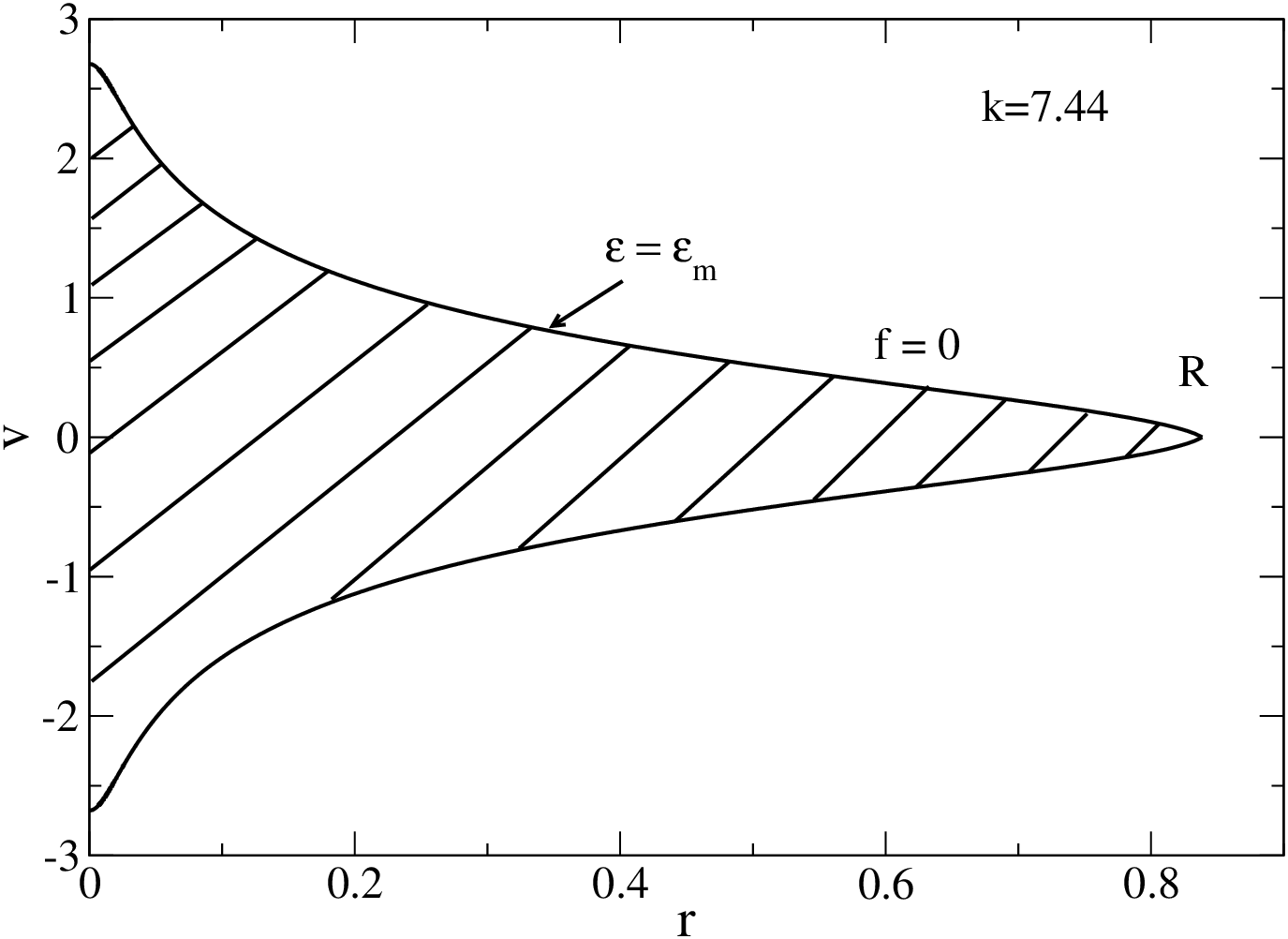}
\caption{Phase-space portrait of the King model for $k=k_{MCE}=7.44$ (see below). The velocity is normalized
by $(2/\beta)^{1/2}$ and the radial distance by $r_0$. The envelope where $f=0$, corresponding to $\epsilon=\epsilon_m$,
is given by $w_m=\sqrt{\chi(\zeta)}$ in scaled variables.}
\label{pskMCE}
\end{center}
\end{figure}

For the King model, the functions $I_n(z)$  defined in the general case by Eq. (\ref{fde7}) can be written, after an integration by parts, as
\begin{equation}
I_n(z)=\frac{8\pi e^z}{2n+1} \int_0^{\sqrt{z}} e^{-w^2} w^{2n+2}\, dw.
\label{df5}
\end{equation}
These functions may be expressed in terms of the error function. Their asymptotic behaviors for small and large values of $z$ are easily obtained. For $z\rightarrow 0$, we get
\begin{equation}
I_n(z)\sim \frac{8\pi}{(2n+1)(2n+3)}z^{(2n+3)/2},
\label{df6}
\end{equation}
so that $I_1(z)\sim ({8\pi}/{15})z^{5/2}$ and $I_2(z)\sim ({8\pi}/{35})z^{7/2}$. For $z\rightarrow +\infty$, we get
\begin{equation}
I_n(z)\sim \frac{4\pi e^z}{2n+1}\Gamma\left (n+\frac{3}{2}\right ),
\label{df8}
\end{equation}
so that $I_1(z)\sim \pi^{3/2}e^z$ and $I_2(z)\sim ({3}/{2})\pi^{3/2}e^z$. The density profile of the King model is given by Eq. (\ref{fde10}) where $\chi$ is the solution of the differential equation (\ref{fde13})-(\ref{fde14}) with the function $I_1(z)$ defined by Eq. (\ref{df5}). The density profile vanishes at a radius $r=R$ corresponding to the tidal radius. The phase space portrait of the King model is represented in Fig. \ref{pskMCE}.

\subsection{The equation of state}
\label{sec_eos}

The equation of state $p_T(\rho)$ of the King model is defined by the parametric equations (\ref{fde8}) and (\ref{ene3}) where $I_1(z)$ and $I_2(z)$ are given  by Eq. (\ref{df5}).

For $\chi\rightarrow +\infty$, we find that
\begin{equation}
\rho\sim A\left (\frac{2}{\beta}\right )^{3/2} \pi^{3/2}e^{\chi},\quad p\sim \frac{1}{3} A\left (\frac{2}{\beta}\right )^{5/2} \frac{3}{2}\pi^{3/2}e^{\chi},
\label{eos1}
\end{equation}
leading to the isothermal equation of state
\begin{equation}
p\sim \frac{\rho}{\beta}.
\label{eos2}
\end{equation}
This equation of state is valid at high densities.

For $\chi\rightarrow 0$, we find that
\begin{equation}
\rho\sim A\left (\frac{2}{\beta}\right )^{3/2} \frac{8\pi}{15}\chi^{5/2},\quad p\sim \frac{1}{3} A\left (\frac{2}{\beta}\right )^{5/2} \frac{8\pi}{35}\chi^{7/2},
\label{eos3}
\end{equation}
leading to the polytropic equation of state
\begin{equation}
p\sim \frac{1}{7}\left (\frac{15}{4\pi A\beta}\right )^{2/5}\rho^{7/5}.
\label{eos4}
\end{equation}
This equation of state is valid at low densities.

\begin{figure}[!h]
\begin{center}
\includegraphics[clip,scale=0.3]{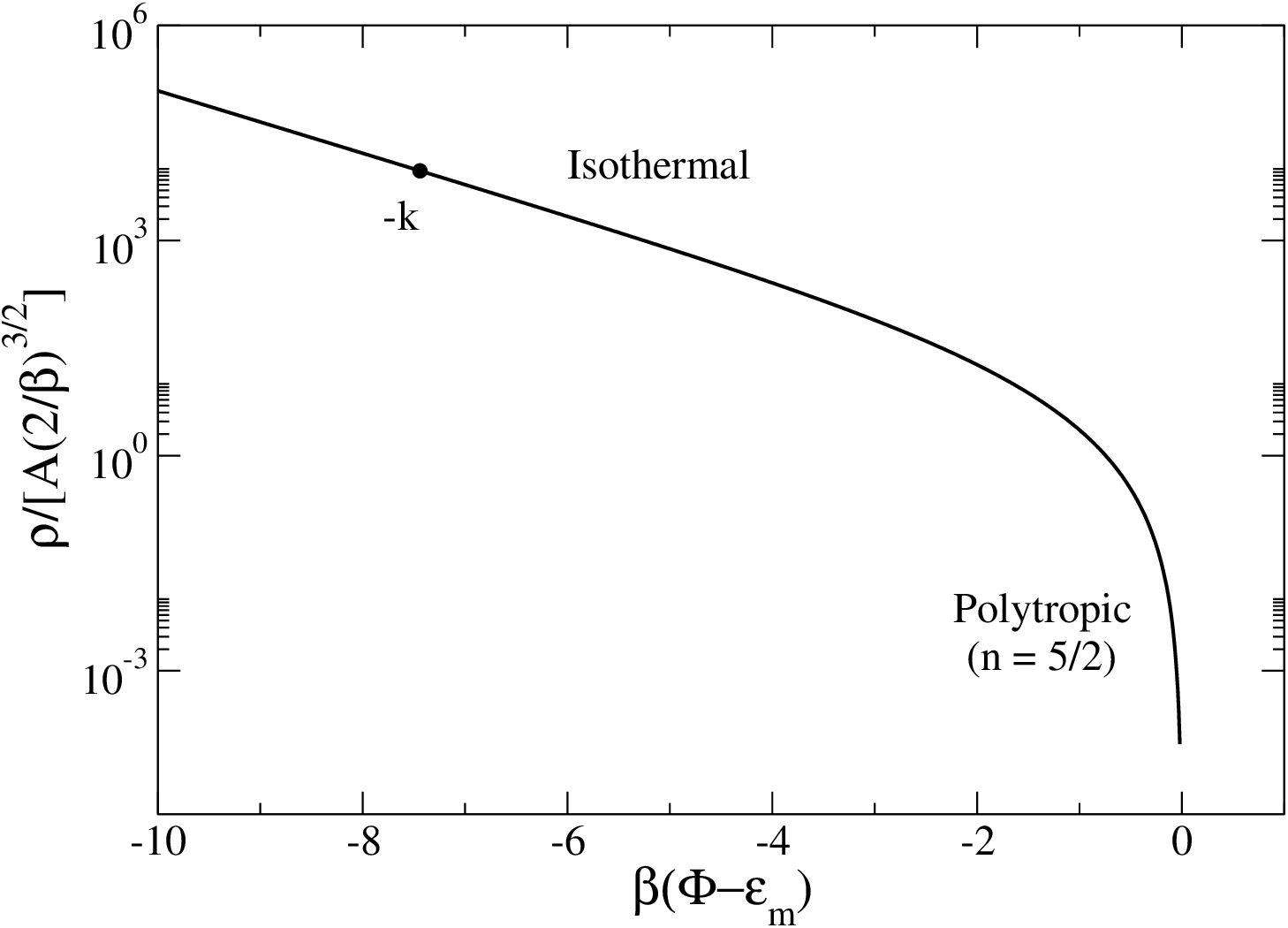}
\caption{The density $\rho(\Phi)$ in scaled variables showing the isothermal
core, the isothermal halo, and the polytropic envelope.}
\label{rhophi}
\end{center}
\end{figure}

For $\Phi\rightarrow -\infty$ the density is related to the
gravitational potential by the Boltzmann distribution $\rho(\Phi)\propto
e^{-\beta\Phi}$ and for $\Phi\rightarrow \epsilon_m$ the density is related to
the gravitational potential by the distribution $\rho(\Phi)\propto
(\epsilon_m-\Phi)^{5/2}$ corresponding to a polytrope of index $n=5/2$. The
relation $\rho(\Phi)$ is represented in Fig. \ref{rhophi}.

\subsection{The polytropic limit $k\rightarrow 0$}
\label{sec_poly}

In the limit $k\rightarrow 0$, the function $\chi$ is always small, so we can use the approximation (\ref{df6}) everywhere. As a result, the King model is equivalent to a pure polytrope ($p=K\rho^{1+1/n}$) of index $n=5/2$ and polytropic constant $K=(1/7)\left ({15}/{4\pi A\beta}\right )^{2/5}$. Defining $\theta=\chi/k$ and $\xi=\zeta/\sqrt{k}$, we find that the differential equation (\ref{fde13}) reduces to the Lane-Emden equation
\begin{equation}
\frac{1}{\xi^2}\frac{d}{d\xi}\left (\xi^2\frac{d\theta}{d\xi}\right )=-\theta^{5/2}
\label{poly1}
\end{equation}
\begin{equation}
\theta(0)=1,\qquad \theta'(0)=0,
\label{poly2}
\end{equation}
corresponding to a polytrope $n=5/2$ \cite{chandra}. Solving this equation numerically, we obtain $\xi_1=5.36$ and $\theta'_1=-7.63\, 10^{-2}$.  Using the theory of polytropes, we can analytically obtain the mass-radius relation and the expression of the energy. This allows us to obtain an analytical expression of the series of equilibria $\tilde\beta(\tilde E)$ for $k\rightarrow 0$. We proceed as follows.

Using the virial theorem $2K+W=0$, the total energy $E=K+W$ is given by
\begin{equation}
E=\frac{W}{2}.
\label{poly3}
\end{equation}
The potential energy of a polytrope of index $n<5$ is
\begin{equation}
W=-\frac{3}{5-n}\frac{GM^2}{R}.
\label{poly4}
\end{equation}
Specializing to the index $n=5/2$, and using Eq. (\ref{poly3}), we obtain
\begin{equation}
E=-\frac{3}{5}\frac{GM^2}{R}.
\label{poly5}
\end{equation}
The mass-radius relation of a polytrope of index $n<5$ is
\begin{equation}
M^{(n-1)/n}R^{(3-n)/n}=\frac{K(n+1)}{G(4\pi)^{1/n}}\omega_n^{(n-1)/n},
\label{poly6}
\end{equation}
where $\omega_n\equiv -\xi_1^{(n+1)/(n-1)}\theta'_1$. For $n=5/2$, we have $\omega_{5/2}\equiv -\xi_1^{7/3}\theta'_1=3.83$. Using the expression of $K$  in terms of $A$ and  $\beta$ given above, Eq. (\ref{poly6}) takes the form
\begin{equation}
M^{3}R=\frac{\lambda}{G^5A^2\beta^2}
\label{poly7}
\end{equation}
with $\lambda={225}\omega_{5/2}^3/({8192\pi^4})=1.58\, 10^{-2}$. Combining Eqs. (\ref{poly5}) and (\ref{poly7}), we get
\begin{equation}
E=-\frac{3G^6M^5A^2\beta^2}{5\lambda}.
\label{poly9}
\end{equation}
Introducing the dimensionless temperature, the dimensionless energy, and the dimensionless radius defined by Eqs.  (\ref{temp5}), (\ref{ene7}) and (\ref{rad1}), Eqs. (\ref{poly7}) and (\ref{poly9}) lead to
\begin{equation}
\tilde{R}=\frac{\lambda (8\pi\sqrt{2})^2}{\tilde\beta^2}=\frac{20.0}{\tilde\beta^2},
\label{poly10b}
\end{equation}
\begin{equation}
\tilde\beta=8\pi\left (\frac{10\lambda}{3}\right )^{1/2}(-\tilde E)^{1/2}=5.77 \, (-\tilde E)^{1/2}.
\label{poly10}
\end{equation}
The radius and the energy are related by $\tilde{R}=-3/(5\tilde{E})$. According to Eqs. (\ref{temp5}) and (\ref{df6}),  we also have
\begin{equation}
\tilde\beta=\left (\frac{15}{8\pi}\right )^{2/3}(-\xi_1^2 \theta'_1)^{4/3}\, k^{1/3}=2.02\,  k^{1/3}
\label{poly11}
\end{equation}
from which we get $\tilde{R}=4.90\, k^{-2/3}$ and $\tilde{E}=-0.123\, k^{2/3}$.
These relations are valid for $k\rightarrow 0$, hence for
$\tilde\beta\rightarrow 0$, $\tilde E\rightarrow 0$, and $\tilde{R}\rightarrow
+\infty$. We also note that
$\epsilon\rightarrow -3/5$ and $\eta\sim -\xi_1 \theta'_1 k\sim 0.409\, k\rightarrow 0$.

\subsection{The isothermal limit $k\rightarrow \infty$}
\label{sec_iso}

In the limit $k\rightarrow +\infty$, the function $\chi$ is always large, except close to the tidal radius, so we can use the approximation (\ref{df8}) in almost all the cluster. As a result, the King model is almost equivalent to an isothermal sphere ($p=\rho/\beta$). Defining $\psi=k-\chi$ and $\xi=\zeta$, we find that the differential equation (\ref{fde13}) reduces almost everywhere to the Emden equation
\begin{equation}
\frac{1}{\xi^2}\frac{d}{d\xi}\left (\xi^2\frac{d\psi}{d\xi}\right )=e^{-\psi},
\label{iso1}
\end{equation}
\begin{equation}
\psi(0)=0,\qquad \psi'(0)=0,
\label{iso2}
\end{equation}
corresponding to the isothermal sphere \cite{chandra}.

\section{The proper thermodynamic treatment (fixed  $A$)}
\label{sec_proper}

In this section, we develop the proper thermodynamic treatment of the
King model associated to the maximization problems (\ref{vp2}) and
(\ref{vp3}). As explained in Sec. \ref{sec_gen}, in order to solve
these maximization problems, we must work at fixed $A$, not at fixed
$R$. Therefore, the thermodynamical parameters must be normalized by
$A$, not by $R$. Accordingly, the thermodynamical parameters denoted
$\beta$, $E$ and $R$ in this section correspond to the dimensionless
parameters $\tilde{\beta}$, $\tilde{E}$ and $\tilde{R}$ defined by
Eqs. (\ref{temp5}), (\ref{ene7}) and (\ref{rad1}). On the other hand,
$S$ and $J$ refer to $S/M$ and $J/M$.

\begin{figure}[!h]
\begin{center}
\includegraphics[clip,scale=0.3]{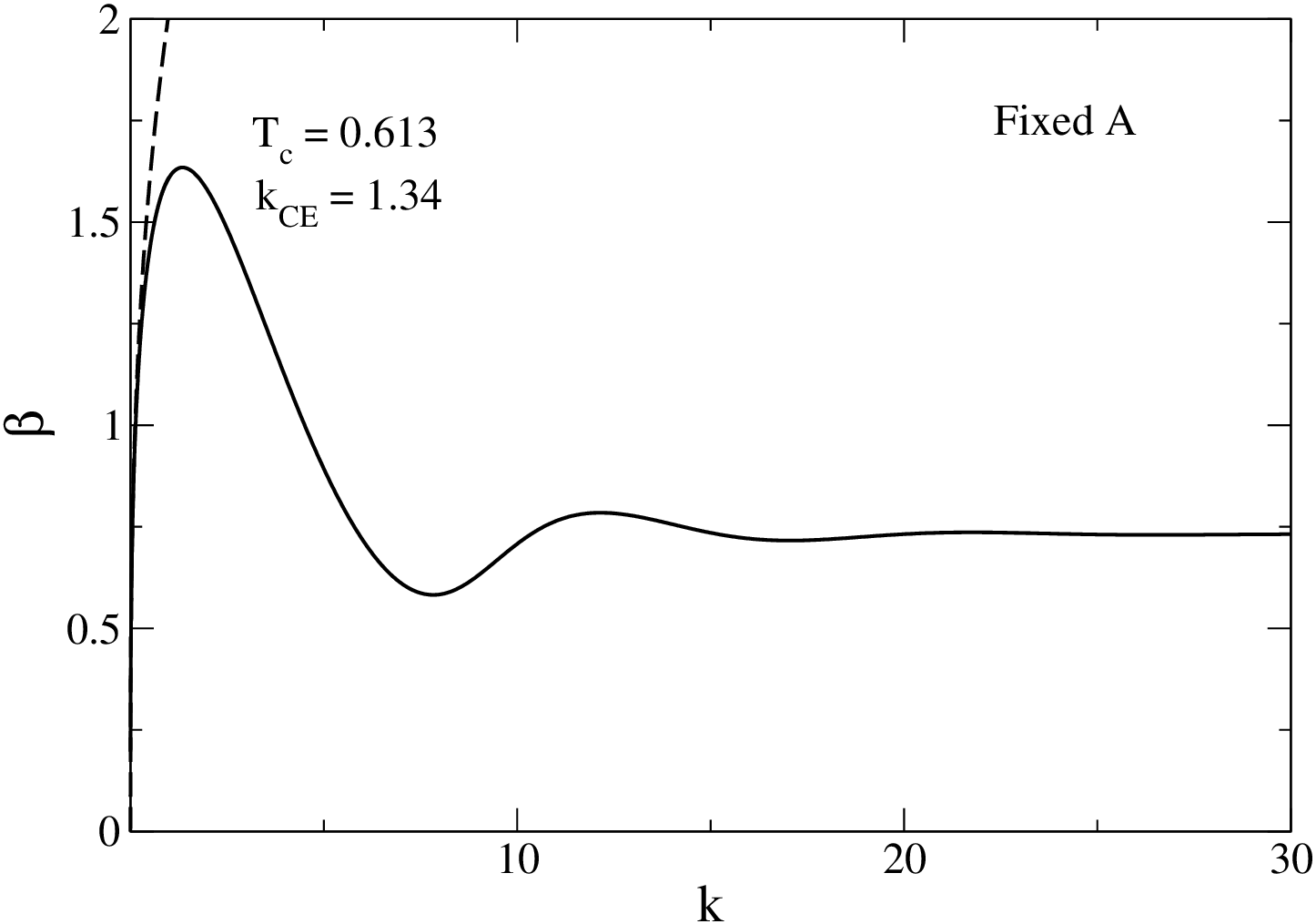}
\caption{Series of equilibria giving the inverse temperature $\beta$ as a
function of the concentration parameter $k$ for the classical King model. The
dashed line corresponds to the
analytical formula obtained in the polytropic
approximation.}
\label{abA}
\end{center}
\end{figure}

\begin{figure}[!h]
\begin{center}
\includegraphics[clip,scale=0.3]{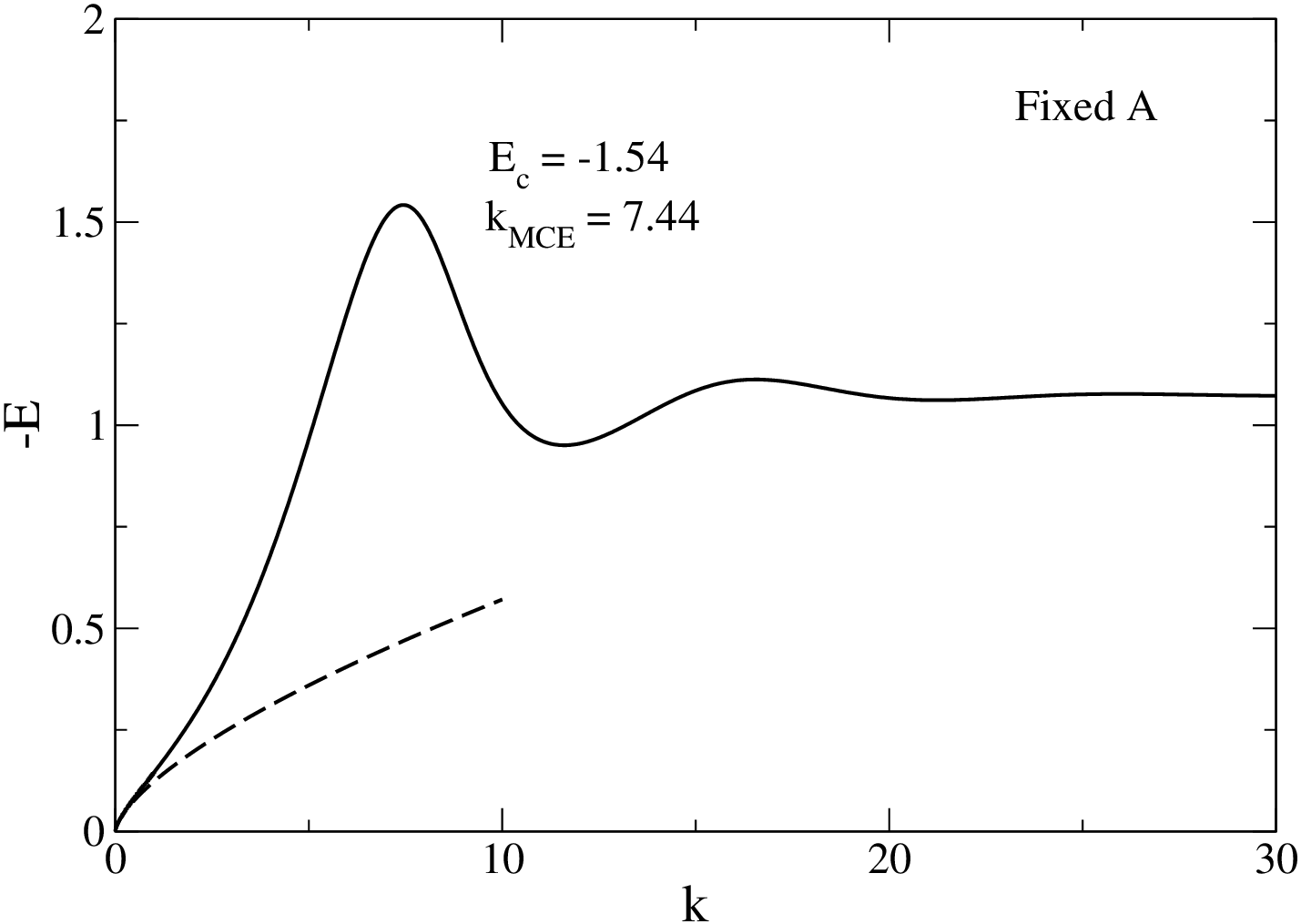}
\caption{Series of equilibria giving the energy $-E$ as a function of the
concentration parameter $k$ for
the classical King model.
The  polytropic approximation is valid for relatively low values of $E$, before
the inflection point of the curve $E(k)$ occurring at about $k\sim 0.6$ and
$E\sim -0.1$.}
\label{aeA}
\end{center}
\end{figure}

In Figs. \ref{abA} and \ref{aeA}, we plot the inverse temperature $\beta$ and the energy $-E$ as a function of the normalized central potential $k$ parameterizing the series of equilibria.

For small $k$, the system is equivalent to a polytrope of index $n=5/2$ and the
functions $\beta(k)$ and $E(k)$ are approximately given by Eqs. (\ref{poly10})
and (\ref{poly11}) represented as dashed lines in Figs. \ref{abA} and \ref{aeA}.
 For large $k$, the system is similar to the isothermal sphere ($n=+\infty$). As
for a classical isothermal sphere confined within a box (see, e.g.,  Figs. 3 and 5 in
\cite{pt}), the curves $\beta(k)$ and $E(k)$ present damped oscillations about some asymptotes $\beta=\beta_{\infty}$ and $E=E_{\infty}$. For the King model, $\beta_{\infty}=0.731$ and $E_{\infty}=-1.07$. The temperature has a first peak at
($k_{CE}=1.34$, $\beta_c=1.63$) and the energy has a first peak at ($k_{MCE}=7.44$,
$E_c=-1.54$). For box-confined isothermal spheres, we have $k_{CE}^{box}=3.47$ and $k_{MCE}^{box}=6.56$ where $k_{box}=\beta(\Phi(R)-\Phi_0)$.

\begin{figure}[!h]
\begin{center}
\includegraphics[clip,scale=0.3]{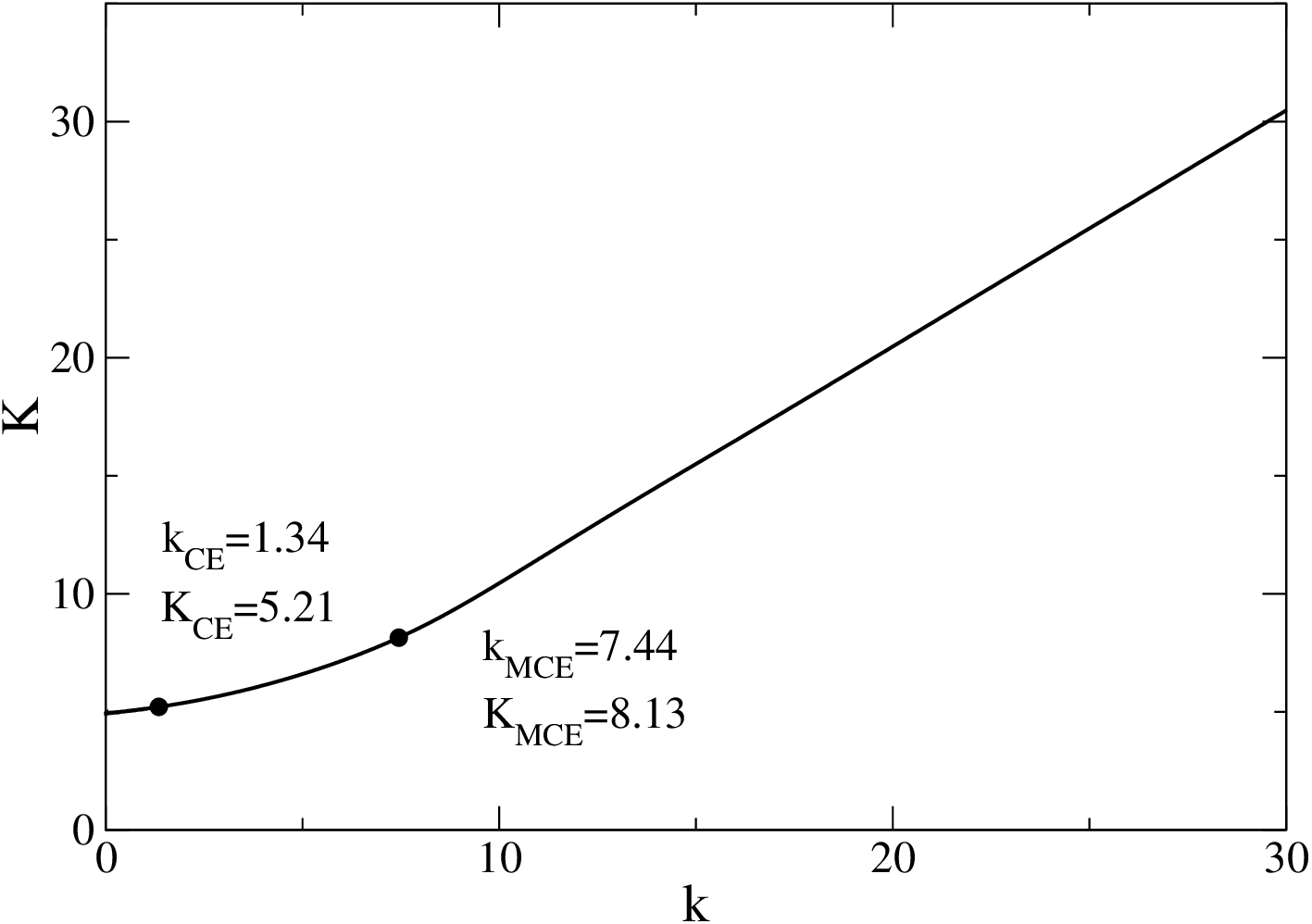}
\caption{Relation between the parameters $k$ and ${\cal K}$. The bullets indicate the limits of canonical and microcanonical stability.}
\label{kK}
\end{center}
\end{figure}

Instead of parameterizing the series of equilibria by $k$, we can use the parameter ${\cal K}$ defined
in Sec. \ref{sec_k}. For the King model, the function ${\cal K}(k)$ was computed
by Katz \cite{katzking}. It is recalled in Fig. \ref{kK} for completeness. From
this figure, we find that ${\cal K}_{CE}=5.21$ and ${\cal K}_{MCE}=8.13$. Using
the results of Sec. \ref{sec_kingclass}, we can easily establish that  ${\cal
K}(0)=(7/2)(1-\xi_1\theta'_1)=4.93$ and ${\cal K}(k)\sim k$ for $k\rightarrow
+\infty$.

\begin{figure}[!h]
\begin{center}
\includegraphics[clip,scale=0.3]{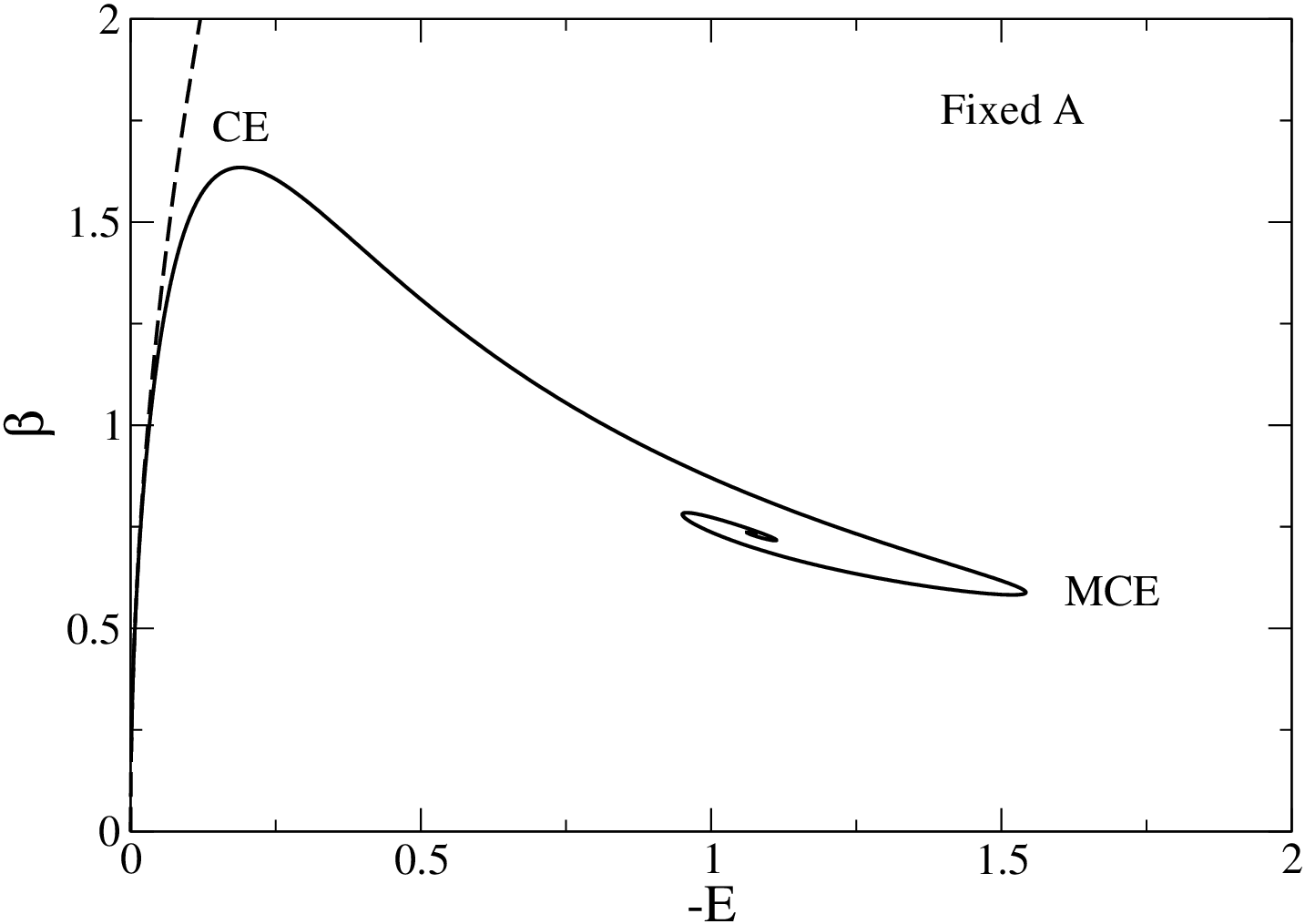}
\caption{Series of equilibria  (parameterized by $k$)  giving the inverse
temperature $\beta$ as a function of the energy $-E$ for the classical King
model.}
\label{ebA}
\end{center}
\end{figure}

The series of equilibria  $\beta(-E)$ is plotted in Fig. \ref{ebA}.  This curve
updates the one given by Katz \cite{katzking} that was drawn by hand for large
values of $k$. This curve has a snail-like structure (spiral) similar to the
series of equilibria of classical isothermal spheres confined within a box (see,
e.g.,  Fig. 1 in \cite{pt}). We note, however, that the energy is always
negative in the present case. This is a consequence of the virial theorem
(\ref{ene1}) for a self-confined system. By contrast, for box-confined
isothermal spheres, there is an additional term in the virial theorem due to the
pressure against the boundary so the energy can be either positive or negative.
The concentration parameter $k$ increases along the series of equilibria. For
small $k$, the system is equivalent to a polytrope of index $n=5/2$ and the
function $\beta(E)$ is approximately given by Eq. (\ref{poly10}) represented as
a dashed line in Fig. \ref{ebA}. This is valid for $E\rightarrow 0^-$ and
$T\rightarrow +\infty$. For large $k$, the system approaches an isothermal
sphere ($n=+\infty$) and the series of equilibria spirals about the limit point
($E_{\infty}$, $\beta_{\infty}$).

In MCE, where the control parameter is the energy $E$, there exist equilibrium
states only for $E>E_c$ with $E_c=-1.54$ (first turning point of energy). The
critical energy $E_c$ is the equivalent of the Emden energy for box-confined
isothermal spheres \cite{emden}. For $E<E_c$ there is no equilibrium state and
the system undergoes a gravothermal catastrophe. For classical particles, this
leads to a singularity corresponding to a tight binary surrounded by a hot
halo.\footnote{This is the most probable structure in MCE. Indeed, we can
increase indefinitely the entropy $S$ of a self-gravitating system at fixed mass
and energy by approaching two particles at a very close distance to each other
and redistributing the released energy in the halo 
in the form of kinetic energy (see Appendix A of \cite{sc2002}). The binary has
a small mass $2m\ll M$ but a huge potential energy $E_{binary}\rightarrow
-\infty$. Since the total energy is fixed in MCE, the kinetic energy
(temperature) of the halo $T\rightarrow +\infty$ and, consequently, the entropy
$S\sim \frac{3}{2}Nk_B\ln T\rightarrow +\infty$. Since the halo is ``hot'', it
has the tendency to extend at large distances. It can be shown \cite{sc2002}
that the divergence of entropy is maximum when the mass in the core is the
smallest, corresponding, in the discrete case, to a binary. We can also
formulate this argument in terms of the density of states \cite{paddy}. The
density of states of a self-gravitating system diverges for $N\ge 3$ because we
can form a pair with a binding energy tending to $-\infty$ and transfer this
energy to the kinetic energy of the other stars which grows to $+\infty$ to
maximize the volume of phase space they explore.} In CE, where the control
parameter is the temperature $T$, there exist equilibrium states only for
$T>T_c$ with $T_c=0.613$ (first turning point of temperature). The critical
temperature $T_c$ is the equivalent of the Emden temperature for box-confined
isothermal spheres \cite{emden}. For $T<T_c$ there is no equilibrium state and
the system undergoes an isothermal collapse. For classical particles, this leads
to a singularity corresponding to a Dirac peak containing all the
mass.\footnote{This is the most probable structure in CE. Indeed, we can
increase indefinitely the free energy $J$ of a self-gravitating system at fixed
mass by collapsing all the particles at the same point (see Appendix B of
\cite{sc2002}). It can be shown \cite{sc2002} that the divergence of free energy
is maximum when the mass in the core is the largest. We can also formulate
this argument in terms of the partition function \cite{paddy}. The partition
function of a self-gravitating system diverges for $N\ge 2$ when all the
particles are concentrated at the same point \cite{kiessling}.}

We now investigate the stability of the classical King distributions according to the maximization problems (\ref{vp2}) and (\ref{vp3}). The stable part of the series of equilibria in each ensemble defines the caloric curve.

We first consider the canonical ensemble (\ref{vp3}) in which the control parameter is the temperature $T$. For $T\rightarrow +\infty$ the system is stable in CE since it is equivalent to a polytrope with an index $n=5/2$ smaller than the critical value $n=3$ in CE \cite{cspoly}. Using the Poincar\'e theory, we conclude that the series of equilibria is stable until the first turning point of temperature CE and that it becomes unstable after that point. In other words, the King distribution is a maximum of free energy at fixed mass for $k<k_{CE}$ and a saddle point of free energy at fixed mass for $k>k_{CE}$. Since the series of equilibria always rotates clockwise, a mode of stability is lost at each turning point of temperature, so the system is more and more unstable as $k$ increases.

We now consider the microcanonical ensemble (\ref{vp2}) in which the control parameter is the energy $E$. For $E\rightarrow 0$ the system is stable in MCE since it is equivalent to a polytrope with an index $n=5/2$ smaller than the critical value $n=5$ in MCE \cite{cspoly}. Using the Poincar\'e theory, we conclude that the series of equilibria is stable until the first turning point of energy MCE and that it becomes unstable after that point. In other words, the King distribution is a maximum of entropy at fixed mass and energy for $k<k_{MCE}$ and a saddle point of entropy at fixed mass and energy for $k>k_{MCE}$. Since the series of equilibria always rotates clockwise, a mode of stability is lost at each turning point of energy, so the system is more and more unstable as $k$ increases.

Accordingly, there exist a region of ensemble inequivalence between points CE and MCE in Fig. \ref{ebA}, i.e. for configurations with $k_{CE}<k<k_{MCE}$, where $k_{CE}=1.34$ and $k_{MCE}=7.44$ (we check, in passing, that $k_{CE}<k_{MCE}$ since a canonical equilibrium is always a microcanonical equilibrium).  This part of the series of equilibria is stable in MCE (entropy maxima at fixed mass and energy) but unstable in CE (saddle points of free energy at fixed mass). It corresponds to configurations with negative specific heat $C=dE/dT<0$. We know that such configurations are forbidden in CE while they are allowed in MCE. These results are very similar to those obtained for box-confined isothermal spheres (see, e.g., the reviews \cite{paddy,katzrevue,ijmpb}).

Since there is no global  maximum of free energy at fixed mass for classical
self-gravitating systems (see footnote 16),  the configurations with $k<k_{CE}$
in CE are only metastable (local maxima of free energy at fixed mass).
Similarly, since there is no global entropy maximum at fixed mass and energy for
classical self-gravitating systems (see footnote 15), the configurations with
$k<k_{MCE}$ in MCE are only metastable (local maxima of entropy at fixed mass
and energy). However, the probability to cross the barrier of free energy in CE,
or the barrier of entropy in MCE, and leave a metastable state, is a very rare
event as it scales as $e^{-N}$ \cite{metastable,ijmpb}. For self-gravitating
systems with a large number of particles (for example globular clusters contain
about $N=10^6$ stars and the number of particles in dark matter halos is much
larger) this probability is totally negligible. Therefore, in practice,
metastable states are stable states \cite{metastable,ijmpb}. In this sense,
self-gravitating systems described by the King model  with  $k<k_{CE}$ in CE and
with   $k<k_{MCE}$ in MCE can be considered to be at statistical equilibrium,
even if there is no statistical equilibrium state in a strict sense. Their
lifetime is controlled by evaporation and gravitational collapse as discussed in
the Introduction.

\begin{figure}[!h]
\begin{center}
\includegraphics[clip,scale=0.3]{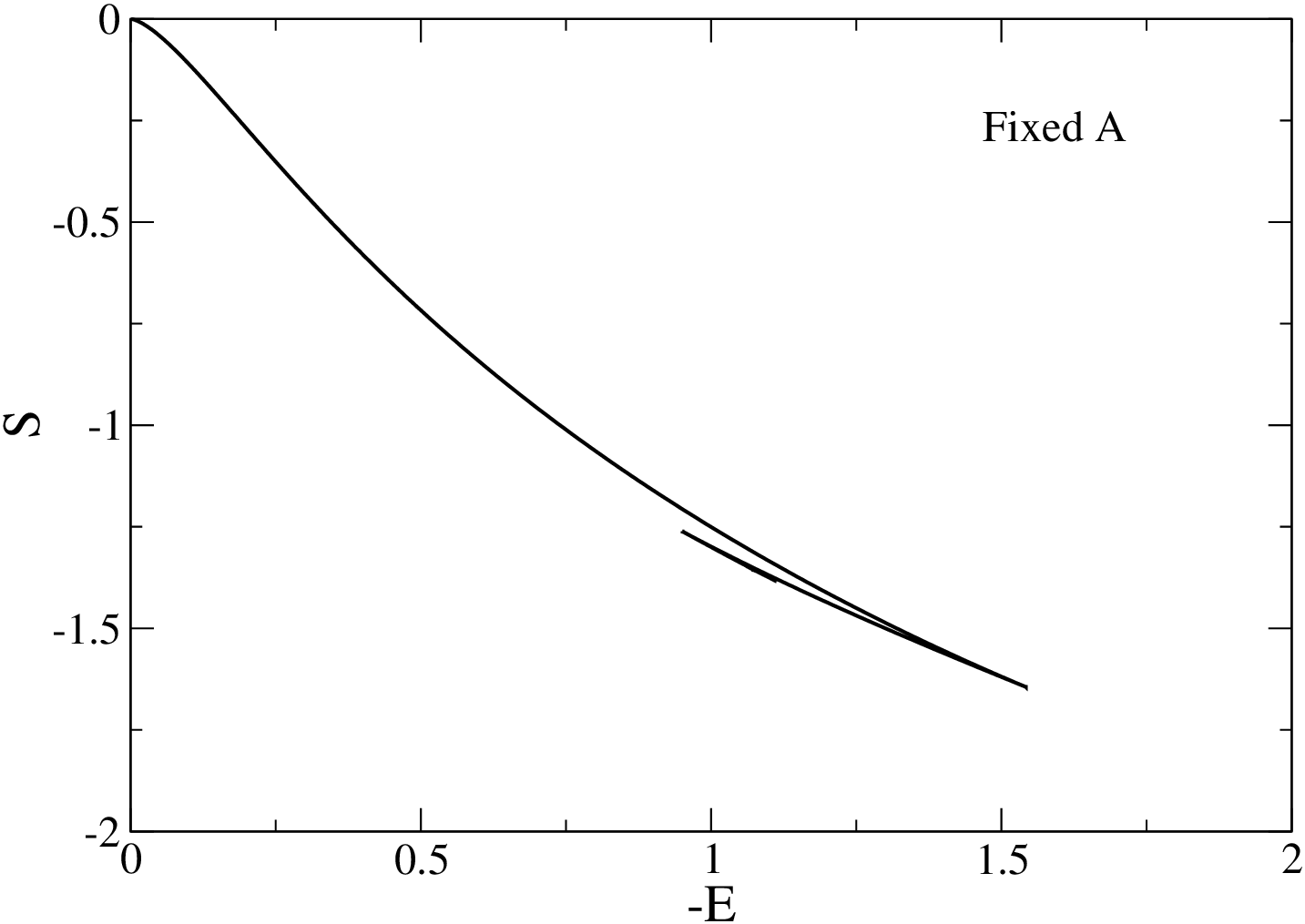}
\caption{Entropy versus energy for the classical King model.}
\label{esA}
\end{center}
\end{figure}

\begin{figure}[!h]
\begin{center}
\includegraphics[clip,scale=0.3]{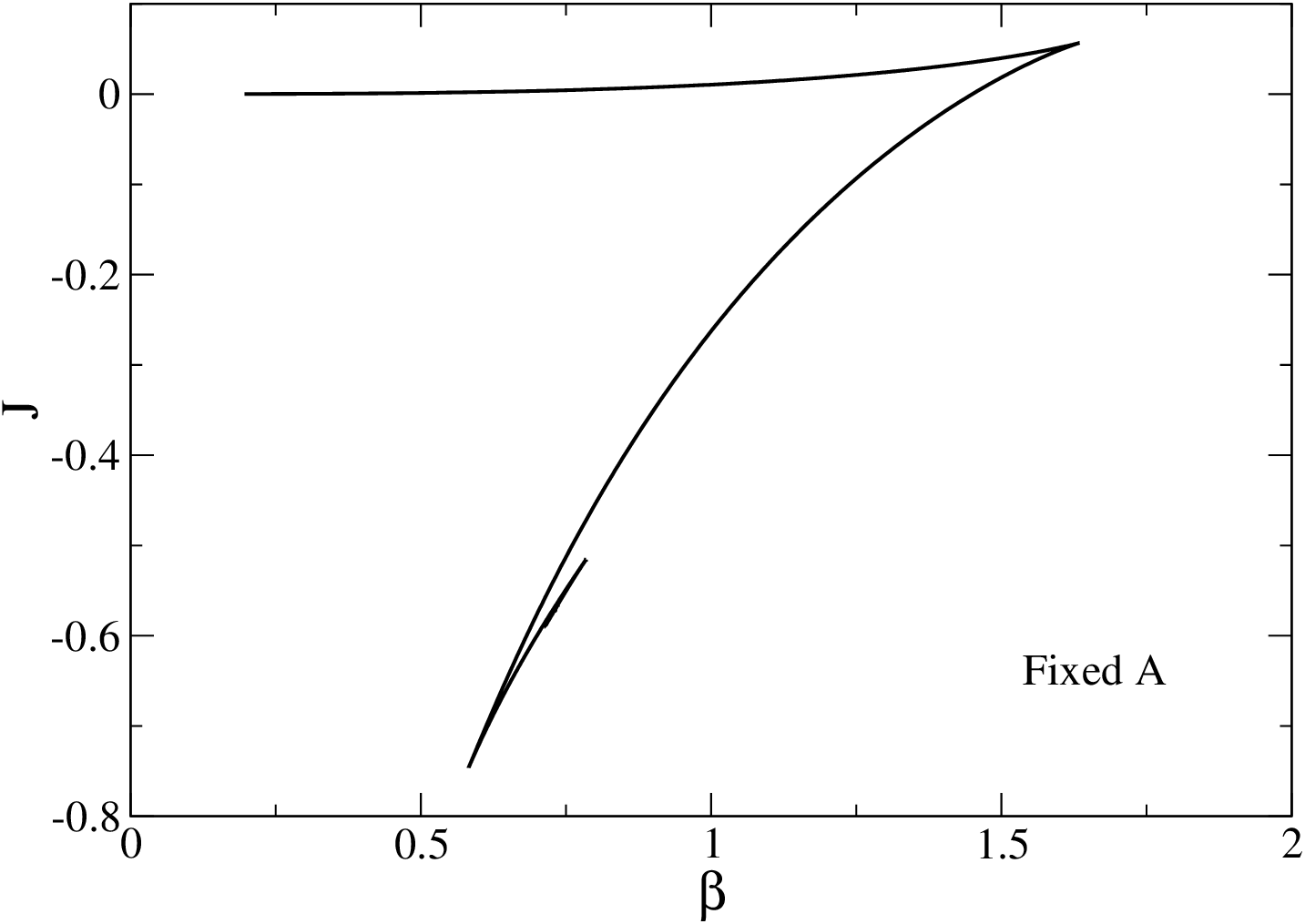}
\caption{Free energy versus inverse temperature for the classical King model.}
\label{bjA}
\end{center}
\end{figure}

The physical caloric curve in CE corresponds to the part of the series of equilibria represented in Fig. \ref{ebA} up to point CE and the physical caloric curve in MCE corresponds to the part of the series of equilibria represented in Fig. \ref{ebA} up to point MCE. They are made of long-lived metastable states.

\begin{figure}[!h]
\begin{center}
\includegraphics[clip,scale=0.3]{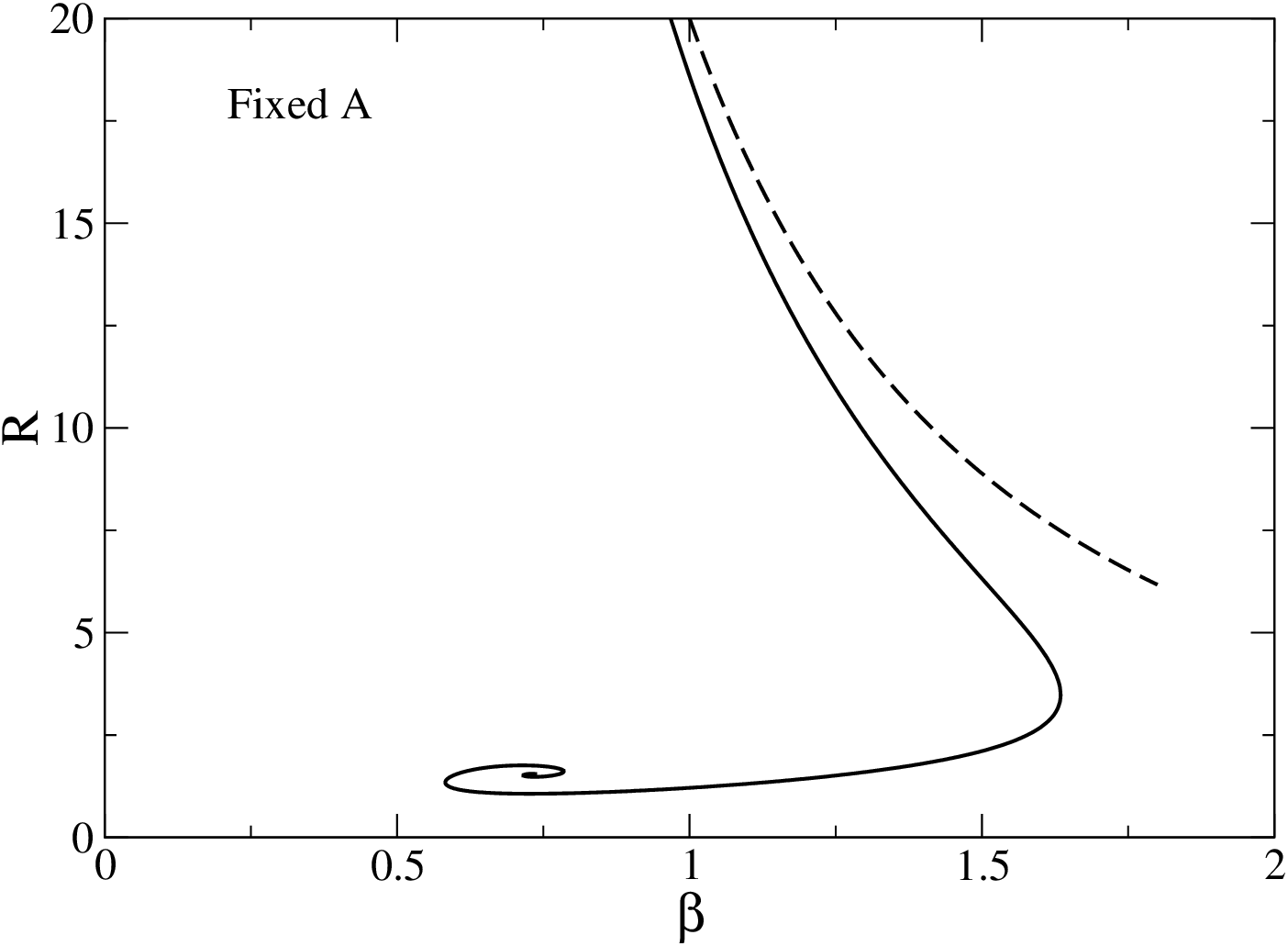}
\caption{Tidal radius versus inverse temperature for the classical King model.}
\label{betaRayon}
\end{center}
\end{figure}

\begin{figure}[!h]
\begin{center}
\includegraphics[clip,scale=0.3]{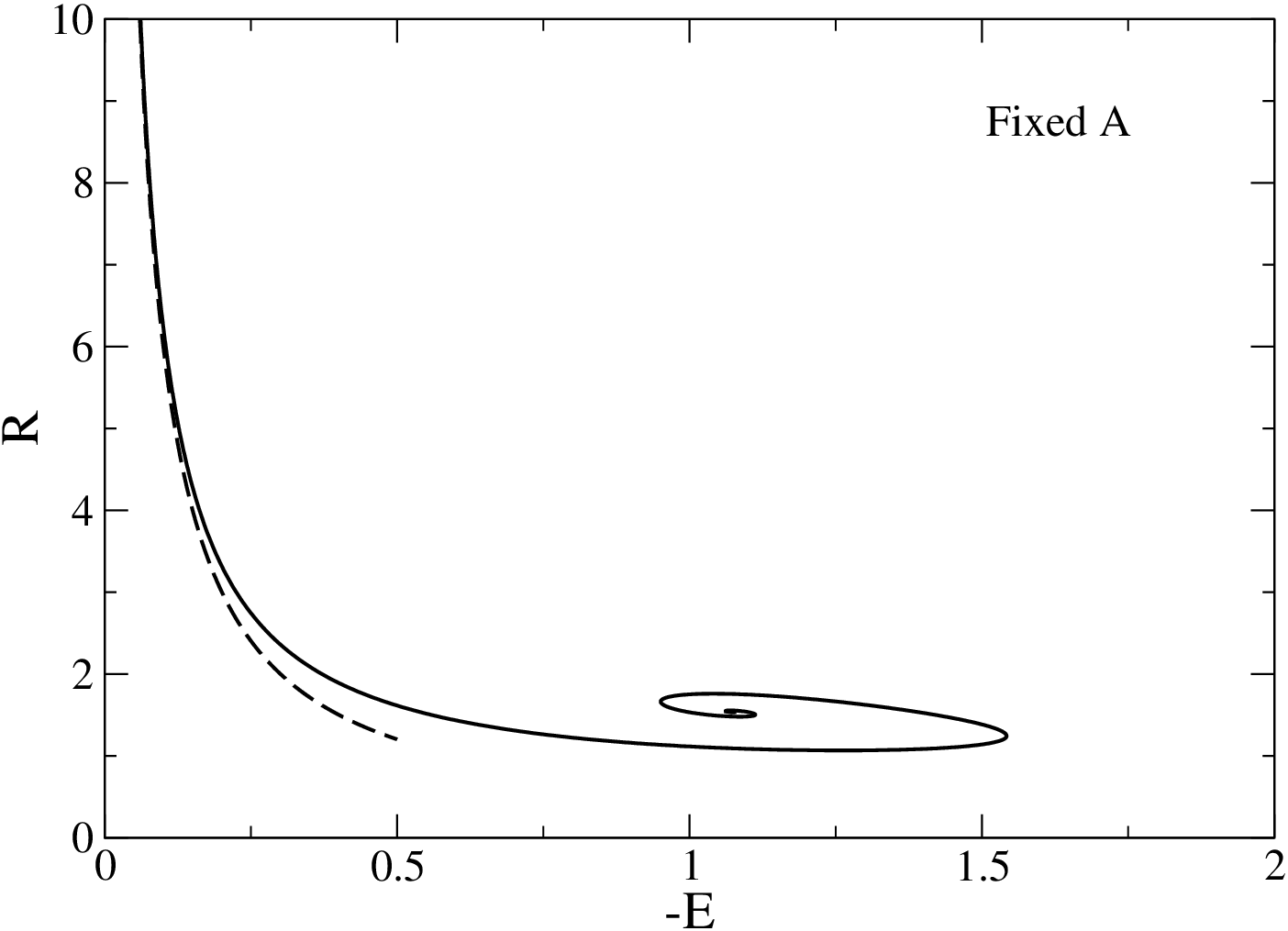}
\caption{Tidal radius versus energy for the classical King model.}
\label{energyRayon}
\end{center}
\end{figure}

\begin{figure}[!h]
\begin{center}
\includegraphics[clip,scale=0.3]{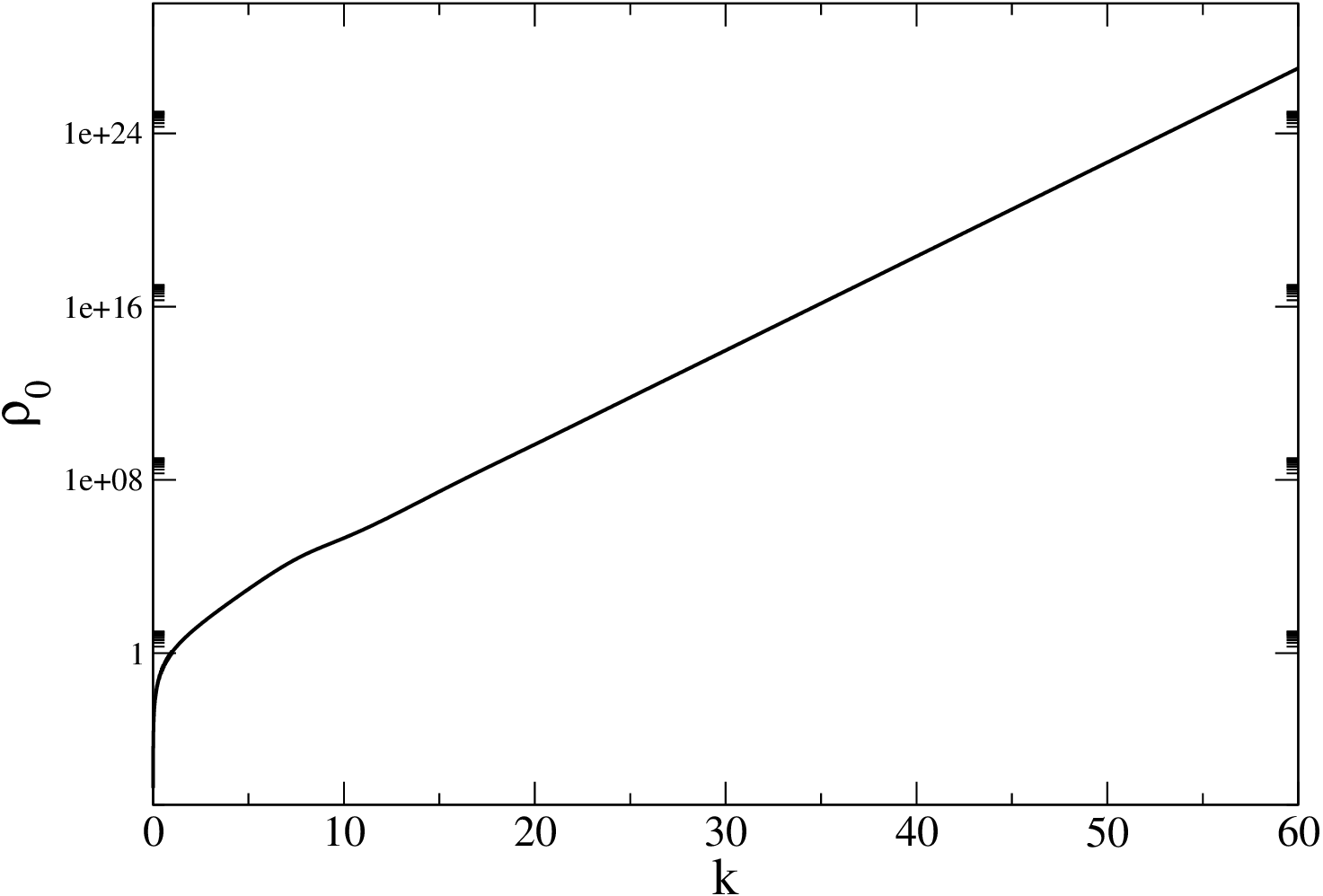}
\caption{Central density as a function of
$k$ in semi-log scales.}
\label{kdensity0}
\end{center}
\end{figure}

In Figs. \ref{esA} and \ref{bjA}, we plot the entropy $S$ as a function of the
energy $-E$ in MCE and the free energy $J$ as a function of the inverse
temperature $\beta$ in CE. Since $\delta S=\beta\delta E$ (for a fixed mass $M$)
in MCE, we find that $S(k)$ is extremum when $E(k)$ is extremum. Similarly,
since $\delta J=-E\delta\beta$ (for a fixed mass $M$) in CE, we find that $J(k)$
is extremum when $\beta(k)$ is extremum. This explains the ``spikes'' observed
in Figs \ref{esA} and \ref{bjA}. Similar spikes are found for box-confined
isothermal spheres in Newtonian gravity (see Figs. 4 and 6  in \cite{pt}) and
for box-confined self-gravitating systems described by a linear equation of
state in general relativity (see Fig. 5 in \cite{gr2008}). The series of
equilibria becomes unstable after the first spike in each ensemble. This is in
agreement with the fact that the states on the unstable branches (after the
first spike) have lower entropy or lower free energy than the states on the
stable branch (before the first spike).

In Fig. \ref{betaRayon},  we plot the tidal radius $R$ as a function of
the inverse temperature $\beta$ in CE. For $T\rightarrow +\infty$ the
tidal radius tends to $+\infty$ and it decreases as $T$ decreases. At
the critical temperature $T_c$ the value of the tidal radius is $R_{CE}=3.50$. This is
the minimum stable value of the radius in CE (as we have seen previously, the part
of the curve situated after the turning point of temperature is unstable). In
Fig. \ref{energyRayon}, we plot the tidal radius $R$ as a function of
the energy $-E$ in MCE. For $E\rightarrow 0^-$ the tidal radius tends
to $+\infty$. As $E$ decreases, the radius first decreases up to the
value $R_{min}=1.07$ (reached at $E=-1.26$) then increases. At the
critical energy $E_c$ the value of the tidal radius is $R_{MCE}=1.24$.

In Fig. \ref{kdensity0}, we plot the central density normalized by  $32\pi A^2
G^3 M^2$ as a function of $k$. This curve is monotonic so that the parameter $k$
can be considered as a measure of the central density. Using the results of Sec.
\ref{sec_kingclass}, we can easily establish that $\tilde\rho_0\sim (8\pi/15)^2
k^2/(-\xi_1^2\theta'_1)^2\sim 0.584\, k^2$ for $k\rightarrow 0$ and
$\tilde\rho_0\sim \pi^{3/2}e^k/\tilde\beta_{\infty}^{3/2}\sim 8.91\, e^k$ for
$k\rightarrow +\infty$.

Finally, even if we have not represented the curves $\tilde\epsilon_m(k)$ and ${\tilde \sigma}_0^2(k)$ for brevity, we give their asymptotic values. We find that $\tilde\epsilon_m\sim -0.202 \, k^{2/3}$ and ${\tilde \sigma}_0^2\sim 0.141\, k^{2/3}$ for $k\rightarrow 0$ and we find that $\tilde\epsilon_m\rightarrow -0.651$ and ${\tilde \sigma}_0^2\rightarrow 1.37$ for $k\rightarrow +\infty$.

\section{The effect of fixing  $R$ instead of $A$}
\label{sec_improper}

For box-confined self-gravitating classical isothermal spheres, the temperature
and the energy are normalized by the box radius $R$ (see, e.g., \cite{ijmpb}).
This is the proper normalization in that context because the box radius is a
fixed quantity. By analogy, we could normalize the temperature and the energy of
the classical King model by the tidal radius $R$ (the radius at which the
density drops to zero). This normalization was considered by Lynden-Bell and
Wood \cite{lbw} and, more recently, by Casetti and Nardini \cite{nardini}.
However, as already noted by Katz \cite{katzking}, this normalization is not
correct for a thermodynamical analysis. Indeed, when we study the maximization
problems  (\ref{vp2}) and (\ref{vp3}), we must consider that $A$, not $R$, is
fixed. It is only under this condition that the theory of Poincar\'e applies and
that the turning points of energy and temperature correspond to a change of 
thermodynamical stability in MCE and CE according to the maximization problems
(\ref{vp2}) and (\ref{vp3}). If we fix $R$ instead of $A$, the turning points of
energy and temperature do not correspond to a change of thermodynamical
stability in MCE and CE.  It is not clear whether these turning points signal
another form of instability.

We can give several arguments why $A$ should be kept fixed instead of $R$ (see
also the arguments given in the Appendix of Katz \cite{katzking}): (i)
Basically, we must fix $A$ because it explicitly enters in the expression of the
entropy functional defined by Eqs. (\ref{vp1}) and (\ref{df4}). In order to
apply the theory of Poincar\'e, all the parameters that appear in the entropy
functional must be fixed  along the series of equilibria; they act as external
parameters; (ii) It is only when $A$ is regarded as given that  a
self-gravitating system described by a distribution function of the form 
(\ref{fde1})-(\ref{fde2}) has a well-defined barotropic equation of state
$p_T(\rho)$ as discussed in Sec. \ref{sec_eosb}; (iii) According to Eq.
(\ref{chem1}),  fixing $R$ is equivalent to fixing $\epsilon_m$. However, we
expect that the tidal radius and the escape energy change along the series of
equilibria as we vary the energy or the temperature. Therefore, on a
mathematical and physical point of view, it is  more relevant to fix $A$ rather
than $R$.

To make the difference between the two prescriptions clear, we consider in this section the effect of fixing $R$ instead of $A$. Accordingly, in this section, the thermodynamical parameters $\beta$ and $E$ correspond to the dimensionless parameters $\eta$ and $\epsilon$ defined by Eqs.  (\ref{temp3}) and (\ref{ene6}). On the other hand, $S$ and $J$ still refer to $S/M$ and $J/M$ (their expressions are unchanged whether $A$ or $R$ is fixed).

\begin{figure}[!h]
\begin{center}
\includegraphics[clip,scale=0.3]{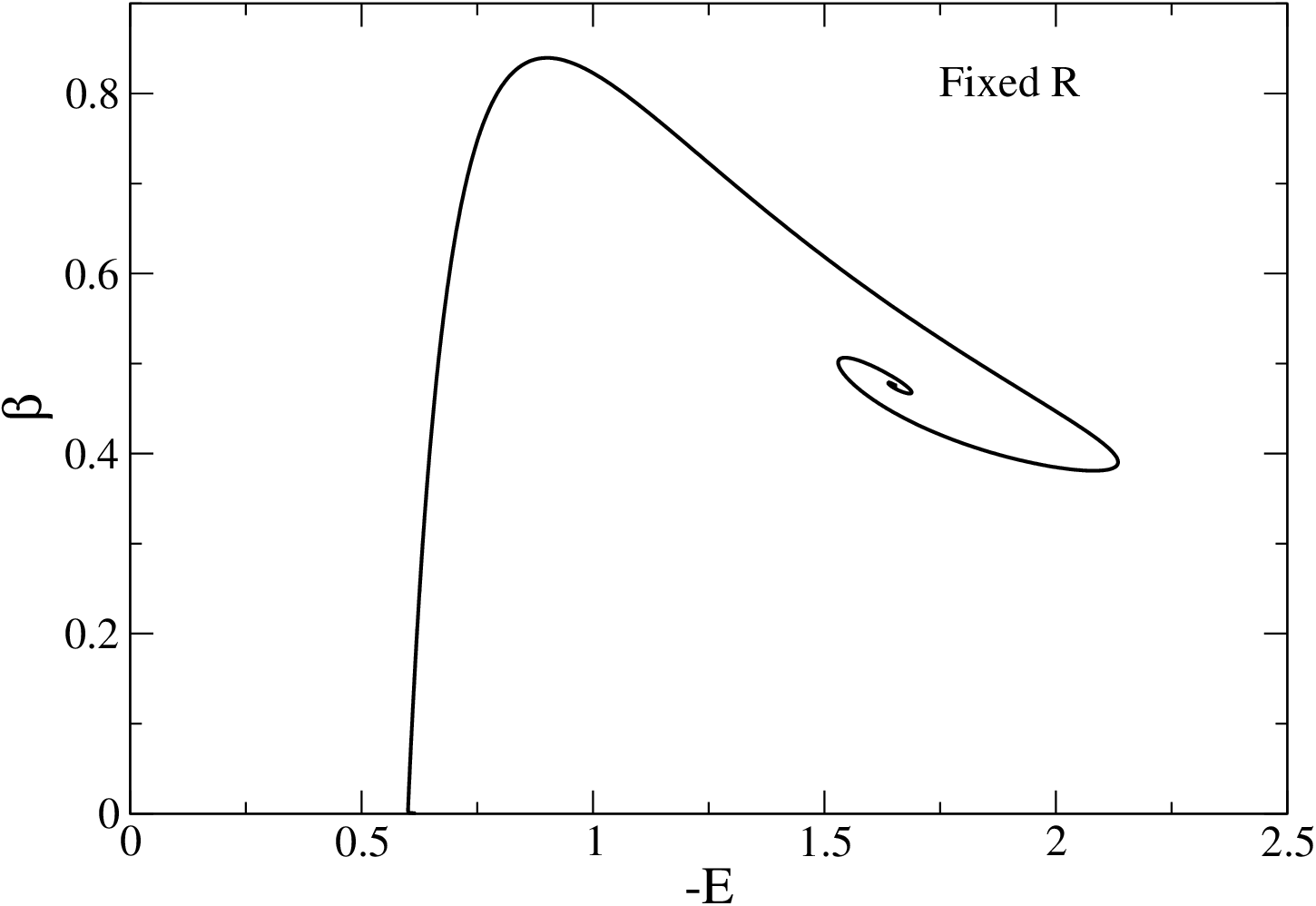}
\caption{Series of equilibria  $\beta(-E)$ for the classical King model when $R$ is fixed instead of $A$.}
\label{ebR}
\end{center}
\end{figure}

In Fig. \ref{ebR} we plot the series of equilibria $\beta(-E)$ for fixed $R$.
This curve is obtained from Eqs.  (\ref{temp3}) and (\ref{ene6}) by varying $k$
from $0$ to $+\infty$. It starts from $(-E,\beta)=(3/5,0)$ (see Eq.
(\ref{poly5})) and makes a spiral for large values of $k$ about
the point $(E_{\infty},\beta_{\infty})=(-1.65,0.476)$.
The series of equilibria  presents a first turning point of temperature at
$k_{CE}'=3.98$, $\beta_c'=0.840$  and a first turning point of energy at
$k_{MCE}'=8.50$, $E_c'=-2.13$. However, as
discussed above, the meaning of these turning points regarding the stability of
the system is unclear. At least, they indicate that, when $R$ is fixed, there is
no equilibrium below $E'_c$, or above $\beta'_c$. We also note that there is no
possible equilibrium for $E>-3/5$ when $R$ is fixed.

\begin{figure}[!h]
\begin{center}
\includegraphics[clip,scale=0.3]{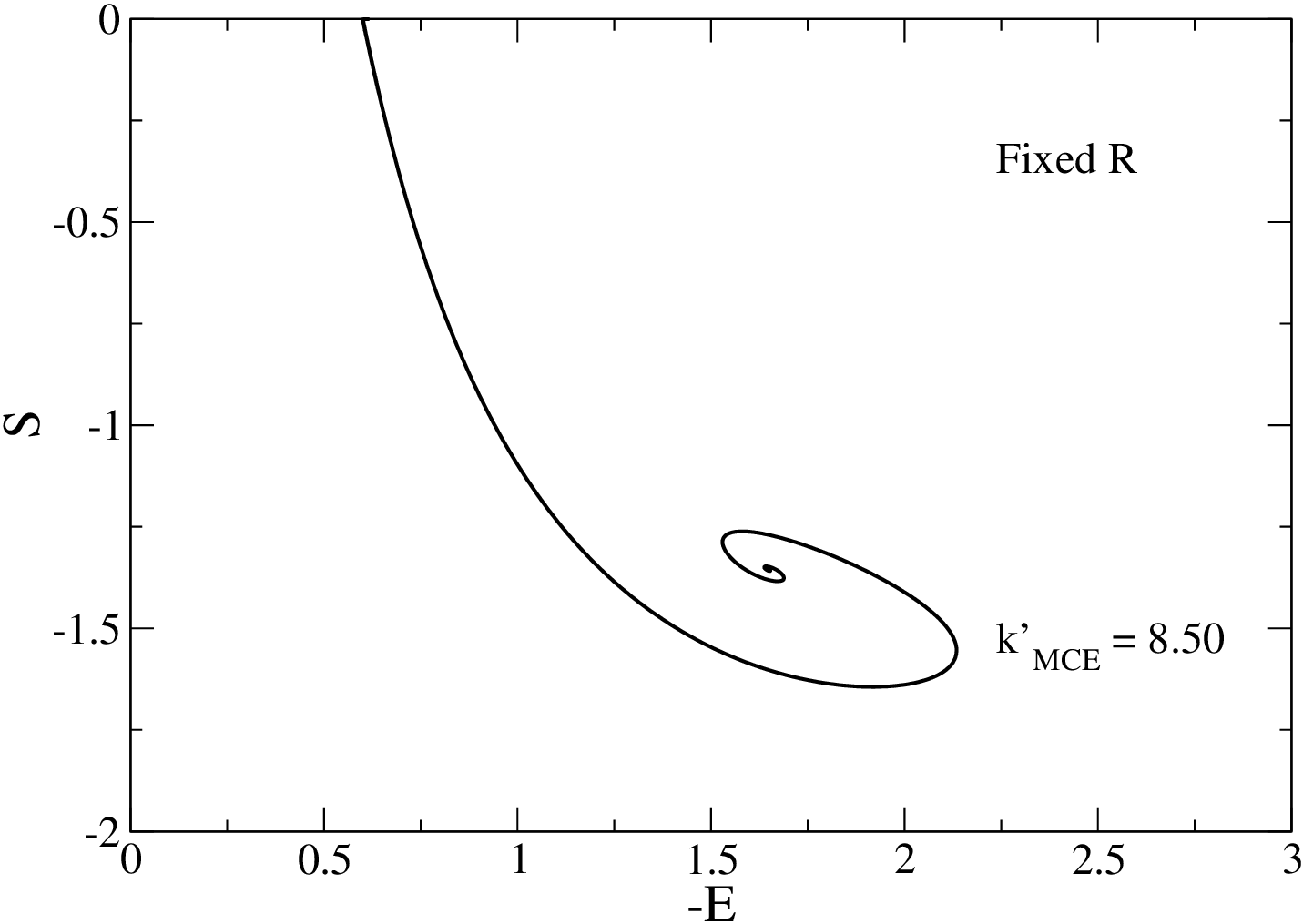}
\caption{Entropy versus energy plot for the classical King model when $R$ is
fixed instead of $A$. This Figure reveals a paradox when $R$ is
fixed because the configurations with $k<k_{MCE}'$ (that could be
expected to be entropy maxima) have a lower entropy than the configurations with
$k>k_{MCE}'$ (that are unstable).}
\label{esR}
\end{center}
\end{figure}

\begin{figure}[!h]
\begin{center}
\includegraphics[clip,scale=0.3]{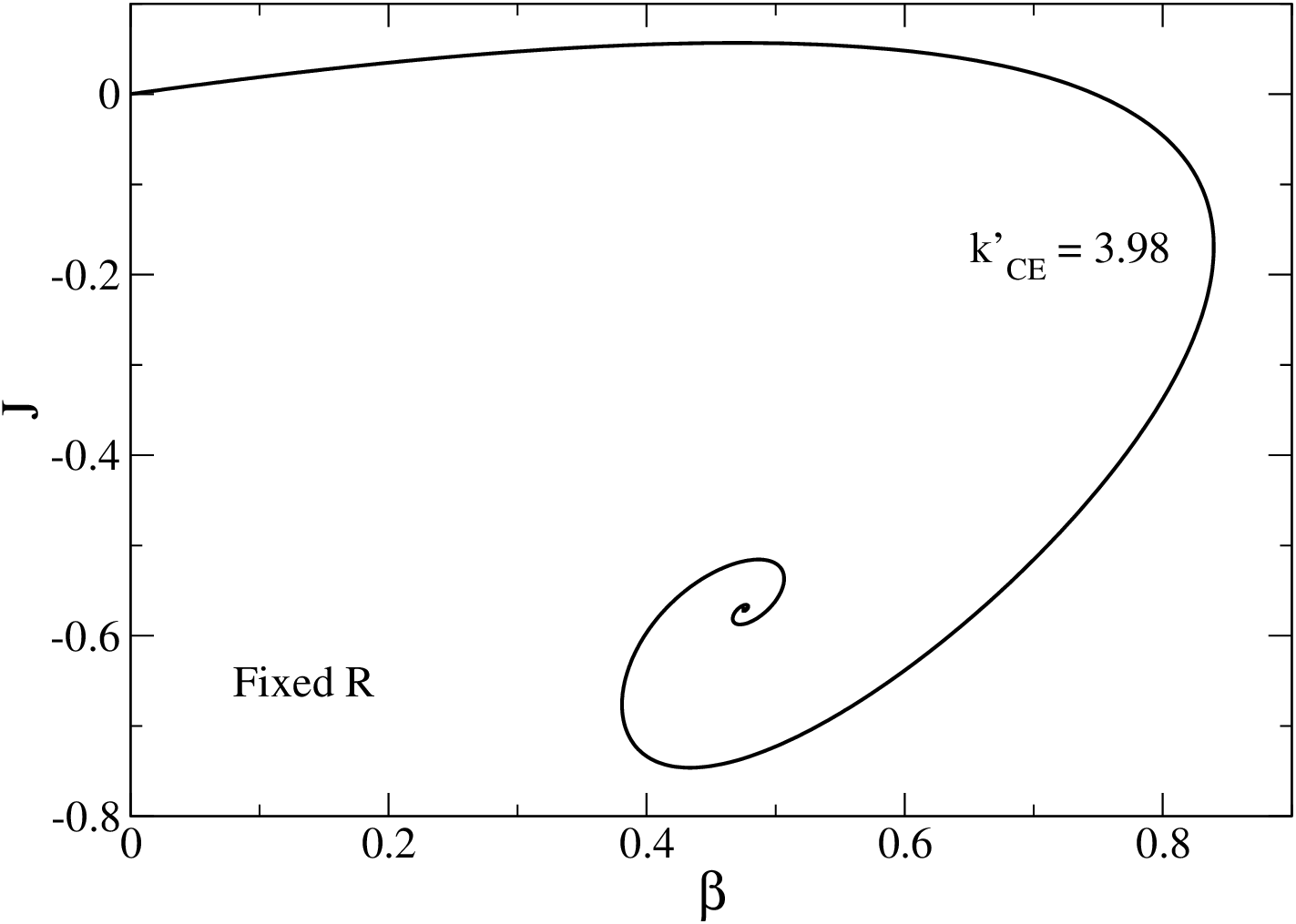}
\caption{Free energy versus inverse temperature plot for the classical King
model when $R$ is fixed instead of $A$.}
\label{bjR}
\end{center}
\end{figure}

\begin{figure}[!h]
\begin{center}
\includegraphics[clip,scale=0.3]{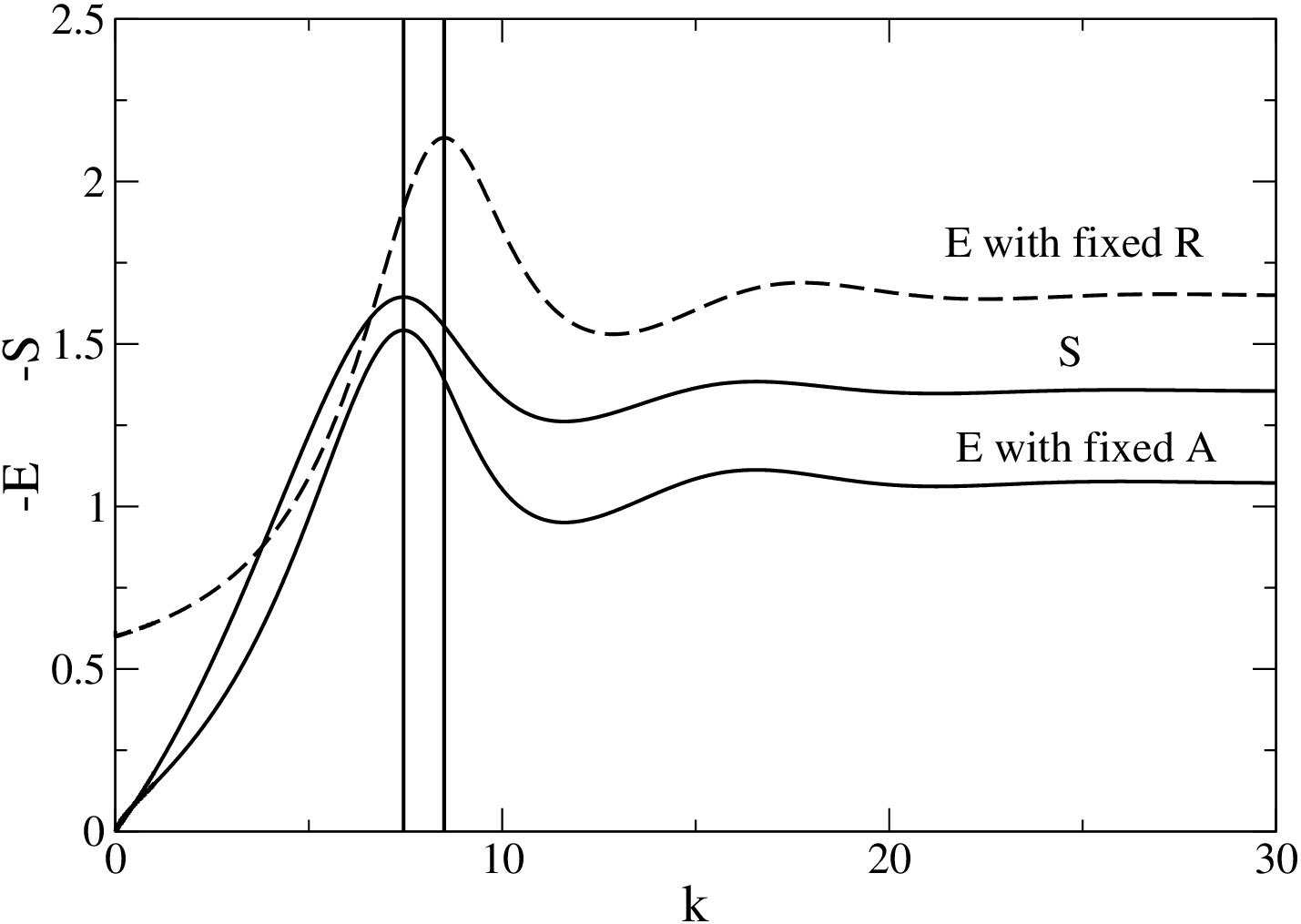}
\caption{Entropy $S$, energy $\tilde{E}$ (normalized by $A$), and energy $\epsilon$ (normalized by $R$) as a function of $k$.}
\label{EetS}
\end{center}
\end{figure}

Since  $\delta S\neq \beta \delta E$ and $\delta J\neq -E\delta\beta$ when $R$ is fixed instead of $A$, the entropy $S(k)$ and the energy $E(k)$ on the one hand, and the free energy $J(k)$ and the inverse temperature $\beta(k)$ on the other hand, do not have their extrema at the same values of $k$. As a result, the curves $S(E)$ and $J(\beta)$ present turning points instead of spikes (compare Figs. \ref{esR} and \ref{bjR} to Figs. \ref{esA} and \ref{bjA}).

Fig. \ref{EetS} recapitulates the difference between fixing $A$ or $R$. First, we note that the curve $S(k)$ is the same in the two cases. When $A$ is fixed,  the energy $\tilde E(k)$ and the entropy  $S(k)$ have their extrema at the same points. When $R$ is fixed, the extrema of $\epsilon(k)$ and $S(k)$ are different.

\section{Comparison of the classical King model with the observations of dark matter halos}
\label{sec_comp}

In order to compare a specific theoretical model of dark matter halos to observations, it is necessary to introduce quantities that are directly measurable. In the present section, we introduce such quantities. We define them for an arbitrary distribution function so they can be applied to various models in future works. Then, we explicitly calculate these quantities for the classical King model and compare the results to observations.

\subsection{The halo radius and the tidal radius}
\label{sec_hr}

\begin{figure}[!h]
\begin{center}
\includegraphics[clip,scale=0.3]{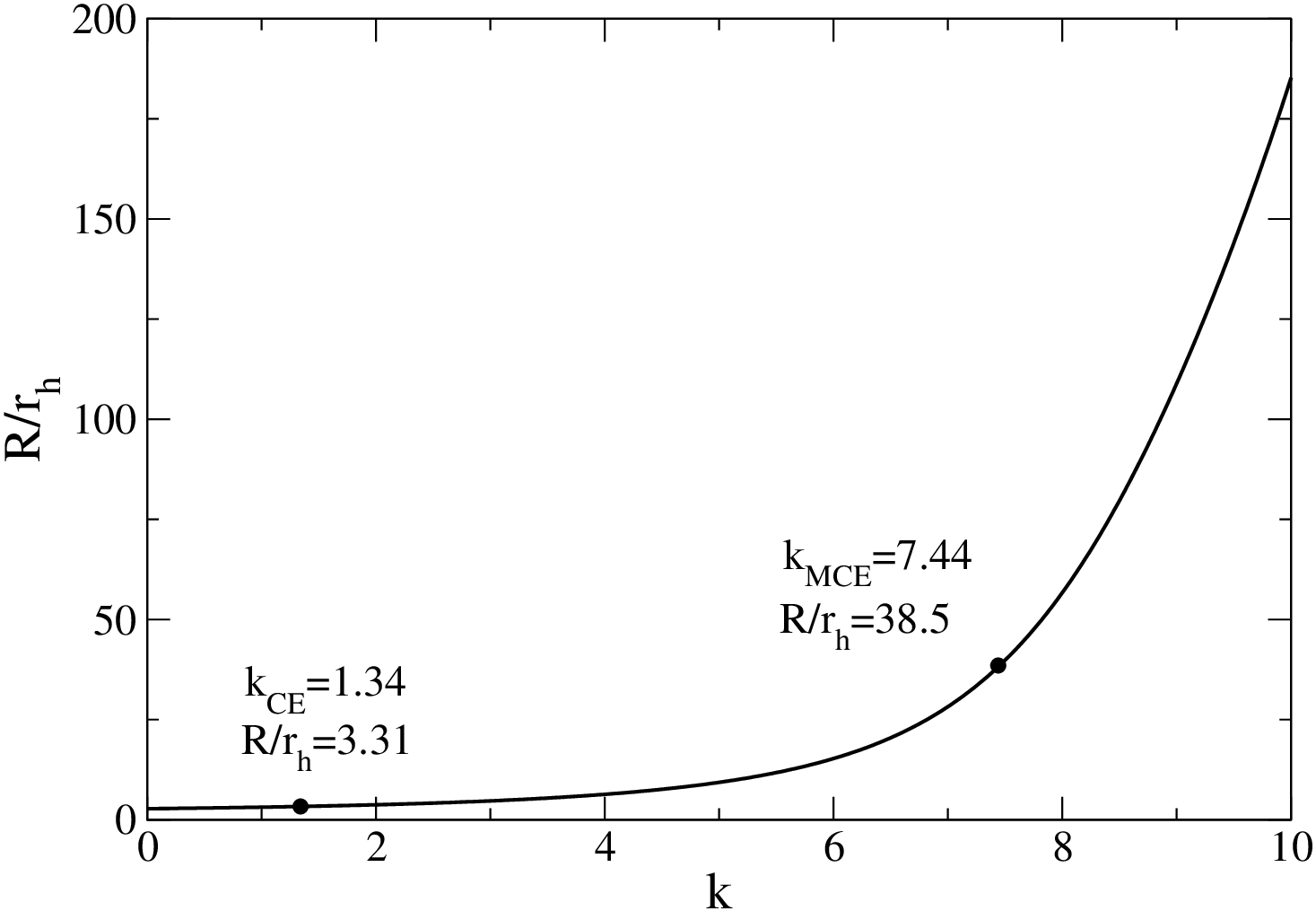}
\caption{Tidal radius normalized by the halo radius $R/r_h$ as a function of
$k$. For $k=k_{CE}=1.34$, one finds that $R/r_h=3.31$. For $k=k_{MCE}=7.44$, one finds that $R/r_h=38.5$. }
\label{RsurRh}
\end{center}
\end{figure}

We consider a spherical cluster described by  a distribution function of the form $f(\epsilon)$ with $f'(\epsilon)<0$. The density profile of the system is given by Eq. (\ref{fde10}). Following de Vega and Sanchez \cite{vega,vega2,vega3,vega4,vega5}, we define the halo radius $r_h$ such that $\rho(r_h)/\rho_0=1/4$. The dimensionless halo radius $\zeta_h$ is therefore determined by the equation
\begin{equation}
\frac{I_1\lbrack \chi(\zeta_h)\rbrack}{I_1(k)}=\frac{1}{4}.
\label{hr1}
\end{equation}
This is a function $\zeta_h(k)$ of the the variable  $k$ parameterizing the series of equilibria. The halo radius is then given by  $r_h=r_0\zeta_h$ where $r_0$ is defined by Eq. (\ref{fde12b}). The radial distance normalized by the halo radius can be written as $r/r_h=\zeta/\zeta_h$. The tidal radius normalized by the halo radius is given by
\begin{equation}
\frac{R}{r_h}=\frac{\zeta_1(k)}{\zeta_h(k)}\equiv {\cal R}(k).
\label{hr2}
\end{equation}
The function ${\cal R}(k)$ is plotted in Fig. \ref{RsurRh} for the classical
King model. For $k\rightarrow 0$, we find that ${\cal R}\rightarrow {\cal
R}(0)=2.75$ (this asymptotic value can be directly obtained from the study of
the Lane-Emden equation (\ref{poly1}) from which we get $\xi_h=1.945$ and
$\xi_1=5.36$) and for $k\rightarrow +\infty$, we find that ${\cal R}\rightarrow
+\infty$. We note that the tidal radius normalized by the halo radius increases
monotonically with $k$ while the tidal radius normalized by $A$ initially
decreases with $k$ and finally makes damped oscillations (see Sec.
\ref{sec_proper}).

\subsection{The density profile}
\label{sec_dens}

\begin{figure}[!h]
\begin{center}
\includegraphics[clip,scale=0.3]{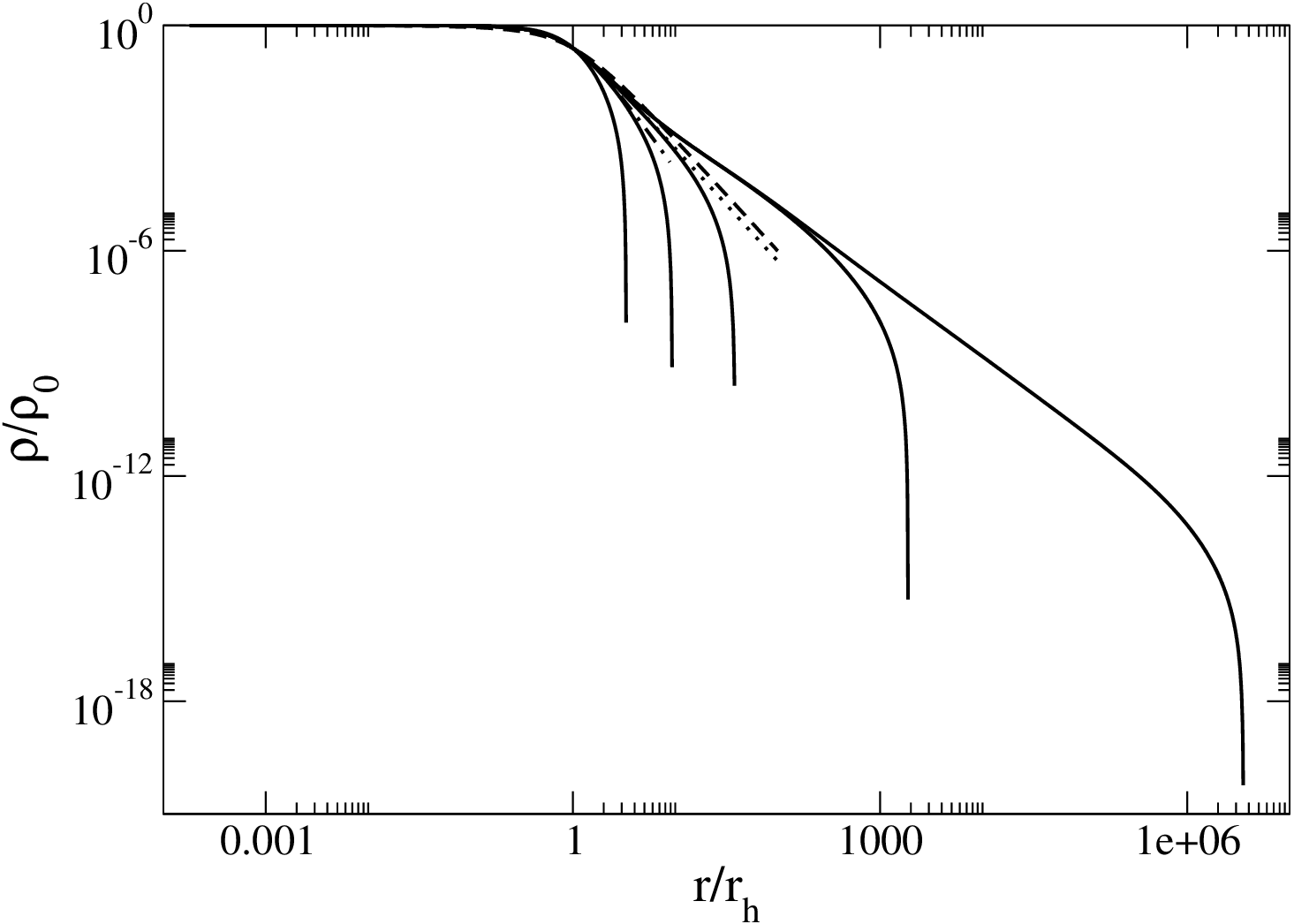}
\caption{Normalized density profiles of the
classical King model in logarithmic scales for (left to right) $k=k_{CE}=1.34$
($E=-0.188$,
$\beta=1.63$), $k=5$ ($E=-0.965$, $\beta=0.893$), $k=k_{MCE}=7.44$ ($E=-1.54$,
$\beta=0.589$), $k=15$ ($E=-1.09$, $\beta=0.735$), and $k=30$
($E=-1.07$, $\beta=0.732$). Dotted line: modified Hubble profile. Dashed-dotted line: isochrone profile. Dashed line: Burkert profile.}
\label{densityLOG}
\end{center}
\end{figure}

\begin{figure}[!h]
\begin{center}
\includegraphics[clip,scale=0.3]{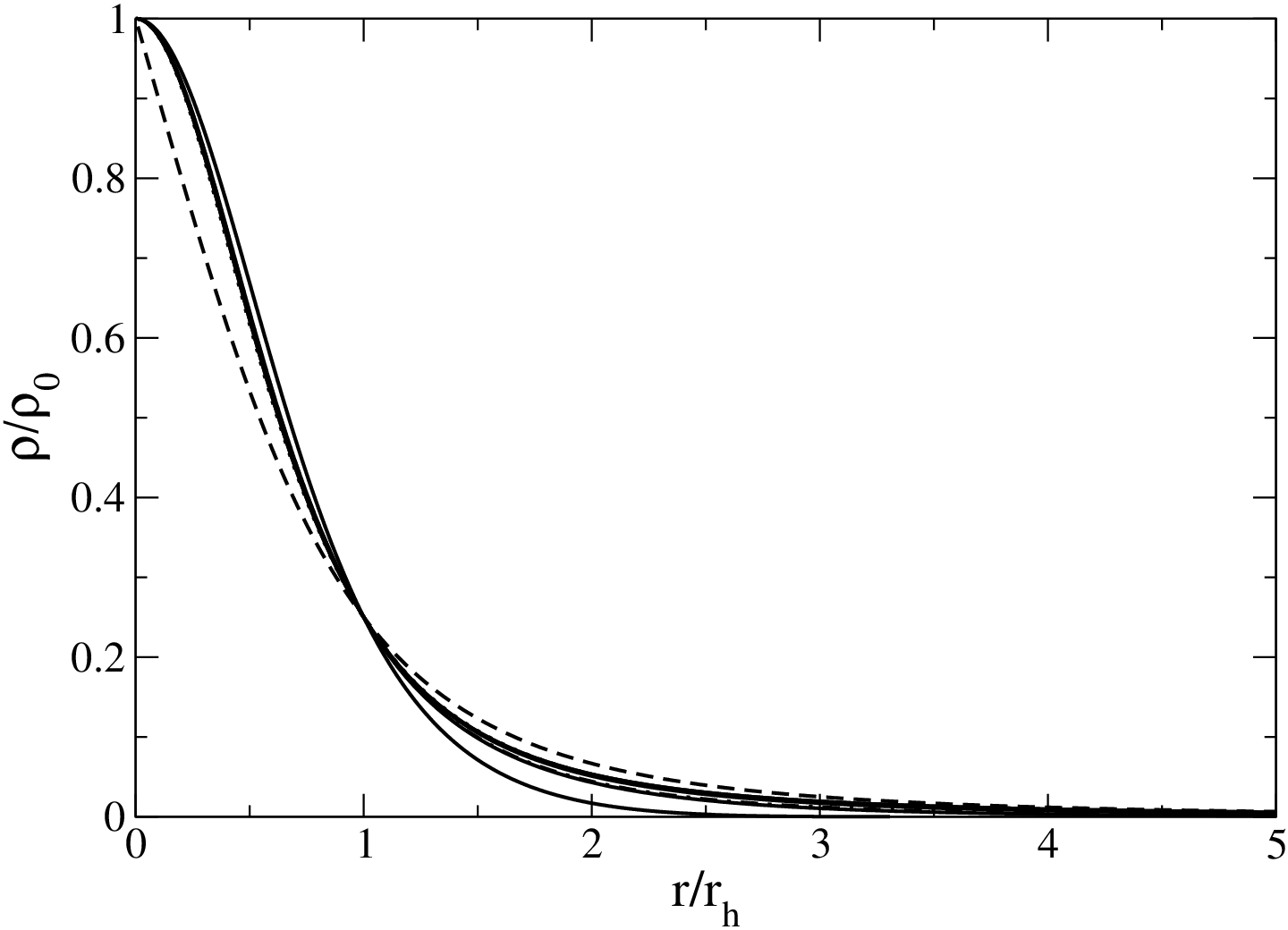}
\caption{Normalized density profile of the classical King model in linear
scales for (bottom to top) $k=1.34$, $k=5$, $k=7.44$, $k=15$, and $k=30$. }
\label{densityLIN}
\end{center}
\end{figure}

The density profile normalized by the central density is given by
\begin{equation}
\frac{\rho(r)}{\rho_0}=\frac{I_1\lbrack \chi(\zeta)\rbrack}{I_1(k)}.
\label{dens1}
\end{equation}
The normalized density profile $\rho(r)/\rho_0$ corresponding to the classical
King model is plotted as a function of the normalized radial distance $r/r_h$ in
Figs. \ref{densityLOG} and \ref{densityLIN} in 
logarithmic and linear scales respectively for different values of $k$. Roughly
speaking, for a given value of $k$, the core and the halo of the distribution
are isothermal (provided that $k$ is sufficiently large) while the envelope is
polytropic with an index $n=5/2$. This is because the density is high in the
core and the halo, and low in the envelope (see Sec. \ref{sec_eos}). It is the
polytropic nature of the envelope that confers to the system a finite radius.
Indeed, a purely isothermal system extends to infinity and has infinite mass. By
contrast, a polytrope with index $n=5/2$ has a compact support.

The proportion of the isothermal region with respect to the polytropic
one depends on $k$.

For $k\rightarrow 0$, the density profile almost coincides with that of a polytrope of index $n=5/2$ (in Figs.  \ref{densityLOG} and \ref{densityLIN}, the King profile with $k=k_{CE}=1.34$ is indistinguishable from a pure $n=5/2$ polytrope). In that case, the tidal radius $R$ is of the order of the halo radius $r_h$.

For $k\rightarrow +\infty$, the tidal radius $R$ is 
rejected to $+\infty$ and the density profile approaches the profile of the
classical isothermal sphere except at very large distances $r\sim R$ where the
density drops to zero, ensuring a finite mass. For $r_h\ll r\ll R$, the density
decreases algebraically as $r^{-\alpha}$ with $\alpha=2$. Actually, the density
profile exhibits damped oscillations about the $r^{-2}$ profile\footnote{These
oscillations give rise  to those of $\beta(k)$ and $E(k)$ in Figs. \ref{abA} and
\ref{aeA} leading to the spiral $\beta(E)$ of Fig. \ref{ebA}. Therefore, the
onset of gravitational collapse in CE and MCE (associated with the turning
points of temperature and energy) can be traced back to the oscillations of the
density profile.}  as for the classical isothermal sphere (see, e.g.,  Fig. 7 in
\cite{ijmpb}). However, our study  shows that the profiles with $k>k_{MCE}=7.44$
are thermodynamically unstable. Therefore, the oscillations of the density
profile are not physically relevant.

As $k$ decreases, the effective slope of the density profile increases. For $k=k_{MCE}=7.44$
the density profile decreases approximately as $r^{-\alpha}$ for $r_h\ll r\ll R$ with an effective
slope $\alpha\sim 3$. For $k=5$ the density profile
has an effective slope $\alpha\sim 4$. In MCE, the King model is stable ($k<k_{MCE}$) as long as the
effective slope $\alpha$ is approximately larger than $3$. In CE, the King model is stable ($k<k_{CE}$)
only when it is close to the $n=5/2$ polytrope.

The dotted line represents the modified Hubble profile (see Appendix \ref{sec_mh}) that has a
slope $\alpha=3$ \cite{bt}. It fits
well the core of the isothermal sphere for $r<1.63 r_h$ \cite{bt}. It also fits
well the
King model with $k\sim k_{MCE}$ up to
$\sim 5 r_h$. The dashed-dotted line represents H\'enon's isochrone profile (see Appendix \ref{sec_isochrone}) that
has a slope $\alpha=4$ \cite{isochrone}. It fits
well the  King model with $k\sim 5$ up to $\sim 2 r_h$. The dashed line
represents the Burkert profile   corresponding
to the observations of dark matter halos \cite{observations}. It has a slope $\alpha=3$ (see Sec.  \ref{sec_dm}).

In MCE, the isochrone profile is stable, the modified Hubble profile is close to the limit of marginal stability, and the classical isothermal profile is unstable. In CE, the isochrone profile, the modified Hubble profile, and  the classical isothermal profile are all unstable.

\begin{figure}[!h]
\begin{center}
\includegraphics[clip,scale=0.3]{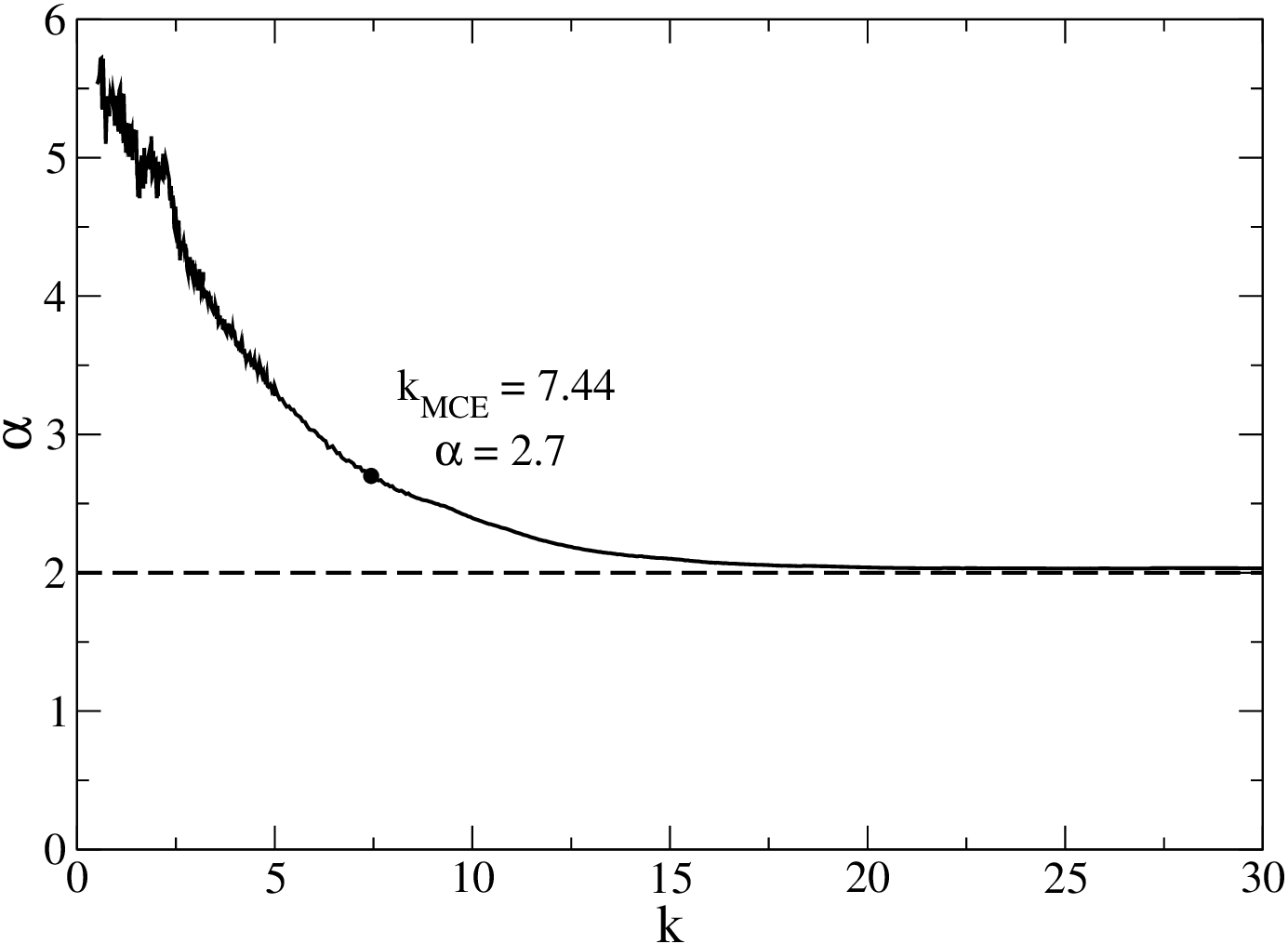}
\caption{Effective slope of the density profile in the halo as a
function of the concentration parameter $k$. The point of microcanonical marginal stability $k_{MCE}=7.44$
corresponds to $\alpha=2.7$ close to $3$. The slope $\alpha=4$ corresponds to
$k=3.3$.}
\label{kalpha}
\end{center}
\end{figure}

Violent relaxation usually generates a density profile 
with a core-halo structure. The density in the halo decreases as $r^{-\alpha}$
with $\alpha=4$ \cite{henonVR,albada,roy,joyce}. This is the same exponent as
H\'enon's isochrone profile. This also corresponds to a King model with a
concentration parameter $k\sim 5$ that is stable in MCE (but unstable in CE). We
argue in Appendix A of Paper II that the concentration parameter $k(t)$
increases monotonically with time because of collisions\footnote{For globular
clusters, collisions refer to weak long-range interactions (two-body encounters)
and for dark
matter halos they refer to strong short-range interactions.} and evaporation
until
an instability takes place at $k_{MCE}$. Such an evolution is shown numerically
by Cohn \cite{cohn} in the case of globular clusters. Since the effective slope
$\alpha$ of the density profile decreases with $k$ (see Fig. \ref{kalpha}), we
conclude that $\alpha(t)$ decreases monotonically with time. In MCE, the King
profile is stable as long as $\alpha(t)\ge 3$ (i.e. $k(t)\le k_{MCE}=7.44$) and
it
becomes unstable afterwards. In  CE, the King profile destabilizes at
$k(t_*)=k_{CE}=1.34$ before even producing an effective power law.

\subsection{The circular velocity}
\label{sec_circ}

\begin{figure}[!h]
\begin{center}
\includegraphics[clip,scale=0.3]{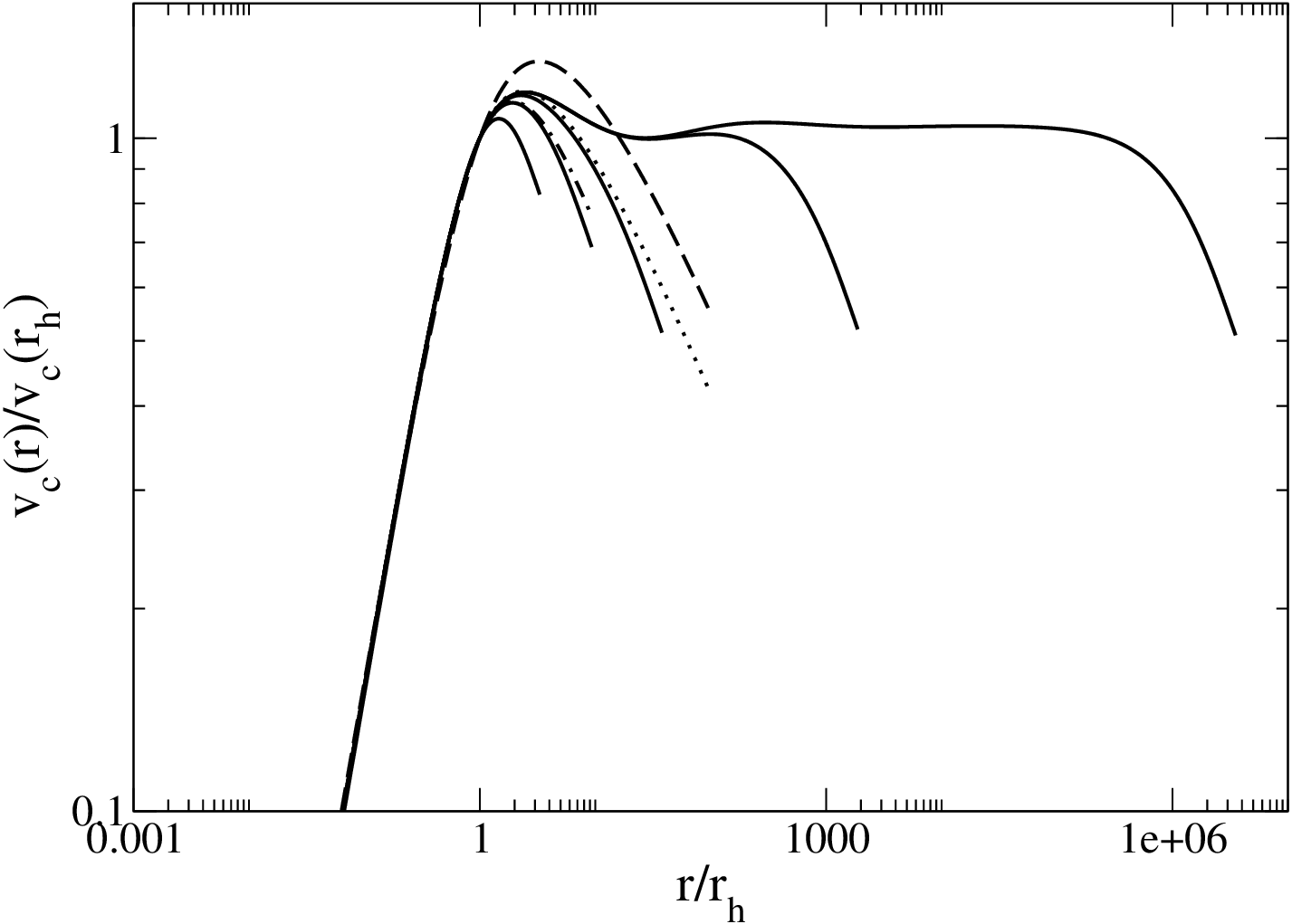}
\caption{Normalized rotation curve of the classical King model in logarithmic
scales for (left to right) $k=1.34$, $5$, $7.44$, $15$, and $30$.}
\label{vitesseLOG}
\end{center}
\end{figure}

\begin{figure}[!h]
\begin{center}
\includegraphics[clip,scale=0.3]{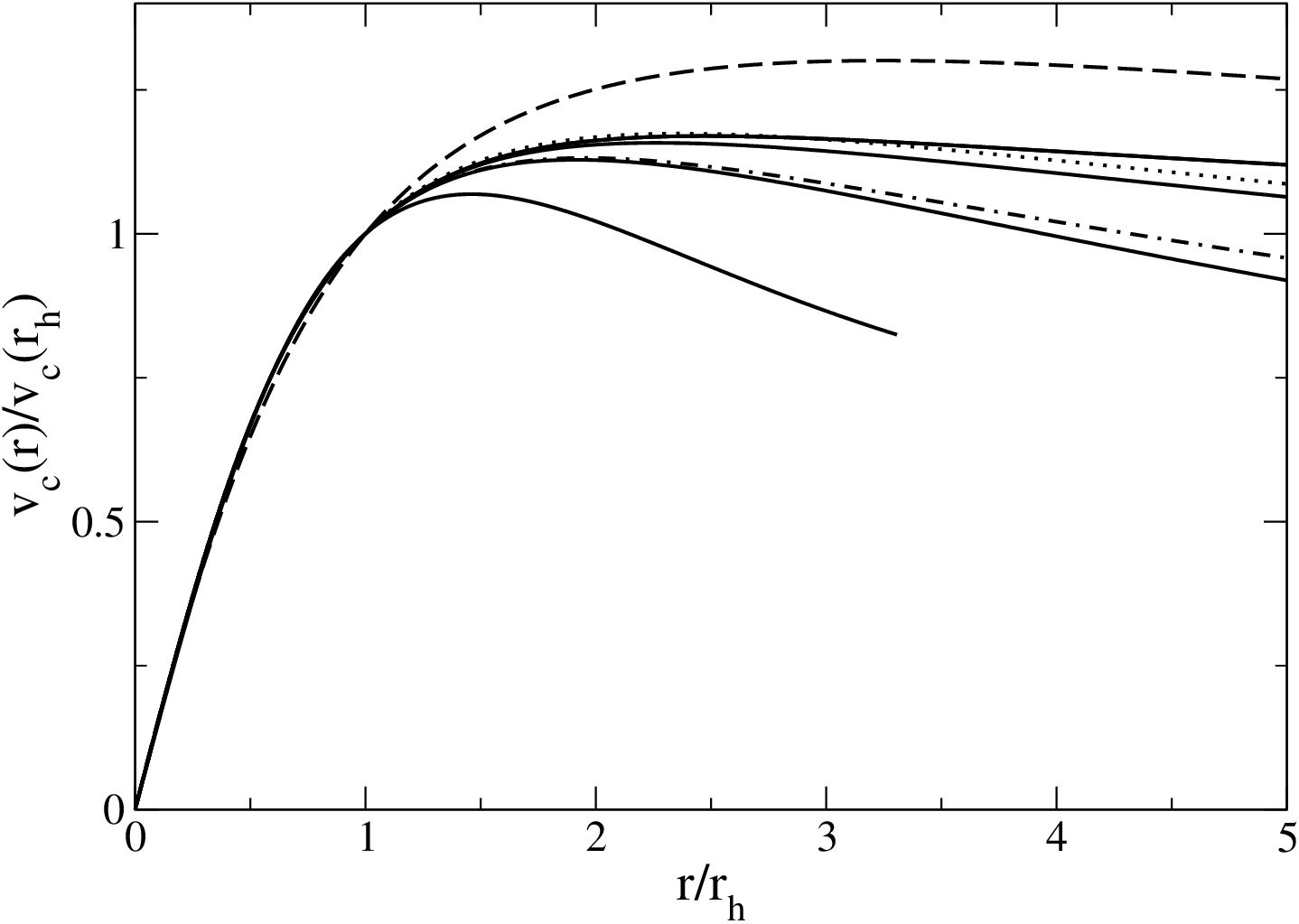}
\caption{Normalized rotation curve of the classical King model in
linear scales for (bottom to top) $k=1.34$, $5$, $7.44$, $15$, and $30$. }
\label{vitesseLIN}
\end{center}
\end{figure}

The circular velocity is defined by Eq. (\ref{circb3}). The value of the circular velocity at the halo radius is $v_c(r_h)=\sqrt{{GM_h}/{r_h}}$, where  $M_h=M(r_h)$ is the halo mass. Using Eq. (\ref{circb4}),  the circular velocity normalized by its value at $r=r_h$ is given by
\begin{equation}
\frac{v_c(r)}{v_c(r_h)}=\sqrt{\frac{\zeta\chi'(\zeta)}{\zeta_h \chi'(\zeta_h)}}.
\label{circ2}
\end{equation}
The normalized circular velocity $v_c(r)/v_c(r_h)$ corresponding to the classical King model is plotted as a function of the normalized radial distance $r/r_h$ in Figs. \ref{vitesseLOG} and \ref{vitesseLIN} in logarithmic and linear scales respectively for different values of $k$. We first note that the rotation curve does not sensibly depend on the value of $k$ in the range $0\le r\le r_h$. By contrast, differences appear for $r\ge r_h$.

For $k\rightarrow 0$, the system is close to a polytrope of index $n=5/2$ and the tidal radius $R$ is of the order of the halo radius $r_h$. For $r>R$, the density of the dark matter halo is equal to zero so the rotation curve has a Keplerian profile (not represented).

For $k\rightarrow +\infty$, the density decreases as $r^{-2}$ at large
distances, like for the classical isothermal sphere, leading to a flat rotation
curve. Actually, the rotation curve presents damped oscillations about the
plateau (due to the oscillations of the density profile) that are clearly
visible in logarithmic scales. However, the
profiles with $k>k_{MCE}=7.44$ are thermodynamically unstable so these
oscillations are not physically relevant. In addition, real halos do not
extend at such large distances where these oscillations would appear (if they
were relevant).

For smaller values of $k$, the density decreases more rapidly than $r^{-2}$ at
large distances and consequently the rotation curve decreases with the distance.
For $k\sim k_{MCE}$, the rotation curve presents a maximum close to the halo
radius $r_h$ before decreasing. This is in agreement with the observations (see
Sec. \ref{sec_dm}). The modified Hubble profile fits relatively
well the King profile with $k=k_{MCE}$ up to the tidal radius $R=38.5\, r_h$.
Similarly, the isochrone profile fits relatively well the King profile with
$k=5$ up to the tidal radius $R=9.33\, r_h$. Therefore, the fit is better
on the rotation curves than on the density profiles.

\subsection{The velocity dispersion}
\label{sec_dis}

\begin{figure}[!h]
\begin{center}
\includegraphics[clip,scale=0.3]{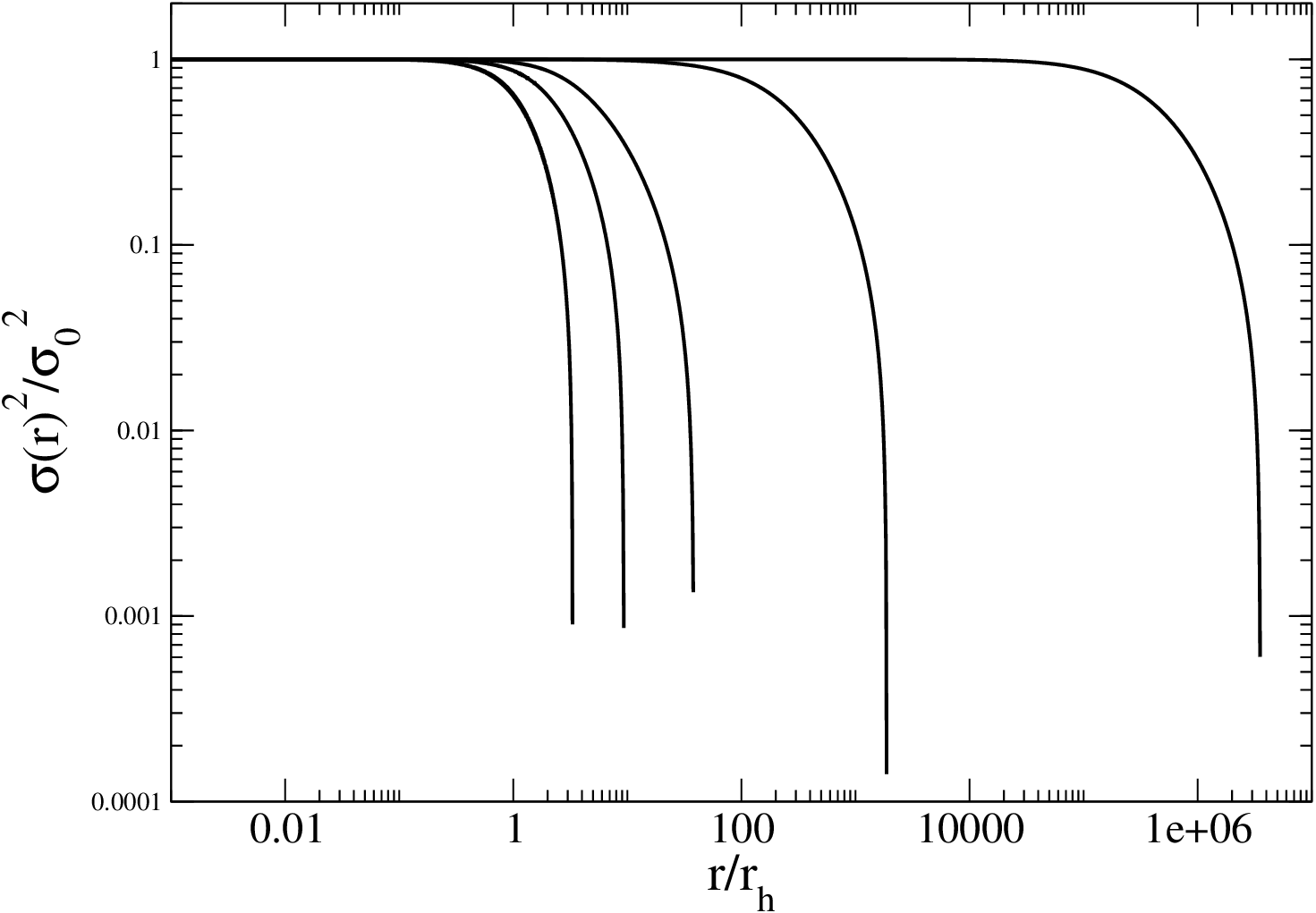}
\caption{Normalized velocity dispersion of the classical King model in logarithmic scale for (left to right) $k=1.34$, $5$, $7.44$, $15$, and $30$.}
\label{temperatureLOG}
\end{center}
\end{figure}

\begin{figure}[!h]
\begin{center}
\includegraphics[clip,scale=0.3]{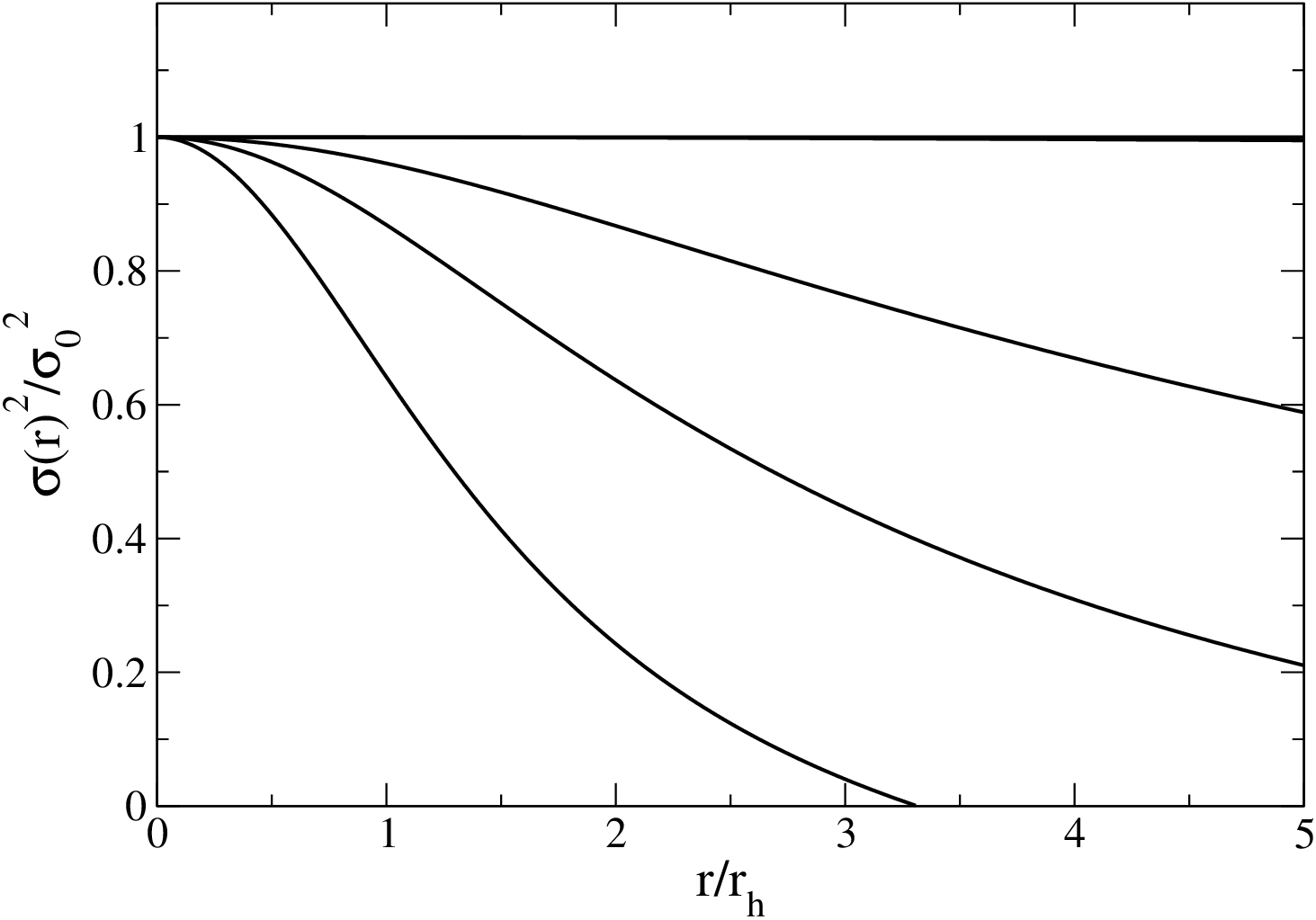}
\caption{Normalized velocity dispersion of the classical King model in linear scale for (bottom to top) $k=1.34$, $5$, $7.44$, $15$, and $30$.}
\label{temperatureLIN}
\end{center}
\end{figure}

The local velocity dispersion of a spherically symmetric distribution function $f(\epsilon)$ is defined by Eq. (\ref{dis1}). According to Eq. (\ref{dis2}), the velocity dispersion profile normalized by the central velocity dispersion is given by
\begin{equation}
\frac{\sigma^2(r)}{\sigma_0^2}=\frac{I_2[\chi(\zeta)] I_1(k)}{I_1[\chi(\zeta)]I_2(k)}.
\label{dis4b}
\end{equation}
The normalized velocity dispersion profile $\sigma^2(r)/\sigma_0^2$
corresponding to the classical King model is plotted as a function of the
normalized radial distance $r/r_h$ in Figs. \ref{temperatureLOG} and
\ref{temperatureLIN} in logarithmic and linear scales respectively for different
values of $k$. For sufficiently large $k$, these curves clearly show 
the isothermal region where the velocity dispersion is almost uniform
(coinciding with the temperature $T$) and the polytropic region where the
velocity dispersion decreases rapidly with the distance. For $k\rightarrow 0$,
the system almost coincides with a polytrope of index $n=5/2$ and the velocity
dispersion is far from being uniform. Actually, it is related to the density
profile by $\sigma^2(r)=K\rho^{2/5}(r)$ where $K$ is the polytropic constant
defined in Sec. \ref{sec_poly}. As $k$ increases, the velocity dispersion
becomes more and more uniform in the inner region of the distribution that
extends at larger and larger radii. For $k\rightarrow +\infty$, the system is
almost isothermal except at very large distances, close to the tidal radius $R$.
For $k\sim k_{MCE}$, the system is isothermal  for $r<2\, r_h$ and polytropic
for $2\, r_h<r<R=38.5\, r_h$.
Using Eq. (\ref{dis3}), we find that the ratio between the central velocity dispersion and the temperature behaves as  $\sigma_0^2/T\sim (2/7)k$ for $k\rightarrow 0$ while $\sigma_0^2/T\rightarrow 1$ for $k\rightarrow +\infty$ as expected.

\subsection{The functions $F$ and $G$}
\label{sec_dim}

\begin{figure}[!h]
\begin{center}
\includegraphics[clip,scale=0.3]{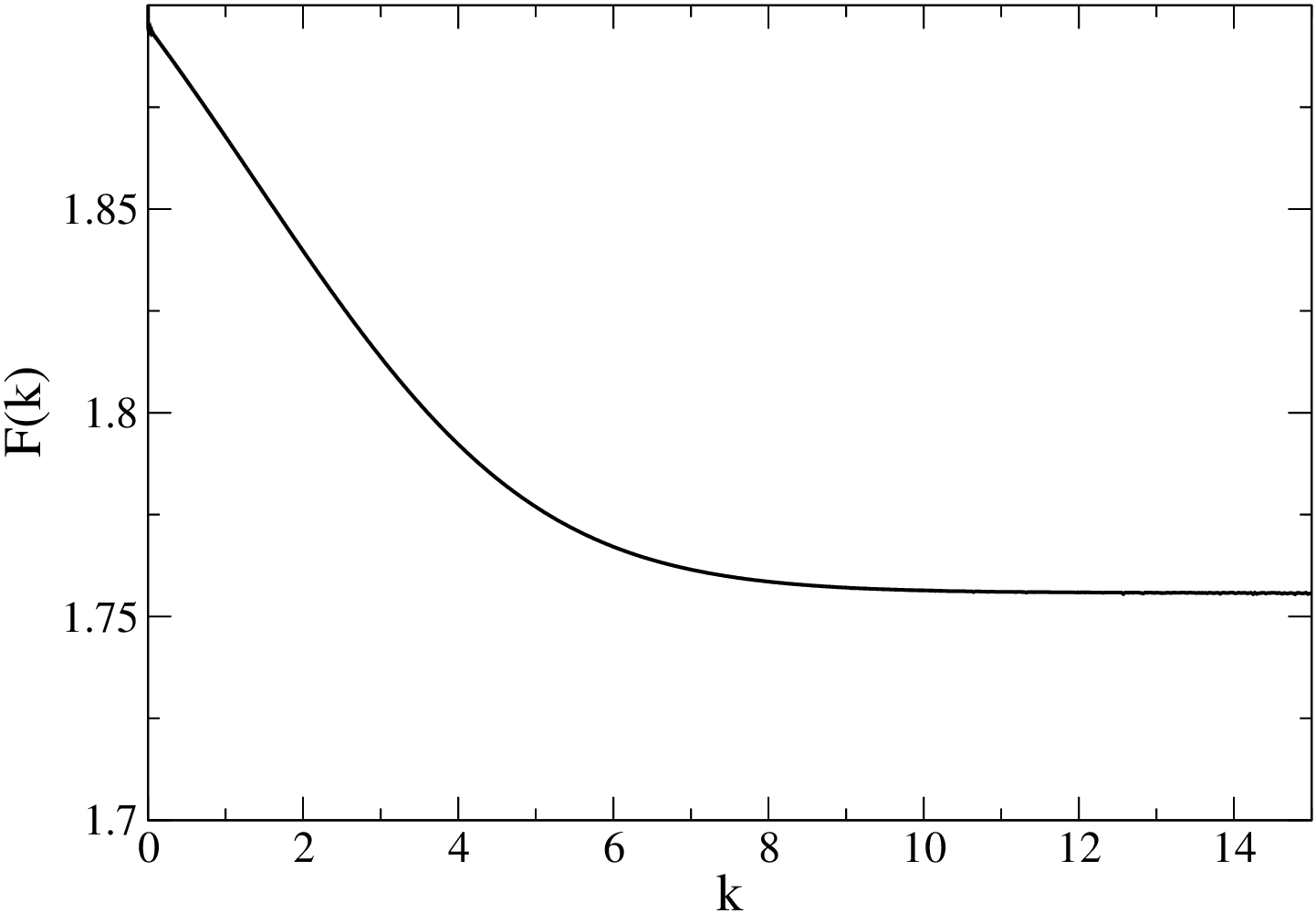}
\caption{The function $F(k)$.}
\label{Q}
\end{center}
\end{figure}

Applying Eq. (\ref{temp2}) at $r=r_h$, we find that the halo mass $M_h$ normalized by $\rho_0 r_h^3$ is given by
\begin{equation}
\frac{M_h}{\rho_0r_h^3}=-4\pi\frac{\chi'[\zeta_h(k)]}{\zeta_h(k)}\equiv F(k).
\label{dim1}
\end{equation}
This is a function $F(k)$ of the parameter $k$ parameterizing the series of
equilibria. For the classical King model, this function is plotted in Fig.
\ref{Q}. Its asymptotic values can be obtained analytically. For $k\rightarrow
0$, the system reduces to a pure polytrope of index $n=5/2$ and one finds
$F(0)=1.89$ (using $\xi_h=1.945$ and $\theta'_h=-0.293$ obtained from the study
of the Lane-Emden equation (\ref{poly1})-(\ref{poly2})). For $k\rightarrow
+\infty$, the system tends to the classical isothermal sphere and one finds
$F(+\infty)=1.76$ (using $\xi_h=3.63$ and $\psi'_h=0.507$ obtained from the
study of the Emden equation (\ref{iso1})-(\ref{iso2})).    We see in Fig.
\ref{Q} that the function $F(k)$ decreases monotonically between these two
values. We note that this function does not change much as a function of $k$ so
that it has an almost ``universal'' value $\sim 1.8$.
For $k=k_{MCE}$, corresponding to the stability threshold in MCE, we get $F(k_{MCE})=1.76$.

\begin{figure}[!h]
\begin{center}
\includegraphics[clip,scale=0.3]{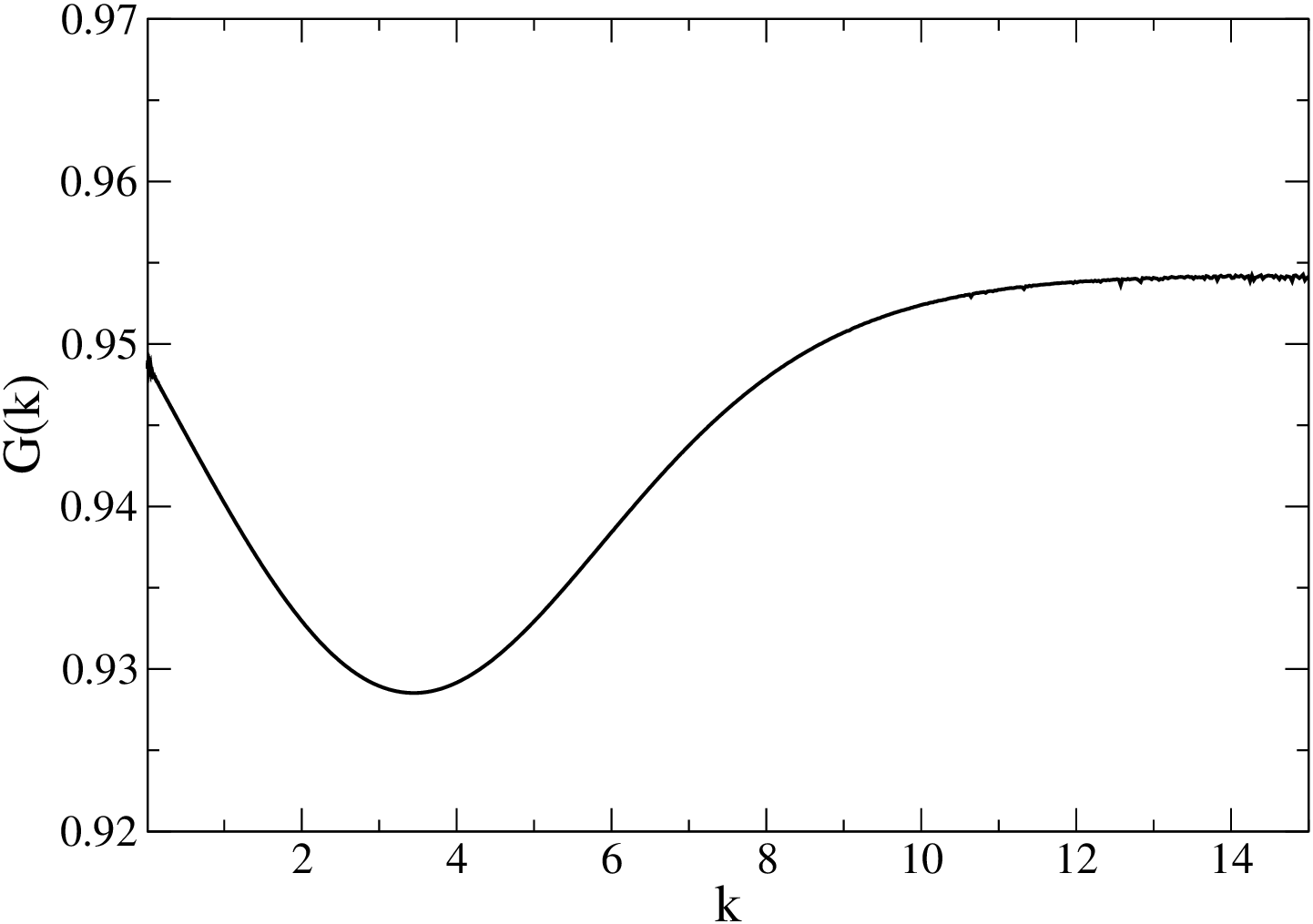}
\caption{The function $G(k)$.}
\label{F}
\end{center}
\end{figure}

Applying Eq. (\ref{fde12}) at $r=r_h$ and using Eq.  (\ref{fde12b}), we obtain the expression of the temperature normalized by $G\rho_0 r_h^2$:
\begin{equation}
\frac{1}{\beta G\rho_0 r_h^2}=\frac{4\pi}{\zeta_h^2(k)}.
\label{dim2}
\end{equation}
Combining  Eqs. (\ref{dis3}) and (\ref{dim2}), we find that the central velocity dispersion normalized by $G\rho_0 r_h^2$ is given by
\begin{equation}
\frac{\sigma_0^2}{G\rho_0r_h^2}=\frac{8\pi}{3}\frac{1}{\zeta_h^2(k)}\frac{I_2(k)}{I_1(k)}\equiv G(k).
\label{dim3}
\end{equation}
This is a function $G(k)$ of the parameter $k$ parameterizing the series of
equilibria. For the classical King model, this function is plotted in Fig.
\ref{F}. As for the function $F(k)$, its asymptotic values can be obtained
analytically. For $k\rightarrow 0$, one finds $G(0)=0.949$ and for $k\rightarrow
+\infty$, one finds $G(+\infty)=0.954$. We see in Fig. \ref{F} that the function
$G(k)$ first decreases, reaches a minimum $G_{min}=0.929$ at $k=3.45$, then
increases towards its asymptote. We note that this function does not change much
as a function of $k$ so that it has an almost ``universal'' value $\sim 0.95$.
For $k=k_{MCE}$, we get $G(k_{MCE})=0.946$.

We  note that $v_c^2(r_h)/G\rho_0r_h^2=M_h/\rho_0r_h^3=F(k)$. Therefore, using Eqs. (\ref{dim1}) and (\ref{dim3}), we obtain
\begin{equation}
\frac{v_c^2(r_h)}{\sigma_0^2}=\frac{F(k)}{G(k)}.
\label{dim3b}
\end{equation}
This function varies between $1.99$ for $k\rightarrow 0$ and $1.84$ for
$k\rightarrow +\infty$. For $k=k_{MCE}$, it takes the value $1.86$.

\subsection{Application to dark matter halos}
\label{sec_dm}

Some measurable quantities of dark matter halos are the central density
$\rho_0$, the central velocity dispersion $\sigma_0$, the halo radius $r_h$, and
the halo mass $M_h$. They are given for different types of galaxies (dwarf and
large) in Table 1 of Ref. \cite{vega2}. On the other hand, the circular
velocities $v_c(r)$ are known with precision from galaxy observational data. The
observation of the rotation curves of a large number of galaxies shows that the
density of dark matter halos can be represented by a universal profile, called
the Burkert profile \cite{observations}, given by the empirical
law
\begin{equation}
\frac{\rho(r)}{\rho_0}=\frac{1}{(1+x)(1+x^2)},\qquad x=\frac{r}{r_h}.
\label{dm1}
\end{equation}
The Burkert profile decreases at large distances as $r^{-3}$
like the NFW profile \cite{nfw}. This leads to a mass profile  $M(r)$ diverging
logarithmically with $r$. However, contrary to
the NFW profile, the Burkert profile presents a flat core density for
$r\rightarrow 0$ instead of exhibiting a $r^{-1}$ density cusp. Density cusps
are not observed in dark matter halos unless they contain a massive central
black hole.

The rotation curve corresponding to the Burkert profile is
\begin{equation}
v_c^2(r)=2\pi G\frac{\rho_0 r_h^3}{r}\left\lbrack \ln(1+x)-\arctan x+\frac{1}{2}\ln(1+x^2)\right \rbrack.
\label{dm2}
\end{equation}
After normalization by the circular velocity at the halo radius, we get
\begin{equation}
\frac{v_c(r)}{v_c(r_h)}=\frac{1.98}{\sqrt{x}}\left\lbrack \ln(1+x)-\arctan x+\frac{1}{2}\ln(1+x^2)\right \rbrack^{1/2}.
\label{dm3}
\end{equation}

The halo mass is obtained by integrating Eq. (\ref{dm1}) from zero to $r_h$. This yields
\begin{equation}
\frac{M_h}{\rho_0 r_h^3}=1.60.
\label{dm4}
\end{equation}
Alternatively, using the observational data given in Table 1 of \cite{vega2}, we
find that
\begin{equation}
\frac{M_h}{\rho_0 r_h^3}\sim 2.5,\qquad  \frac{\sigma_0^2}{G\rho_0r_h^2}\sim 0.4.
\label{dm5}
\end{equation}
The comparison between Eqs. (\ref{dm4}) and (\ref{dm5}) shows that we should not
give too much importance on the precise value of these quantities. We just note
that the typical values of these quantities deduced from the Burkert profile or
directly  from the observations are consistent with those obtained theoretically
with the classical King model (see Sec. \ref{sec_dim}). Actually, it can be
shown \cite{prep} that many models of dark matter halos yield values of
${M_h}/{\rho_0 r_h^3}$ and $\sigma_0^2/{G\rho_0r_h^2}$ that agree with the
observational results. Therefore, the comparison between theory and observations
for these quantities is not very discriminatory.

The density profiles and the circular velocity profiles obtained
from the King model for different values of the concentration parameter $k$ are
compared with the Burkert profile in Figs. \ref{densityLOG}-\ref{vitesseLIN}. In
the range $0\le r\le r_h$, all the theoretical curves coincide, whatever the
value of $k$, and they are in good agreement with the Burkert
profile.\footnote{We note that the Burkert density profile behaves as
$\rho/\rho_0-1\propto r$ for $r\rightarrow 0$ while the King density profiles
behave as $\rho/\rho_0-1\propto r^2$ for $r\rightarrow 0$  which is the natural
behavior of spherically symmetric systems. We recall that the Burkert profile is
purely empirical so we should not give too much credit to its precise behavior
for $r\rightarrow 0$ (its main property is to have a flat core).
We also note that the difference of behavior between the Burkert profile and the King profiles
for $r\rightarrow 0$  is almost imperceptible on the rotation curves.} This
corresponds to the region where the distribution function is isothermal. This
suggests that the core of dark matter halos is isothermal. Actually, it can be
shown \cite{prep} that many models of dark matter halos yield the same results
in this range of radial distances so the agreement with the Burkert profile for
$r\le r_h$ cannot be considered as a vindication of a particular
theoretical model. By contrast, at larger distances $r>r_h$, the theoretical
rotation curves sensibly depend on $k$ and the comparison with the Burkert
profile gives more stringent constraints on the acceptable models. We note that
the virial radius of dark matter halos is of the order of $10$-$100\, r_h$
\cite{vega3}, so
we have to compare the theoretical profiles with the Burkert profile on
distances greater than $r_h$.

For $k\rightarrow +\infty$, we recover the classical isothermal profile. However, this profile does not agree with the observational Burkert profile at large distances because the density decreases too slowly. For $r_h\ll r\ll R$, the density decreases as $r^{-2}$ instead of  $r^{-3}$ and the rotation curve forms a plateau while the observational rotation curves slightly decrease at large distances. Therefore, the King profiles with a large value of the concentration parameter $k$ are not in agreement with the observations. This is consistent with our theoretical study since we find that the King models with $k>k_{MCE}=7.44$ are thermodynamically unstable.

For $k\rightarrow 0$, the King model is equivalent to a polytrope of index $n=5/2$. This profile does not agree with the Burkert profile because the density drops to zero too rapidly. In addition, the tidal radius is of the order of the halo radius while observational rotation curves extend well beyond the halo radius. Therefore, the King models with a low value of the concentration parameter $k$ are not in agreement with the observations. This is consistent with our theoretical study since we argue in Appendix A of Paper II  that the concentration parameter $k(k)$ increases with time as a result of collisions and evaporation so that sufficiently old halos should have relatively large values of $k$.

The best agreement with the Burkert profile is achieved for the
King models with $k\sim k_{MCE}$. In that case, the density profile can be
approximated by the modified Hubble profile (see Appendix \ref{sec_mh}) that
decreases as $(r/r_h)^{-3}$ like the Burkert profile. The prefactors are
respectively $0.534$ and $1$. The rotation curves corresponding to the modified
Hubble profile and to the Burkert profile have a similar behavior. They achieve
a maximum before decreasing. The maximum is located at
$(r/r_h,v_c/v_c(r_h))=(2.37,1.17)$ for the modified Hubble profile  and at
$(3.25,1.30)$ for the Burkert profile. This difference is in the error bars
of the observations (at least $20\%$).  We also note that the tidal radius of
the King model with $k=k_{MCE}$ is equal to $38.5\, r_h$  which is of the same
order of magnitude as the observational virial radius of dark matter halos.
Therefore, we conclude that the observations of dark matter halos, represented
by the empirical  Burkert profile, can be relatively well-explained by a King
model at, or close to, the limit of microcanonical stability. Strictly
speaking, the density profiles of dark matter halos are not universal since they
depend on the concentration parameter $k$, but it is natural to expect that
most observed halos have a concentration parameter close to $k_{MCE}$, which
explain why their profile is {\it quasi universal}. Indeed, the concentration parameter
cannot be much smaller than $k_{MCE}$ since $k(t)$ increases with time, and it
cannot be larger than $k_{MCE}$ since, at that concentration, the clusters
become thermodynamically unstable and collapse.  Therefore,
large dark matter halos that have not collapsed should have a concentration
parameter of the order of
$k\sim k_{MCE}$, and this happens to be consistent with the observations.

In conclusion, we propose to describe large dark matter halos
by a classical King model at the point of marginal stability in MCE. It can be approximated by a modified Hubble
profile with a slope $\alpha=3$.\footnote{The fact that the modified Hubble
profile
can be interpreted as a King model at the limit of microcanonical stability may
also explain why it gives a good fit to certain elliptical galaxies \cite{bt},
to globular clusters \cite{kingempiric}, and to clusters of galaxies
\cite{kingcosmo}.} This profile approximately  accounts for the observed
rotation curves of dark matter halos up to the tidal radius $R=38.5\, r_h$ which
is of the same order of magnitude as the virial radius of dark matter halos
($\sim 10$-$100\, r_h$). The fact that we observe dark matter halos with a
slope $\alpha=3$ instead of $\alpha=4$ (a typical outcome of collisionless
violent relaxation \cite{henonVR,albada,roy,joyce}) may be an indication that
dark matter is collisional. Indeed, collisions and evaporation increase the
concentration $k(t)$ and decrease the slope $\alpha(t)$ from the initial state
$k_i\sim 5$ and $\alpha_i=4$ ($\sim$ H\'enon's isochrone profile) to the final
state $k_f=k_{MCE}=7.44$ and $\alpha_f\sim 3$ ($\sim$ modified Hubble profile).
The same is true for globular clusters.

\subsection{Universal scaling laws}
\label{sec_law}

It is an empirical fact that the surface density $\Sigma_0=\rho_0 r_h$ is approximately the same for all galaxies \cite{vega3} even if their sizes and masses vary by several orders of magnitudes (see, e.g., Table 1 of \cite{vega2}). Its typical value is $\Sigma_0=120 M_{\odot}/{\rm pc}^2$. As a result, it is convenient to rewrite Eqs. (\ref{dim1}) and (\ref{dim3}) in terms of $\Sigma_0$ instead of $\rho_0$. We get
\begin{equation}
\frac{M_h}{\Sigma_0r_h^2}=F(k),\qquad \frac{\sigma_0^2}{G\Sigma_0r_h}=G(k).
\label{law4}
\end{equation}
Considering   $\Sigma_0$ as a constant,\footnote{Actually, this is not true
for the largest dark matter halos where $\Sigma_0$ can reach values of the order
of $7000\, M_{\odot}/{\rm pc}^2$ instead of $\Sigma_0=120 M_{\odot}/{\rm pc}^2$ (see Table 1 of \cite{vega2}).} these
equations exhibit the scalings $M_h\sim r_h^2$ and $\sigma_0^2\sim r_h$.
Introducing relevant scales, the foregoing relations may be rewritten as
\begin{equation}
\frac{M_h}{M_{\odot}}=F(k)\frac{\Sigma_0}{M_{\odot}/{\rm pc}^2}\left (\frac{r_h}{\rm pc}\right )^2,
\label{law2}
\end{equation}
\begin{equation}
\frac{\sigma_0^2}{({\rm km}/{\rm s})^2}=4.30\, 10^{-3} G(k) \frac{\Sigma_0}{M_{\odot}/{\rm pc}^2}\frac{r_h}{\rm pc}.
\label{law3}
\end{equation}
For the King model, the quantities $F(k)$ and $G(k)$ are plotted in Figs. \ref{Q} and \ref{F}. As we have seen in Secs. \ref{sec_dim} and \ref{sec_dm}, these quantities do not change much with $k$ and have the typical values $1.8$ and $0.95$ respectively. Furthermore, we have explained that the concentration parameter $k$ should be close to $k_{MCE}$. This fixes the prefactors in Eqs. (\ref{law2}) and (\ref{law3}) to the values $1.76$ and $0.946$ respectively.

We emphasize that the scalings (\ref{law2})-(\ref{law3}) do not depend on
the distribution function chosen to model dark matter halos. Only the functions
$F(k)$ and $G(k)$ depend on the model. Furthermore, most models of dark matter
halos give values of  $F(k)$ and $G(k)$ that sensibly have the same order of
magnitude \cite{prep}. Therefore, the observation of the scaling laws
(\ref{law2}) and  (\ref{law3}) cannot be considered as a vindication of a
particular theoretical model.

{\it Remark:} For large dark matter halos, which are non degenerate, the 
central velocity dispersion  $\sigma_0^2=k_B T/m$ represents the ratio of the
temperature of the cluster on the mass of the particles (this is valid for
sufficiently large $k$). It is possible to determine $\sigma_0$ observationally
(see Table 1 of \cite{vega2}). However, since the temperature of the clusters 
is unknown, we cannot determine the mass $m$ of the particles that compose them.
Assuming that dark matter halos are made of fermions, the mass of the fermions
can be obtained only from the observation of dwarf dark matter  halos that are
degenerate \cite{vega,vega2}.

\subsection{Comparison with other works}
\label{sec_compdvs}

\begin{figure}[!h]
\begin{center}
\includegraphics[clip,scale=0.3]{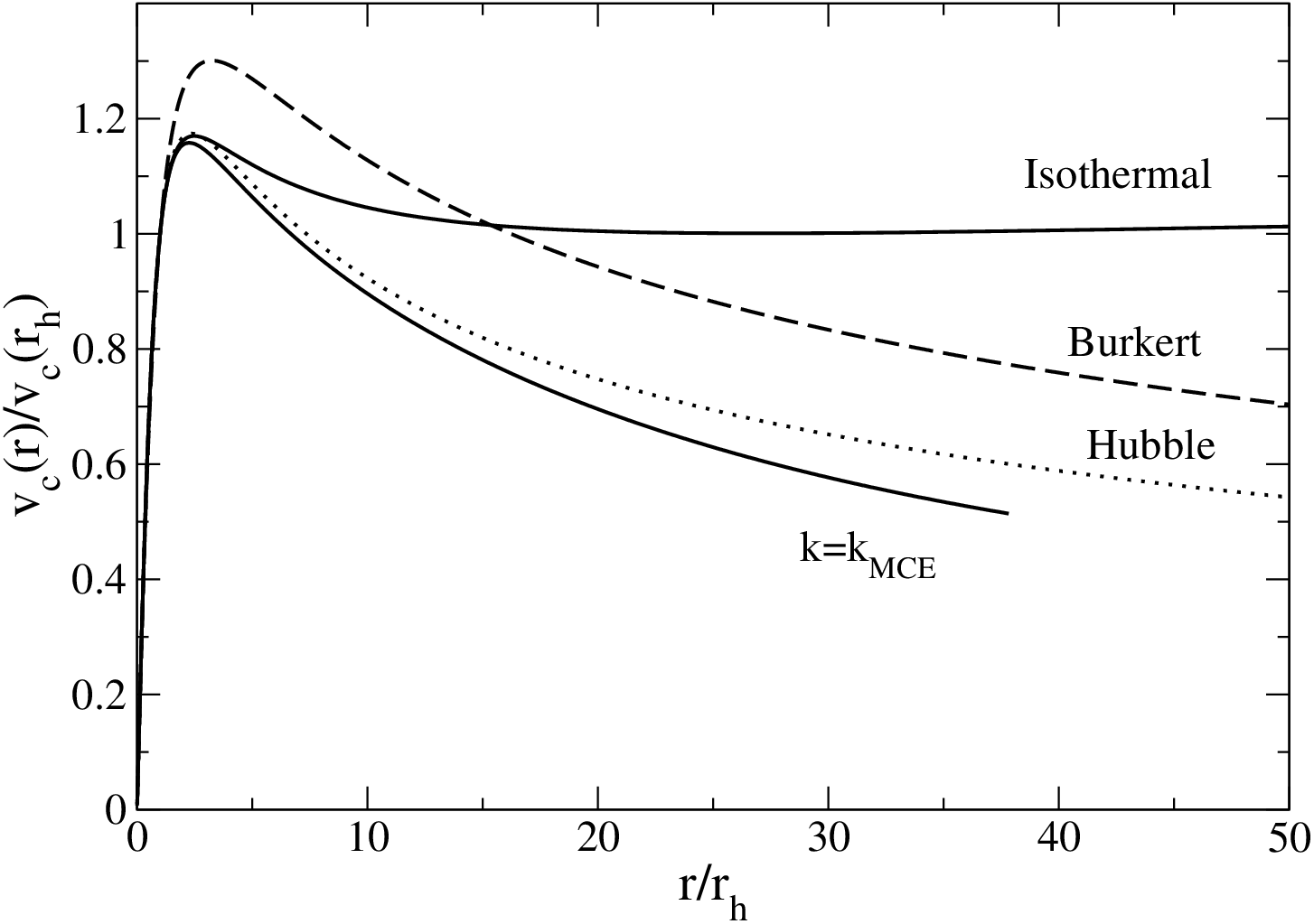}
\caption{Comparison between the observational Burkert profile (dashed line),
the isothermal profile (upper solid line), the King model at the limit of
microcanonical stability (lower solid line), and the modified Hubble profile
(dotted line).}
\label{comparaison}
\end{center}
\end{figure}

In a nice series of papers
\cite{vega,vega2,vega3,vega4,vega5}, de Vega and Sanchez propose to model dark
matter halos as a self-gravitating gas of fermions at finite temperature
described by the Fermi-Dirac-Poisson system (Thomas-Fermi approximation).
This idea is not new since several works in the past already considered 
fermionic dark matter halos at finite temperature (see the Introduction).
However, de
Vega and Sanchez confront this model to observations and give convincing
arguments that the mass of the fermions should be of the order of $2\, {\rm
keV}/c^2$. This mass scale corresponds to warm dark matter (WDM). The dark
matter
particle could be a sterile neutrino. In their first papers
\cite{vega,vega2}, they
argue that the cusp problem and the satellite problem of CDM are solved by
quantum mechanics (Pauli exclusion principle). This is valid for small halos
($M<10^6\, M_{\odot}$) for which quantum effects are important. In particular,
the most compact known dwarf halo (Willman 1)  with mass  $M=0.39 \, 10^6\,
M_{\odot}$ may be considered as a completely degenerate self-gravitating Fermi
gas at zero temperature stabilized by quantum mechanics. However, for large dark
matter halos ($M>10^6\, M_{\odot}$), which constitute most of the available
observational data (see Table I of \cite{vega2}),  quantum effects are
negligible and the classical limit applies.\footnote{Since the classical limit applies to most dark matter halos for which we have observational data, we cannot rule out the possibility
that dark matter is made of bosons instead of fermions. For large
halos, the quantum nature of particles (fermions or bosons) does not play any
role (actually, this depends whether the bosons are self-interacting or not, as discussed in Paper II). The distinction between bosonic and fermionic dark matter can be made only
by considering dwarf halos for which we have only few observational results.
Therefore, the possibility that dark matter is made of bosons should not
be rejected.} Therefore, in Ref. \cite{vega3},
de Vega and Sanchez describe large dark matter halos by the Boltzmann distribution. In that case,
the cusp problem is solved by finite temperature effects, not by quantum
mechanics. de Vega and
Sanchez  argue that large dark matter halos have a universal profile
corresponding to the classical isothermal profile. This profile, which  has a
homologous structure \cite{chandra}, has been considered by many authors in the
past. de Vega and Sanchez show that this profile agrees with the Burkert profile
for $r<r_h$. However, if we continue the comparison at larger distances,  up to
the typical virial radius of dark matter halos ($\sim 10$-$100\, r_h$), severe
disagreements appear between the classical isothermal profile and the Burkert
profile  (see Fig. \ref{comparaison}) as discussed in Sec. \ref{sec_dm}. In particular, the classical isothermal sphere (in addition of being thermodynamically unstable) leads to flat rotation curves while the observed circular velocity decreases with
the distance. Therefore, the classical isothermal profile cannot correctly describe dark
matter halos up to the virial radius. de Vega and Sanchez recognize this problem
since they argue, in their last papers \cite{vega4,vega5}, that a deviation to
isothermality must be present in the halo in order to account for the
observations. To this aim, they introduce a family of empirical density profiles
of the form
\begin{equation}
\frac{\rho(r)}{\rho_0}=\frac{1}{\left\lbrack 1+(4^{2/\alpha}-1)\left (\frac{r}{r_h}\right)^2\right\rbrack^{\alpha/2}}
\label{empiric}
\end{equation}
and mention that these profiles with $\alpha\sim 3$ are appropriate to fit
galaxy observations up to the virial radius. Then, they use the  Eddington
equation \cite{bt} to determine the corresponding distribution function
$f(\epsilon)$. They obtain a rather complicated expression but they manage to
show that the distribution function is isothermal at low energies
and non-isothermal at high energies. However, no justification of the density
profiles (\ref{empiric}) is given, so their approach remains essentially
empirical.

By using a different approach, we arrived at a similar
conclusion from more physical considerations. We proposed to model dark
matter halos by the fermionic King distribution. For large halos, we can neglect
quantum effects and use the classical King distribution.  We argued, on the
basis of kinetic considerations and stability analysis, that dark matter halos
should be described by the King distribution at, or close to, the point of
marginal microcanonical stability. It turns out that, at this point,
the King profile can be approximated by the modified Hubble profile up to the
tidal radius. The modified Hubble profile corresponds to the profile of Eq.
(\ref{empiric}) with $\alpha=3$ (see Appendix \ref{sec_mh}). Our study provides
therefore a physical
justification of the empirical profile with $\alpha\sim 3$ considered by de
Vega and Sanchez \cite{vega5}. Furthermore, it shows that this profile
comes from a
distribution function that is close to the classical King distribution. This
distribution is isothermal (Boltzmannian) at low energies and polytropic (with
index $n=5/2$) at high energies. By
contrast, the physical meaning of the distribution function $f(\epsilon)$ that de Vega and
Sanchez \cite{vega5} obtain from Eq. (\ref{empiric}) by using  the
Eddington formula is less
clear. Therefore, our approach provides a new light on their results.

\section{Conclusion}
\label{sec_conclusion}

In this paper, we have studied in detail the thermodynamical properties of the classical King
model. This model was originally introduced to describe globular clusters \cite{michie,king} but we have proposed
to apply it also to the case of large dark matter halos. The King model is a physically motivated model that can be justified by statistical
mechanics and kinetic theory once the evaporation of high energy particles is properly taken into account. This distribution has a finite mass,
contrary to the classical isothermal distribution, so  there is no need to enclose the system within an artificial  box. This generalization is
important because it allows us to plot the caloric curves and make a rigorous
stability analysis of the different configurations. We have found that the
system is canonically stable for $k<k_{CE}=1.34$ and microcanonically stable
for $k<k_{MCE}=7.44$ (as previously obtained by  Katz \cite{katzking}) where $k$ is the concentration parameter. The classical isothermal distribution, corresponding
to the limit $k\rightarrow +\infty$, is thermodynamically unstable. The
marginal King distribution in MCE turns out to be close to the modified Hubble profile and the marginal King distribution in CE turns out to be close to a polytrope of index $n=5/2$.

We have then compared the classical King model to observations of dark matter halos. In particular, Figs. \ref{densityLOG}-\ref{vitesseLIN} compare the prediction of the King model with the
Burkert profile (dashed line) that fits a large variety of dark matter halos.
The best agreement, up to the virial radius of about $10$-$100\, r_h$, is
obtained for
$k\sim k_{MCE}$, that is to say for the King model at, or close to, the limit of microcanonical stability.
This is physically natural since the concentration parameter $k(t)$ increases with time as a result of collisions and evaporation
until an instability takes place at $k=k_{MCE}$ \cite{cohn}. Therefore, most
halos that have not collapsed should have a concentration parameter close to its
maximum stable value $k_{MCE}$.
At that value, the King distribution generates a density profile that is close
to the modified Hubble profile with a slope $\alpha\sim 3$. The King models with
$\alpha<3$ are unstable. Our study therefore  provides a physical
justification, from the King model, of the empirical density profiles with $\alpha\sim 3$
introduced by de Vega and Sanchez
\cite{vega4,vega5}. Furthermore, it shows that statistical mechanics properly
modified to take evaporation into account may provide a good approach to understand
the structure of dark matter halos.

The agreement between  the modified Hubble profile and the Burkert
profile is very good in the core for $r\le r_h$. Therefore, it
appears that the core of dark matter halos is isothermal. This thermalization
may be due to a collisionless violent relaxation or to a
collisional relaxation, as explained in the Introduction.
We emphasize that finite temperature effects produce flat core densities and,
therefore, solve the problems of the CDM model
without the need to advocate quantum mechanics (that is relevant only for dwarf halos).
On the other hand, the agreement between the modified Hubble profile and the Burkert profile is less good in the halo for $r\ge r_h$. We can have two points of view.
We can argue that the difference lies in the error bars of observations so that the modified Hubble profile provides
an equally good, or even better, description of dark matter halos than the Burkert profile. Indeed, we have shown in Sec. \ref{sec_dm}
that the two profiles are qualitatively similar, differing from each other by
$20$-$30\%$, which is in the error bars of the observations.
Of course, the superiority of the modified Hubble profile over the Burkert
profile is that it can be justified physically  as an approximation of  the
King model at the limit of microcanonical stability while the Burkert profile
is a purely empirical model (fit) deduced from the observations.  Alternatively, we
can adopt a completely different point of view and argue that the King model is
not relevant to describe dark matter halos.
Indeed, the difference between the modified Hubble profile and the Burkert
profile may be considered to be too strong. In that point of view, we should
recall that the King model is usually justified for collisional self-gravitating
systems, governed by the Landau equation, undergoing a continuous evaporation
\cite{king}. If dark matter halos are purely collisionless, governed by the
Vlasov equation, we may argue that there  is no continuous evaporation
justifying the King distribution.
In this point of view, the structure of dark matter halos may result  from an  incomplete
collisionless violent relaxation \cite{lb}, described by other types of
distribution functions (different from the King model), as in the case of
stellar systems \cite{bt}. Although the cores of dark matter
halos appear to be isothermal
(possibly justified by the statistical theory of violent relaxation \cite{lb}), their halo
is not totally relaxed (in the sense of Lynden-Bell). The same observation is
made for elliptical galaxies.  Models of incomplete violent relaxation are,
unfortunately, difficult to develop \cite{bertin1,bertin2,hjorth}. We note that
the King model can also provide a model of incomplete violent
relaxation \cite{mnras,dubrovnik}. It is important to know if dark matter
halos are collisionless or collisional. The fact that the
density
profiles of dark matter halos decrease as $r^{-3}$ instead of $r^{-4}$ (the
typical outcome of violent relaxation \cite{henonVR,albada,roy,joyce}), and
the presence of black holes at the center of the halos (see below), suggests
that collisions play a certain role in dark matter halos.\footnote{Actually,
these results can also be understood in the case where dark matter is
collisionless. Isolated collisionless self-gravitating systems such as
elliptical galaxies have a density profile decreasing as $r^{-4}$
\cite{bt}. This can be understood as a result of incomplete violent relaxation
\cite{henonVR,albada,roy,joyce,bertin1,bertin2,hjorth}. If dark matter halos are
collisionless we must explain why their density profile decreases as $r^{-3}$
instead of $r^{-4}$. A possibility is that they are subjected to an external
stochastic forcing due to their environment. This stochastic forcing may have an
effect similar to collisions. It may generate a density profile decreasing as
$r^{-3}$. It may also trigger the formation of a central black hole.}

We finally conclude on some speculations concerning the
evolution of dark matter halos, assuming that they are collisional and described
by the King model. For the classical King model,  equilibrium states exist in
MCE only above
a critical energy $E_c$ and for a concentration parameter $k<k_{MCE}$.
Because of collisions and evaporation, the energy $E(t)$ of a self-gravitating system slowly decreases
during its evolution  while its concentration parameter $k(t)$ increases. When $E(t)$
passes below $E_c$, there is no equilibrium state anymore and the system
undergoes a gravitational collapse (gravothermal catastrophe). This corresponds to
a saddle-node bifurcation. As explained previously, large dark matter halos that
are observed in the universe are expected to be close to marginally stable King
distributions with $k\sim k_{MCE}$. However, some halos may have reached the
instability threshold and have undergone gravitational collapse. If the halos
are made of fermions, the collapse stops when their core becomes degenerate as a
consequence of the Pauli exclusion principle.\footnote{As discussed in the Introduction,
gravitational collapse  may also be arrested by the formation of a BEC if dark matter
is made of bosons.} Therefore, complete collapse is
arrested by quantum mechanics. To study the phase
transition between a non-degenerate gaseous sphere and a degenerate compact
object, we can use the fermionic King model. This is the subject of Paper II. It is shown that
 gravitational collapse leads to the formation of a  degenerate compact object (fermion ball)
with a much smaller mass and radius than the original halo, accompanied by the
expulsion of a hot and massive envelope. Indeed, by collapsing, the fermion ball
releases an enormous energy that heats the envelope. As a result, the envelope
is ejected, and dispersed, at very large distances so that, at the end, only the
degenerate nucleus remains.
This process is reminiscent of the formation of red-giants and to the
supernova explosion phenomenon, but it occurs on a cosmological scale and is
considerably much slower (of the order of the Hubble time). This could be a
mechanism\footnote{This is not the only mechanism. Dwarf halos are
thought to result from the Jeans instability of a spatially homogeneous
primordial gas. Then, they merge to form larger structures during hierarchical
clustering. However, it is not impossible that large halos having reached the
point of gravothermal instability collapse again to form smaller structures.} of
formation of dwarf dark matter halos that are completely degenerate.

One important result of our study in Paper II is that large dark matter halos
cannot harbor a
fermion ball, unlike the proposition that has been made in the past \cite{viollier}, because the ``nucleus-halo'' structures that have been considered by these authors are unreachable: they
correspond to  saddle points of entropy at fixed mass and energy. Therefore, it
should not be possible to observe a large dark matter halo with a fermion ball.
This may explain why black holes at the center of galaxies are favored over
fermion balls \cite{nature,reid}. These black holes could be formed by the
mechanism discussed by Balberg {\it et al.} \cite{balberg} if dark matter is
collisional. Because of collisions, the concentration parameter $k(t)$ increases
until the point of gravothermal catastrophe $k_{MCE}$. During the gravothermal
catastrophe, as the central concentration and central temperature increase, the
system undergoes a dynamical (Vlasov) instability of general relativistic origin
and collapses into a black hole. During this process, only the core collapses.
This creates a black hole of large mass\footnote{For weakly interacting
systems such as
globular clusters, the gravothermal catastrophe leads to a singularity (binary
$+$ hot halo) that has an infinite density but zero mass \cite{cohn}. For
strongly collisional systems such as the core of dark matter halos, the
gravothermal
catastrophe leads to a black hole with a large mass \cite{balberg}. This
difference is important to emphasize.} without affecting the structure of the
halo. Therefore, this process leads to large halos compatible with the Burkert
profile for $r>0$ but harboring a central black hole at $r=0$.

However, the fermionic scenario should not be abandoned. Indeed, the structure of dark matter halos crucially depends on their size through the value of the degeneracy parameter $\mu$ as discussed in Paper II. Several configurations are possible making the study of the fermionic King model very rich. The system can be non degenerate (large halos), partially degenerate (intermediate size halos), or completely degenerate (dwarf halos). Therefore, we can have core-halo configurations with a wide diversity of nuclear concentration. This may account for the diversity of dark matter halos observed in the universe.

\appendix

\section{The modified Hubble profile}
\label{sec_mh}

The modified Hubble profile is given by \cite{bt}:
\begin{equation}
\frac{\rho(r)}{\rho_0}=\frac{1}{\left\lbrack 1+\left (\frac{r}{r_0}\right )^2\right\rbrack^{3/2}},
\label{mh1}
\end{equation}
where
\begin{equation}
r_0=\sqrt{\frac{9\sigma^2}{4\pi G\rho_0}},
\label{mh2}
\end{equation}
is the King radius (or core radius) with $\sigma^2=k_B T/m$. The modified Hubble profile provides a good fit of the density profile of the isothermal sphere for $r\le 2r_0$ \cite{bt}. However, the profiles differ at larger distances.  For $r\rightarrow +\infty$, the modified Hubble profile decreases as $r^{-3}$ while the density of the isothermal sphere decreases as $r^{-2}$. The halo radius, defined in Sec. \ref{sec_hr}, is given by
\begin{equation}
r_h=\sqrt{a}r_0,\qquad a=4^{2/3}-1.
\label{mh3}
\end{equation}
Therefore $r_h=1.23 r_0$. The modified Hubble profile can be rewritten as
\begin{equation}
\frac{\rho(r)}{\rho_0}=\frac{1}{(1+a x^2)^{3/2}},\qquad x=\frac{r}{r_h}.
\label{mh4}
\end{equation}
This is a particular case of the family of density profiles defined by Eq. (\ref{empiric}) with $\alpha=3$. The corresponding rotation curve is
\begin{equation}
v_c^2(r)=4\pi G\frac{\rho_0 r_h^3}{r}\left\lbrack \frac{\sinh^{-1}(\sqrt{a}x)}{a^{3/2}}-\frac{x}{a\sqrt{1+a x^2}}\right \rbrack.
\label{mh5}
\end{equation}
After normalization by the circular velocity at the halo radius, we obtain
\begin{equation}
\frac{v_c(r)}{v_c(r_h)}=2.18\left\lbrack
\frac{\sinh^{-1}(\sqrt{a}x)}{\sqrt{a}x}-\frac{1}{\sqrt{1+a x^2}}\right \rbrack^{1/2}.
\label{mh6}
\end{equation}
The normalized density profile $\rho(r)/\rho_0$ and the normalized circular
velocity profile $v_c(r)/v_c(r_h)$ are plotted as a function of the normalized
distance $r/r_h$ in Figs. \ref{densityLOG}, \ref{densityLIN},
\ref{vitesseLOG} and \ref{vitesseLIN}. They are compared to the Burkert
profiles. The halo mass is obtained by integrating Eq. (\ref{mh4}) from zero to
$r_h$. This yields
\begin{equation}
\frac{M_h}{\rho_0 r_h^3}=1.75.
\label{mh7}
\end{equation}
We also have
\begin{equation}
\frac{\sigma^2}{G\rho_0 r_h^2}=\frac{4\pi}{9a}=0.919.
\label{mh8}
\end{equation}
These values can be compared to those obtained in Sec. \ref{sec_dim} for the King model. They are relatively close to those corresponding to the marginal King model ($k_{MCE}=7.44$). Actually, the modified Hubble profile provides a good fit of the marginal King profile up to $5r_h$ for the density and up to $R=38.5 \, r_h$ for the circular velocity.

The density profile (\ref{mh1}) gives rise to a surface density
profile that is similar to the Hubble-Reynolds law fitting the surface
brightness of many elliptical galaxies. This is why it is called the modified
Hubble profile \cite{bt}. This analytic profile was also introduced  empirically
by King \cite{kingempiric} to fit the observed profiles of globular clusters.
For that reason, it is sometimes called the King profile. A few years later,
King \cite{king} developed a more physical model of globular clusters from a
kinetic theory leading to the distribution function (\ref{df1}) generating a one
parameter family of density profiles. To avoid ambiguity, we refer to the
profile (\ref{mh1}) as the modified Hubble profile and we refer to the
one-parameter family of profiles produced by the King distribution (\ref{df1}) 
as the King profiles. As we have seen, the modified Hubble profile provides a
good fit of the King profile at the point of marginal microcanonical stability.

\section{The isochrone cluster}
\label{sec_isochrone}

The density profile and the circular velocity profile of the isochrone cluster can be written as  \cite{bt}:
\begin{equation}
\rho(r)=\frac{M}{4\pi b^3}\frac{2A+1}{(1+A)^2A^3},
\label{isochrone1}
\end{equation}
\begin{equation}
v_c^2(r)=\frac{GM}{b}\frac{A-1}{(A+1)A},
\label{isochrone2}
\end{equation}
where
\begin{equation}
A=\sqrt{1+\left (\frac{r}{b}\right )^2}
\label{isochrone3}
\end{equation}
and $b$ is the core radius. The central density is given by $\rho_0=3M/(16\pi b^3)$. Therefore, we obtain
\begin{equation}
\frac{\rho(r)}{\rho_0}=\frac{4}{3}\frac{2A+1}{(1+A)^2A^3}.
\label{isochrone4}
\end{equation}
The halo radius, defined in Sec. \ref{sec_hr},  is determined by the condition
\begin{equation}
\frac{1}{4}=\frac{4}{3}\frac{2A_h+1}{(1+A_h)^2A_h^3}.
\label{isochrone5}
\end{equation}
We find $A_h=1.50$. Then, we obtain $r_h/b=\sqrt{A_h^2-1}=1.12$. Finally, we can write
\begin{equation}
\frac{r}{r_h}=\sqrt{\frac{A^2-1}{A_h^2-1}}.
\label{isochrone6}
\end{equation}
Equations (\ref{isochrone4}) and (\ref{isochrone6}) determine the normalized density $\rho(r)/\rho_0$ as a function of the normalized
distance $r/r_h$. These equations are parameterized by $A\ge 1$. The normalized circular velocity profile is given by
\begin{equation}
\frac{v_c^2(r)}{v_c^2(r_h)}=\frac{A-1}{(A+1)A}\frac{(A_h+1)A_h}{A_h-1}.
\label{isochrone7}
\end{equation}
Using $M(r)=r v_c^2(r)/G$, we find that the halo mass is given by
\begin{equation}
\frac{M_h}{\rho_0 r_h^3}=\frac{16\pi}{3}\frac{1}{(A_h+1)^2A_h}.
\label{isochrone8}
\end{equation}
Numerically, we obtain  $M_h/\rho_0 r_h^3=1.77$.

The isochrone cluster was introduced by H\'enon \cite{isochrone}
who determined the condition under which the orbital period of a star depends
only on its energy. We note that the density profile of the isochrone cluster
decreases as $r^{-4}$ at large distances like the density profile of many
elliptical galaxies \cite{bt}. We also recall that the isochrone profile
provides a good fit of the King profile for $k\sim 5$ (more precisely $k\sim
3.3$). We can also compare the
isochrone cluster and the empirical density profile arising from the collapse of
a cold uniform sphere \cite{joyce}:
\begin{equation}
\frac{\rho(r)}{\rho_0}=\frac{1}{1+3x^4},\qquad x=\frac{r}{r_h}.
\label{isochrone9}
\end{equation}
They both decay as $r^{-4}$ at large distances. However, close to the center
$\rho(r)-\rho_0$ behaves as $r^2$ and as $r^4$ respectively. Furthermore, the
profile of Eq.
(\ref{isochrone9}) leads to $M_h/\rho_0 r_h^3=2.17$.

\section{Hydrostatic equilibrium}
\label{sec_hydro}

We consider a distribution function of the form $f=f(\epsilon)$ with  $\epsilon=v^2/2+\Phi({\bf r})$.  The local pressure is defined by Eq. (\ref{ene2}). Taking the gradient of this expression, we get
\begin{eqnarray}
\nabla p=\frac{1}{3}\nabla\Phi \int f'(\epsilon) v^2\, d{\bf v}.
\label{hydro1}
\end{eqnarray}
This expression may be rewritten as
\begin{eqnarray}
\nabla p=\frac{1}{3}\nabla\Phi \int \frac{\partial f}{\partial {\bf v}}\cdot {\bf v} \, d{\bf v}.
\label{hydro2}
\end{eqnarray}
Integrating by parts, we obtain
\begin{eqnarray}
\nabla p=\nabla\Phi \int f \, d{\bf v}.
\label{hydro3}
\end{eqnarray}
Using the expression of the local density given by Eq. (\ref{fde3}), the foregoing equation is equivalent to the condition of hydrostatic equilibrium
\begin{eqnarray}
\nabla p+\rho\nabla\Phi={\bf 0}.
\label{hydro4}
\end{eqnarray}
We also recall that a system described by a distribution function of the form $f=f(\epsilon)$ has a barotropic equation of state $p=p(\rho)$ (see Sec. \ref{sec_eosb}). Dividing Eq. (\ref{hydro4}) by $\rho$, taking its divergence, and using the Poisson equation (\ref{fde10b}), we obtain
\begin{eqnarray}
\nabla\cdot\left (\frac{1}{\rho}\nabla p\right )=-4\pi G\rho.
\label{hydro5}
\end{eqnarray}
This is the fundamental equation of hydrostatic equilibrium for a self-gravitating barotropic gas \cite{chandra}.

We now show that Eq. (\ref{fde13}) can be directly obtained from Eq. (\ref{hydro5}). Taking the gradient of Eq. (\ref{ene3}), using the identity (\ref{fde}), and comparing the resulting expression with Eq. (\ref{fde8}), we obtain
\begin{eqnarray}
\nabla p=\frac{1}{\beta}\rho\nabla\chi.
\label{hydro6}
\end{eqnarray}
Substituting this relation in Eq. (\ref{hydro5}), and using Eq. (\ref{fde8}), we get
\begin{eqnarray}
\Delta\chi=-4\pi G A \beta\left (\frac{2}{\beta}\right )^{3/2}I_1(\chi).
\label{hydro7}
\end{eqnarray}
From Eqs. (\ref{fde9}) and (\ref{fde12b}), we have
\begin{eqnarray}
4\pi G A \beta\left (\frac{2}{\beta}\right )^{3/2}I_1(k)=\frac{1}{r_0^2}.
\label{hydro8}
\end{eqnarray}
Therefore, Eq. (\ref{hydro7}) can be rewritten as
\begin{eqnarray}
r_0^2\Delta\chi=-\frac{I_1(\chi)}{I_1(k)}.
\label{hydro9}
\end{eqnarray}
Introducing the variable defined by Eq. (\ref{fde12}), we recover Eq. (\ref{fde13}).

\end{document}